\documentclass[twocolumn,english,aps,pra,longbibliography,superscriptaddress, floatfix]{revtex4-1}
\usepackage[utf8]{inputenc}
\usepackage{xcolor}
\usepackage{amsthm, amssymb}
\usepackage{amsmath}
\usepackage{braket}
\usepackage{txfonts,stmaryrd}
\usepackage{graphicx}
\usepackage{enumerate}
\usepackage[colorlinks=true,urlcolor=blue,citecolor=blue,linkcolor=blue]{hyperref}

\newcommand{\lzop}{\hat{L}_\textrm{z}}
\newcommand{\lyop}{\hat{L}_\textrm{y}}

\usepackage[normalem]{ulem}

\makeatletter
\def\maketitle{
\@author@finish
\title@column\titleblock@produce
\suppressfloats[t]}
\makeatother

\begin{document}

\title{Squeezing and quantum approximate optimization}
\date{\today}
\author{Gopal Chandra Santra}
\email{gopal.santra@kip.uni-heidelberg.de}
\affiliation{Universit\"at Heidelberg, Kirchhoff-Institut f\"ur Physik, Im Neuenheimer Feld 227, 69120 Heidelberg, Germany}
\affiliation{Pitaevskii BEC Center and Department of Physics, University  of  Trento,  Via Sommarive 14, I-38123 Trento, Italy}

\author{Fred Jendrzejewski}
\affiliation{Universit\"at Heidelberg, Kirchhoff-Institut f\"ur Physik, Im Neuenheimer Feld 227, 69120 Heidelberg, Germany}
\affiliation{Alqor UG (haftungsbeschränkt), Marquardstrasse 46, 60489 Frankfurt am Main, Germany}
\author{Philipp Hauke}
\affiliation{Pitaevskii BEC Center and Department of Physics, University of  Trento,  Via Sommarive 14, I-38123 Trento, Italy}
\affiliation{INFN-TIFPA, Trento Institute for Fundamental Physics and Applications, Trento, Italy}
\author{Daniel J. Egger}
\affiliation{IBM Quantum, IBM Research Europe – Zurich, S\"aumerstrasse 4, CH-8803 R\"uschlikon, Switzerland}
\begin{abstract}
    Variational quantum algorithms offer fascinating prospects for the solution of combinatorial optimization problems using digital quantum computers. However, the achievable performance in such algorithms and the role of quantum correlations therein remain unclear. 
    Here, we shed light on this open issue by establishing a tight connection to the seemingly unrelated field of quantum metrology: 
    Metrological applications employ quantum states of spin-ensembles with a reduced variance to achieve an increased sensitivity, and we cast the generation of such squeezed states in the form of finding optimal solutions to a combinatorial MaxCut problem with an increased precision. 
    By solving this optimization problem with a quantum approximate optimization algorithm (QAOA), we show numerically as well as on an IBM Quantum chip how highly squeezed states are generated in a systematic procedure that can be adapted to a wide variety of quantum machines. 
    Moreover, squeezing tailored for the QAOA of the MaxCut permits us to propose a figure of merit for future hardware benchmarks.
    
\end{abstract}

\maketitle

The Quantum Approximate Optimization Algorithm (QAOA)~\cite{Farhi2014} is a promising approach for solving combinatorial optimization problems using digital quantum computers~\cite{torta2021quantum,headley2020approximating}.
In this framework, combinatorial problems such as the MaxCut and MAX-SAT are mapped to the task of finding the ground state of an Ising Hamiltonian~\cite{Lucas2014, Liang2020, Lee2021}. 
QAOA uses constructive interference to find solution states~\cite{Streif2019}, and it has better performance than finite-time adiabatic evolution~\cite{Wurtz2022}.
However, it remains an outstanding challenge to characterize the role of quantum effects such as entanglement in QAOA.

Here, we show how concepts from quantum metrology shed light onto the influence of squeezing and entanglement in the performance of QAOA.
Illustratively, the connection is established through the insight that (a) the aim of QAOA is to obtain the ground state as precisely as possible, while (b) quantum metrology leverages entanglement between particles to generate states that permit precision beyond the capacities of any comparable classical sensor~\cite{Pezze2009, Pezze2018,degen2017quantum}.
For example, squeezed states can increase sensitivity for detecting phases~\cite{Gross2010}, magnetic fields~\cite{Sewell2012}, and gravitational waves~\cite{barsotti2018squeezed}.
The most sensitive states for phase estimation are Dicke states~\cite{Dicke1954, Pezze2018}, where all qubits are equally entangled.
We substantiate this connection through numerically exact calculations and data gathered on IBM Quantum hardware with up to eight qubits.  
Our analysis shows how the search for an optimal solution to the MaxCut problem on a complete graph through QAOA generates a Dicke state, with squeezing and multipartite entanglement. 
Based on this, we propose the amount of squeezing generated as an application-tailored performance benchmark of QAOA.
Our work thus further strengthens the intimate links between quantum metrology and quantum information processing~\cite{Giovannetti2006,Omran2019,Marciniak2022,arrazola2021quantum}.

In the rest of this paper, we first formally connect the QAOA to the generation of entangled squeezed states, which we then numerically illustrate.
Next, we develop a benchmark tailored for QAOA based on squeezing.
Finally, we assess the ability of superconducting qubits to run QAOA on fully connected problems while simultaneously creating Dicke states and estimate the number of entangled particles.

\textit{Combinatorial optimization on quantum computers.} 
Universal quantum computers can address hard classical problems such as quadratic unconstrained binary optimization (QUBO), which is defined through
\begin{align}\label{eqn:qubo}
    \min_{x\in\{0,1\}^n}x^T\Sigma x\quad\text{with}~\Sigma\in \mathbb{R}^{n\times n}.
\end{align}
In QAOA, the binary variables $x_i$ are mapped to qubits through $x_i=(1-z_i)/2\rightarrow(1-\hat{Z}_i)/2$, where $\hat{Z}_i$ is a Pauli spin operator with eigenstates $\ket{0}$ and $\ket{1}$. The result is an Ising Hamiltonian $\hat{H}_C$ whose ground state is the solution to Eq.~(\ref{eqn:qubo})~\cite{Lucas2014}.
The standard QAOA then applies $p$ layers of the unitaries $\exp(-i\beta_k \hat{H}_M)\exp(-i\gamma_k \hat{H}_C)$, with $k=1,...,p$, to the ground state of a mixer Hamiltonian $\hat{H}_M$, such as $-\sum_i\hat{X}_i$ where $\hat{X}_i$ is a Pauli operator, to create the trial state $\ket{\psi(\boldsymbol\beta,\boldsymbol\gamma)}$.
A classical optimizer seeks the $\boldsymbol\beta=(\beta_1,\dots,\beta_p)$ and $\boldsymbol\gamma=(\gamma_1,\dots,\gamma_p)$ that minimize the energy $\braket{\psi(\boldsymbol\beta,\boldsymbol\gamma)|\hat{H}_C|\psi(\boldsymbol\beta,\boldsymbol\gamma)}$, which is measured in the quantum processor.

\begin{figure*}[htbp!]
    \centering
    \includegraphics[width=1\textwidth, clip, trim= 0 0 30 0]{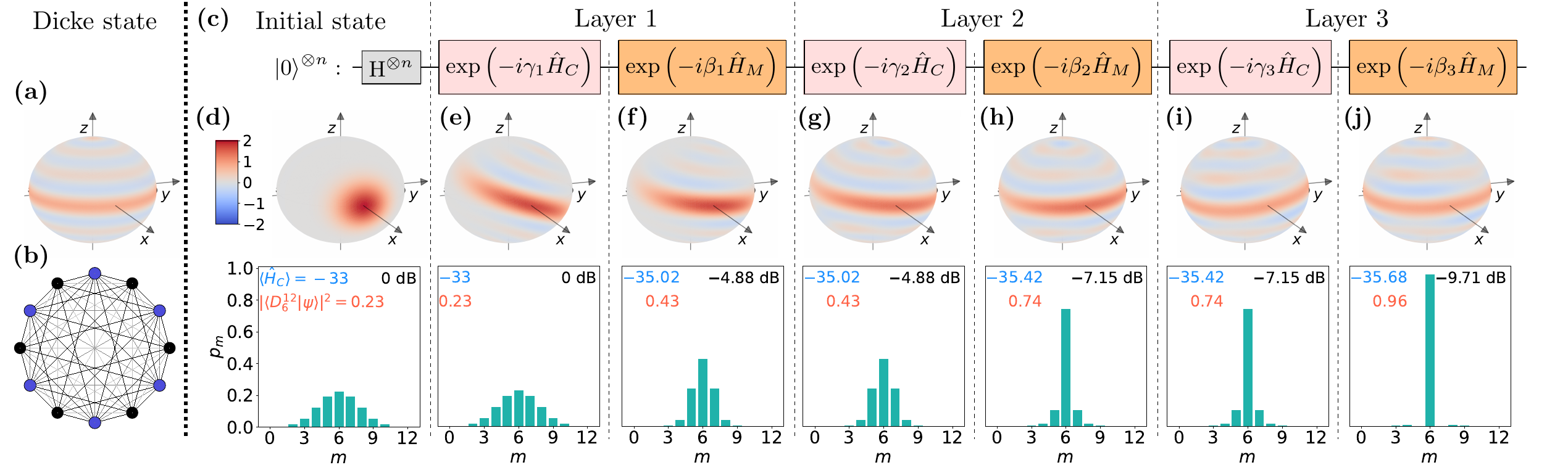}
    \caption{Metrologically useful squeezing generated by a depth-three QAOA for the MaxCut problem in a complete graph of 12 nodes. 
    (a) Wigner quasi-probability distribution of the symmetric Dicke state $D^{12}_6$, an idealized example of a squeezed state and the target for our QAOA.
    (b) Fully connected unweighted graph with the nodes and edges colored according to one of the 924 possible maximum cut configurations. 
    (c) Circuit representation of QAOA with alternating application of cost-function and mixer Hamiltonian.
    The Bloch spheres and histograms from (d) to (j) show the state after the corresponding gate in the optimized QAOA circuit with $(\gamma_i,\beta_i)$ given by $(0.199, 0.127)$, $(0.306, 0.087)$, and $(4.592, 1.518)$ for $i=1$, $2$, and $3$, respectively. 
    Negativity in the Wigner distribution indicates that the states are non-Gaussian~\cite{gross2007non}.
    The squeezing (in black), energy expectation $\langle \hat{H}_C \rangle$ (in blue), and overlap probability density $|\langle D_6^{12}|\psi\rangle|^2$ with the target Dicke state (in orange) are shown inside each histograms.}
    \label{fig:stepQAOA}
\end{figure*}

\textit{Squeezing and quantum combinatorial optimization\label{sec:qaoa}.} 
Squeezed states are entangled states with a reduced variance of one observable at the cost of an increased variance in non-commuting observables. 
A large body of experimental work exists addressing the generation of squeezing in various platforms ~\cite{esteve2008squeezing,purdy2013strong,Strobel2014,muessel2015twist,Xu2022,Marciniak2022}.
Squeezing can also detect multipartite entanglement among qubits ~\cite{sorensen2001many,Korbicz2005,korbicz2006,guhne2009entanglement,toth2007detection}.  

In our setting, we are interested in squeezing within an ensemble of $n$ qubits (whose symmetric subspace can be seen as a qudit with length $\ell=\frac{n}{2}$). 
Consider a coherent state, such as the collective superposition $\ket{+}^{\otimes n}$, where $\ket{+}=(\ket{0}+\ket{1})/\sqrt{2}$. This state has a variance of $\sigma^2_\text{css}=\frac{n}{4}$, commonly called the shot-noise, in the collective observable $\lzop=\frac{1}{2}\sum_{i=1}^{n}\hat{Z}_i$. 
By evolving $\ket{+}^{\otimes n}$, e.g., under the non-linear one-axis-twisting (OAT) operator $\lzop^2 =\frac{1}{4}(n+ \sum_{i\neq j} \hat{Z}_i\hat{Z}_j)$, the state is stretched over the collective Bloch sphere. The direction with reduced variance can be transferred to the $z$ coordinate by rotating the state around the $x$-axis with $\hat{L}_x=\frac{1}{2}\sum_{i=1}^n\hat{X}_i$~\cite{kitagawa1993squeezed,Wang2001,Strobel2014}. 
The resulting $n$ particle state is called number squeezed along $z$ when the observed variance is below $\sigma^2_\text{css}$, i.e., if the squeezing parameter
\begin{align} \label{eq: squeezing}
 \mathcal{S}~[{\rm dB}]=10\log_{10}\left(\frac{\text{Var}(\lzop)}{\sigma^2_\text{css}}\right)
\end{align}
is negative~\cite{kitagawa1993squeezed,ma2011quantum}.

In a quantum circuit representation, it becomes apparent that these steps coincide with a single-layer QAOA sequence, see Fig.~\ref{fig:stepQAOA}(c): 
(i) The application of Hadamard gates to $\ket{0}^{\otimes n}$ initialize the system in $\ket{+}^{\otimes n}$, the ground state of the mixer Hamiltonian $\hat{H}_M$. 
(ii) The evolution under the OAT operator corresponds to applying the unitary $\exp(-i\gamma_1 \hat H_C)$ with the cost function $\hat H_C\propto\lzop^2$. On the qubit level, this corresponds to controlled-z gates generated by $\hat{Z}_i\hat{Z}_j$ between all qubits $i$ and $j$.
(iii) The rotation around the $x$-axis to reveal the squeezing corresponds to the unitary $\exp(-i\beta_1 \hat{H}_M)$, i.e., an application of the mixer.

The above cost function $\hat H_C\propto\lzop^2$ is a special instance of the MaxCut problem.  
MaxCut aims at bipartitioning the set of nodes $V$ in a graph $G(V,E)$ such that the sum of the weights $\omega_{i,j}$ of the edges $(i,j)\in E$ traversed by the cut is maximum, i.e.
\begin{align}
    \max_{z\in\{-1,1\}^n}\frac{1}{2}\sum_{(i,j)\in E}\omega_{i,j}(1-z_iz_j).
\end{align}
Consider the problem instance with $\omega_{i,j}= 1,\forall (i,j)$, i.e., an unweighted fully connected graph labelled $\mathcal{G}_n$, see Fig.~\ref{fig:stepQAOA}(b). 
Dividing $V$ into two sets of 
size as equal as possible creates a maximum cut.
For even $n$, the set of all maximum cuts corresponds to the symmetric Dicke state~\cite{Dicke1954}
\begin{align}\label{dicke}
    D_k^{n}={\binom{n}{k}}^{-1/2}\sum_i P_i\left(\ket{1}^{\otimes k}\otimes\ket{0}^{\otimes (n-k)}\right)\,,
\end{align}
with $k=\frac{n}{2}$.
Here, $P_i(\cdot)$ denotes a permutation of all states with $k$ particles in $\ket{1}$ and $n-k$ particles in $\ket{0}$.
For odd $n$, the set of all maximum cuts corresponds to $(D^n_{\lfloor n/2\rfloor}+D^n_{\lceil n/2\rceil})/\sqrt{2}$.
These states are maximally squeezed along $z$ and are the most useful for metrological applications~\cite{Pezze2018}.
The QAOA cost function Hamiltonian to minimize in this problem is $\hat{H}_C=\frac{1}{2}\sum_{i<j}(\hat{Z}_i\hat{Z}_j-1)= \hat{L}_z^2 -\frac{n^2}{4}\mathbb{I}$.
Therefore, QAOA tasked to find the maximum cut of a fully connected unweighted graph will maximize the squeezing.
This relation thus translates analog metrological protocols~\cite{Strobel2014} to a digital quantum processor. 
In addition, by formulating the constraints that an arbitrary Dicke state $D_k^n$ imposes on the spins as a QUBO, we can generate $D_k^n$ for arbitrary $k$~\cite{NoteXmarked}.

\textit{QAOA as generator of squeezing.}
To illustrate this connection between QAOA and squeezing, we numerically simulate a system with $n=12$ qubits and follow the usual QAOA protocol, using $\hat{H}_C=\hat{L}_z^2-\frac{n^2}{4}\mathbb{I}$, $\hat{H}_M=-2\hat{L}_x$, and $p=3$~\cite{NoteXmarked}. 
We depict the generated collective qudit state using the Wigner quasi-probability distribution as well as the probability distribution over the spin eigenvalues $\{m=\langle \hat L_z \rangle +\frac{n}{2}\}$ at each step, see Fig.~\ref{fig:stepQAOA}(d)-(j).
Each application of $\hat{H}_C$ stretches the Wigner distribution making it resemble an ellipse with the major axis tilted with respect to the equatorial plane of the qudit Bloch sphere. 
As $[\hat{H}_C,\hat{L}_z]=0$, this operation does not alter the distribution of $\langle \hat{L}_z \rangle$.
Next, the mixer operator rotates the Wigner distribution back towards the equator, thereby transferring the squeezing to the operator $\hat{L}_z$.
After three layers, the final state has an overlap with the symmetric Dicke state of $96\%$ and the squeezing number reaches $\mathcal{S}=-9.71~{\rm dB}$.
Crucially, noiseless QAOA with less layers produces less squeezing, e.g., see depth-one QAOA~\cite{NoteXmarked}.

The squeezing in collective spin observables can further be related to entanglement. We employ three different criteria~\cite{NoteXmarked}: 
(E1) If $\frac{n}{4}> \langle \lzop^2\rangle$, the state violates a bound on separable states~\cite{toth2007detection}.  
Any squeezed state ($\text{Var}(\lzop)<\frac{n}{4}$) at the equator ($\langle \lzop\rangle=0$) is witnessed as entangled by this criterion.  
Here, $\langle \lzop^2\rangle=0.32<\frac{n}{4}=3$, which is close to the minimal value of $0$ achieved by the Dicke state. 
(E2) If the quantum Fisher information for a pure state $\psi$, $F_Q[\psi,\mathcal{O}]=4 \text{Var}(\mathcal{O})$, is larger than $(sk^2+r^2)$, where $s =\lfloor n/k \rfloor$ denotes the integer division of $n$ by $k$, and $r$ is the remainder, at least $(k+1)$ particles are entangled~\cite{hyllus2012fisher,toth2012multipartite}. 
Here, $F_Q[\psi,\lyop]=84.48$ and the final state has multipartite entanglement between at least $9$ out of $12$ particles. 
(E3) Specifically for Gaussian states, one can approximately estimate the number of entangled particles $k$ assuming the identity $F_Q/n\simeq\sigma^2_{\mathrm{css}}/\text{Var}(\lzop)$~\cite{Strobel2014,NoteXmarked}, which here yields $k=11$. We will use this estimate below in the hardware results where direct access to $F_Q$ is not possible. 
With $k+1$-partite entangled states the variance of a phase $\theta$ measured $m$ times satisfies $\Delta^2\theta\geq 1/(mkn)$~\cite{toth2012multipartite}.
QAOA-generated Dicke states take more time to prepare than coherent states but have more entanglement.
Even for the disadvantageous geometry of a linear chain, where entanglement needs to be linearly transported, and very moderate values of $k=2$, one can estimate the QAOA-generated states in current superconducting qubits to achieve a better $\Delta^2\theta$ in less time than coherent states for as many as 60 qubits, and this number rises exponentially with improved $k$.
For hardware with slow repetition rates that natively implement the OAT operator, such as Bose-condensed cold-atoms~\cite{Strobel2014}, it is always advantageous to use QAOA-generated states~\cite{NoteXmarked}.

\textit{Warm-start with squeezed states.}
The QAOA circuit that generates squeezed states can be reused as a circuit to create initial states for QAOAs designed to tackle graphs obtained from perturbations of the fully-connect unweighted graph.
Such states lower the number of optimization parameters needed, making it simpler for the classical optimizer to handle~\cite{NoteXmarked}.

\textit{QAOA-tailored hardware benchmarks.} 
The performance of quantum computing hardware is often measured by metrics such as randomized benchmarking~\cite{Magesan2011, Magesan2012b, Corcoles2013} and quantum process tomograph~\cite{OBrien2004, Bialczak2010}, which focus on gates acting on typically one to two qubits, while Quantum Volume (QV) is designed to measure the performance of a quantum computer as a whole~\cite{Cross2019, Jurcevic2021, Pelofske2022}.
For certain applications, these are complemented by specifically designed benchmarks, e.g., for quantum chemistry~\cite{Arute2020}, generative modelling~\cite{Benedetti2019, DallaireDemers2018}, variational quantum factoring~\cite{Karamlou2021}, Fermi-Hubbard models~\cite{DallaireDemers2020}, and spin Hamiltonians~\cite{Schmoll2017}.

It is of particular interest to identify such application-tailored benchmarks also for variational algorithms, as these employ highly structured circuits. This necessity is well illustrated by considering the QV: the circuit complexity of a $2^n$ QV system is equivalent to a $p=2$ QAOA running on $n$ linearly connected qubits~\cite{NoteXmarked}.
As this shows, QV fails to properly capture the dependency on $p$ as QAOA circuits on complete graphs are deeper than their width.
As Ref.~\cite{Franca2021} shows using entropic inequalities, if the circuit is too deep a classical computer can sample in polynomial time from a Gibbs state while achieving the same energy as the noisy quantum computer. That bound is based on the fidelity of layers of gates, which is, however, often overestimated when built from fidelities of gates benchmarked in isolation, e.g., due to cross-talk~\cite{Weidenfeller2022}.

Since the solution to the MaxCut problem on the fully connected unweighted graph $\mathcal{G}_n$ is known, we propose squeezing as a good hardware benchmark for QAOA to complement other performance metrics. 
From a hardware perspective, although $\mathcal{G}_n$ is a specific problem, its QAOA circuit is representative of the noise of an arbitrary fully-connected QUBO problem since the gates constituting a generic cost function $\exp(-i\gamma_k\hat{H}_{C})$ can be implemented with virtual Z-rotations and CNOT gates~\cite{McKay2017}.
The duration of the QAOA pulse-schedule and the absolute amplitude of the pulses are thus independent of the variables $\Sigma$ in the QUBO, see Eq.~(\ref{eqn:qubo}), and the variational parameters $\boldsymbol\gamma$ and $\boldsymbol\beta$.
Therefore, the pulse-schedule of the squeezing circuit will capture the same amount of noise, such as decoherence, unitary gate errors, and cross-talk, as the QAOA circuit for an arbitrary QUBO~\cite{NoteXmarked}.

For our proposed benchmark, we first label the quantum numbers of $\hat{L}_z+\frac{n}{2}$ by $m\in\{0, 1,...,2\ell\}$, which correspond to cuts of size $c(m)=m(n-m)$ on $\mathcal{G}_n$.
We relate squeezing to a QAOA performance metric through the question:
\textit{Given the squeezing $\mathcal{S}$ in the trial state, what is the probability $P_\alpha(n, \mathcal{S})$ of sampling a cut with size $c(m)$ greater than a given $\alpha$-fraction of the maximum cut size $c_\text{max}=n^2/4$?} 
Here, $\alpha$ can be seen as an approximation ratio.
By definition, cuts with $c(m)>\alpha c_\text{max}$ must satisfy $m_-(\alpha)<m<m_+(\alpha)$ for even $n$, where $m_\pm(\alpha)=\frac{n}{2}(1\pm\sqrt{1-\alpha})$.
Under a QAOA trial state $\ket{\psi(\boldsymbol{\beta}, \boldsymbol{\gamma})}$ with a distribution $p_m$ over $m$, see Fig.~\ref{fig:colorplot}(a), the probability to sample cuts larger than $\alpha c_\text{max}$ is thus
\begin{equation}\label{eq:Palpha}
 P_\alpha(n)=\sum_{m=\lceil m_-(\alpha) \rceil}^{\lfloor m_+(\alpha) \rfloor}p_m.
  \end{equation}
We now make the simplifying assumption that the distribution $p_m$ is a Gaussian $\mathcal{N}(\frac{n}{2}, \sigma)$ , where the standard deviation $\sigma$---the only free variable for fixed $n$---is, by definition, in one-to-one correspondence to the squeezing $\mathcal{S}=10 \log_{10}( 4\sigma^2 /n)$.
In summary, the benchmark (i) relates squeezing to the probability of sampling good solutions, a QAOA performance metric, (ii) captures the ability of QAOA to create entangled states, and (iii) is as susceptible to hardware noise as other fully connected QAOA circuits.
\begin{figure}[hbt!]
    \centering
    \includegraphics[width=\columnwidth]{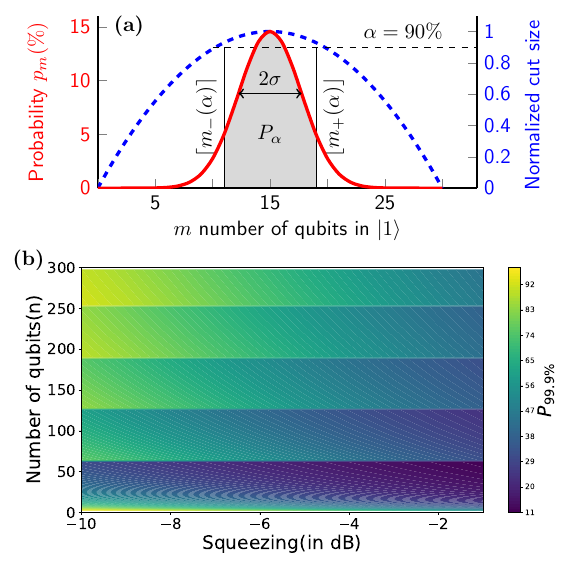}
    \caption{Benchmarking QAOA with squeezing.
    (a) Probability distribution ($p_m$, red solid line) and normalized cut size ($c_m/c_{max}$, blue dashed line) simultaneously plotted against $m=\langle \lzop \rangle+\frac{n}{2}$ for $n=30$. States with normalized cut-size more than $\alpha$ lie in  $m \in [\lceil m_-(\alpha) \rceil, \lfloor m_+(\alpha) \rfloor]$. These yield the shaded area under the probability $p_m$, which is the figure of merit $P_\alpha$ defined in Eq.~(\ref{eq:Palpha}). 
    (b) $P_\alpha$ showing how the probability of sampling high value cuts changes with the squeezing $\mathcal{S}$ and the number of qubits $n$, calculated using trial Gaussian distributions.}
    \label{fig:colorplot}
\end{figure}

We illustrate the benchmark by numerically computing $P_\alpha(n,\mathcal{S})$ as a function of $n$ and the squeezing $\mathcal{S}$ in the Gaussian distribution $p_m(\mathcal{S})$.
Since the ground state of $\mathcal{G}_n$ is highly degenerate, we select a high value of $\alpha$, e.g., $99.9\%$.
At fixed $n$, an increased squeezing (more negative) increases $P_\alpha$,
see Fig.~\ref{fig:colorplot}(b), as cuts with a larger size receive more weight.
In addition, $P_{\alpha}$ has discontinuous jumps at $n_\text{dis.}$, where $z=\lfloor \frac{ n_\text{dis.}}{2}\sqrt{1-\alpha} \rfloor \in \mathbb{Z}^+$~\cite{NoteXmarked}. 
In between discontinuities, $P_\alpha$ diminishes with increasing $n$ because $\sigma$ increases $\propto \sqrt{n}$ for fixed $\mathcal{S}$, which reduces the weight attributed to high value cuts.

\textit{Benchmarking superconducting qubits\label{sec:sc_qubits}.} 
We now evaluate the benchmark on gate-based superconducting transmon qubits~\cite{Krantz2019}.
\begin{figure}
    \centering
    \includegraphics[width=\columnwidth]{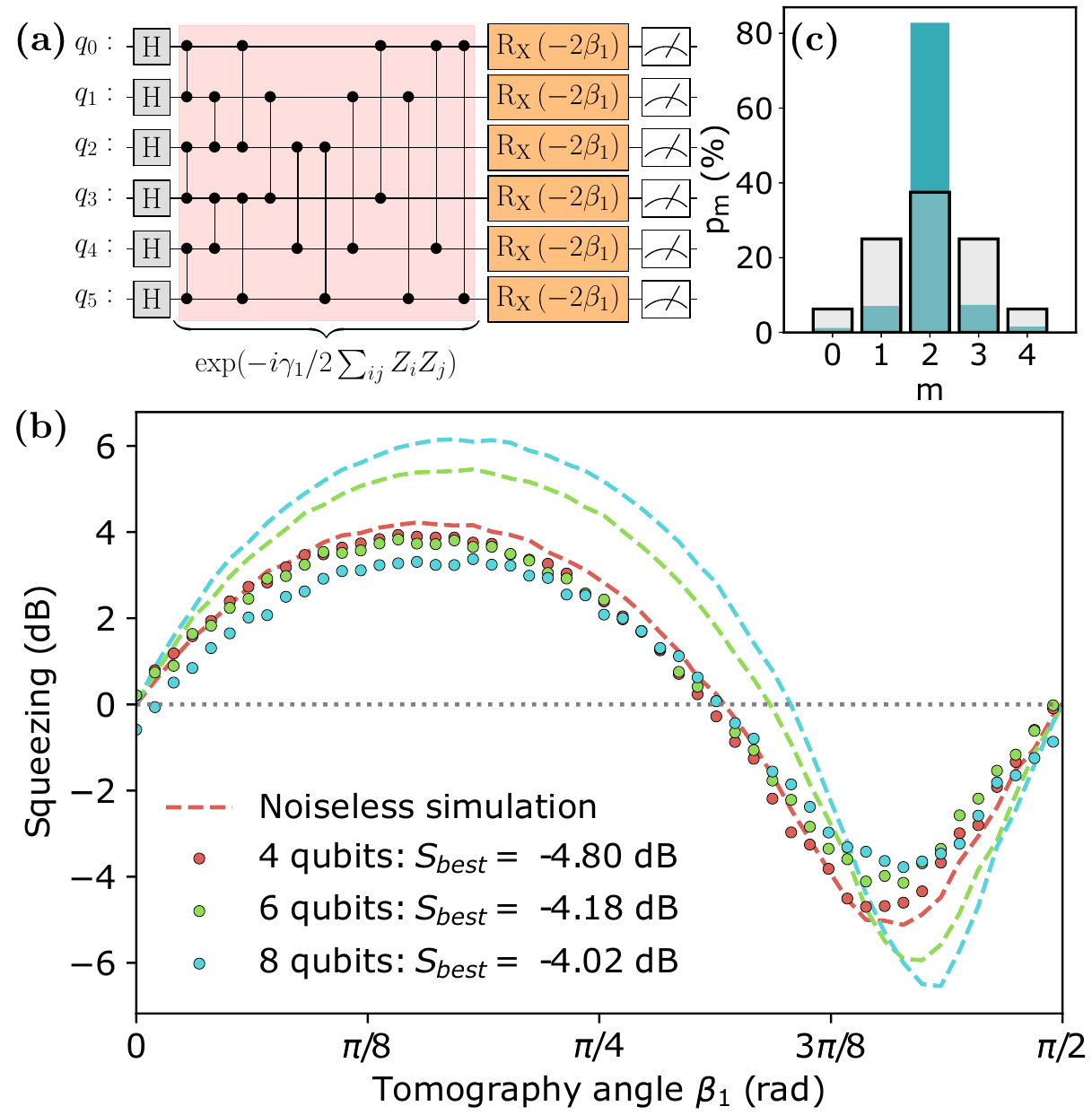}
    \caption{Squeezed states generated on superconducting qubits.
    (a) A quantum circuit implementing a single QAOA-layer for the MaxCut problem on a six-qubit system, producing a state with a reduced variance along the $z$-axis.
    (b) Squeezing measured with the quantum circuit in (a).
    The dashed lines show a noiseless Qasm simulation.
    (c) The $p_m$ distribution of the four-qubit state with $-5.96~{\rm dB}$ squeezing generated by a depth-two QAOA with optimal parameters $\gamma_1= 0.918$, $\gamma_2= -0.257$, $\beta_1 = -0.711$, and $\beta_2=-2.175$.
    The gray histogram shows the state $(H\ket{0})^{\otimes n}$.     }
    \label{fig:measured_squeezing}
\end{figure}
We measure the squeezing on the IBM Quantum system \emph{ibmq\_mumbai} using Qiskit~\cite{Qiskit} for four, six, and eight qubits~\cite{NoteXmarked}.
Since the chosen qubits have a linear connectivity, we use a line swap strategy~\cite{Jin2021,Weidenfeller2022} to create the all-to-all qubit connectivity required by the squeezing circuit, shown in Fig.~\ref{fig:measured_squeezing}(a) for $p=1$. 
This circuit is then transpiled to the cross-resonance-based hardware~\cite{Sheldon2016, Sundaresan2020} employing a pulse-efficient strategy instead of a CNOT decomposition~\cite{Earnest2021} using Qiskit Pulse~\cite{Alexander2020}.
The optimal value of the variational parameter $\gamma$  is found with a noiseless simulation for each $n$.
We use readout error mitigation~\cite{Bravyi2020, Barron2020a}, which on average improves the best measured squeezing by $-0.7\pm 0.1~{\rm dB}$ averaged over all three $n\in\{4,6,8\}$.
At depth one, a sweep of the tomography angle $\beta_1$ reveals a squeezing of $-4.80$, $-4.18$, and $-4.02~{\rm dB}$
whereas noiseless simulations 
reach $-5.14$, $-5.90$, and $-6.56~{\rm dB}$ for $n=4, 6$, and $8$, respectively. These metrological gains are comparable to prior works in trapped ions~\cite{sackett2000experimental,meyer2001experimental,leibfried2003experimental,leibfried2004toward,leibfried2005creation,monz2011}. Given the measured squeezing, we compute a $P_{99.9\%}(n, \mathcal{S})$ of $61.5\%$, $49.1\%$, and $42.6\%$, respectively.
Furthermore, we run a depth-two QAOA on the fully connected four-qubit graph to create a state with a $-5.96~{\rm dB}$ squeezing, see Fig.~\ref{fig:measured_squeezing}(c), which results in $P_{99.9\%}(4, -5.96)=68.2\%$.
These results indicate that the potential to generate squeezing in a four qubit system is limited by the variational form at depth one.
By contrast,  in systems with six and eight qubits, the squeezing generated in practice is limited by the large number of CNOT gates at depth one (40 and 77, respectively).
The criterion (E1) witnesses the generated states in both simulation and hardware as entangled, see Fig.~\ref{fig:entanglement}(a). 
In noiseless simulation of a depth-one QAOA of system sizes $n=4,6,8$, criterion (E2) witnesses at least $4,4,5$ qubit entanglement, respectively. 
In the noisy hardware implementation, estimate (E3) suggests these numbers to still reach $4,3,3$, respectively. 
\begin{figure}[hbt!]
    \centering
    \includegraphics[width=\columnwidth]{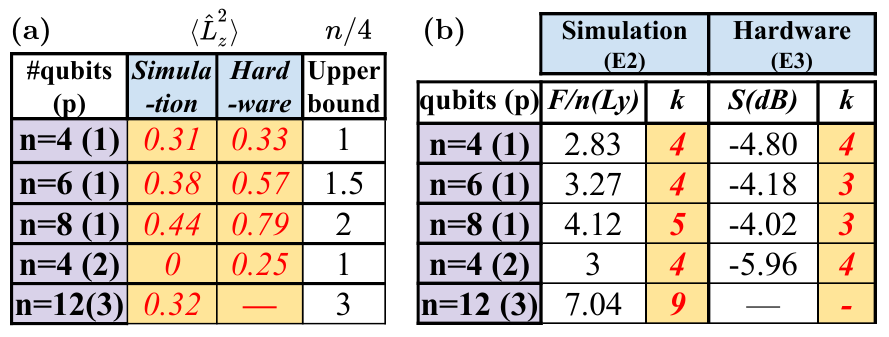}
    \caption{Entanglement from squeezing and quantum Fisher information.  
    (a) The values of $\langle \hat{L}_z^2 \rangle$ (E1) obtained in simulation and hardware are close to 0 indicating that the states are in the vicinity of entangled Dicke states. 
    (b) Number of entangled particles $k$ calculated from $F_Q[\hat{L}_y]$ (E2) in simulation, and estimated for hardware using (E3). }
    \label{fig:entanglement}
\end{figure} 

\textit{Conclusion.}
In summary, the generation of squeezed states that are useful for metrology can be cast as a MaxCut problem, which in turn can be addressed with variational algorithms. The procedure that we illustrated in the creation of a 12 qubit Dicke state can be implemented on universal quantum computing platforms, such as superconducting qubits or trapped ions, as well as on special purpose machines such as BECs trapped in optical tweezers~\cite{Strobel2014}.
Interestingly, an enhancement of squeezing within the multilayer QAOA protocol is not equivalent to simply applying the $\hat{L}_z^2$ operator for a longer period, as the mixer Hamiltonian periodically intervenes~\cite{NoteXmarked}. 
Thus, our results show how variational algorithms may generalize existing protocols and provide systematic guidance for the creation of highly squeezed states for metrology.
By contrast to, e.g., Ref~\cite{Marciniak2022,kaubruegger2019variational,koczor2020variational,meyer2021variational} which uses variational quantum algorithms with a hardware native ansatz to enhance phase sensitivity, the QAOA approach to create squeezing encapsulates the structure of the target state in the variational form which may reduce the number of parameters to optimize.
In a similar vein, we foresee that custom states beyond Dicke states may be generated by QAOA if they can be cast as solutions of a combinatorial optimization problem.
In addition, we suggested squeezing as a QAOA specific hardware benchmark.
This benchmark is both portable across hardware platforms and captures hardware-specific properties such as limited qubit connectivity and cross-talk.

\textit{Acknowledgements.}
FJ and DJE are thankful to the organisers of the Qiskit Unconference in Finland and fruitful discussions with Arianne Meijer as well as Matteo Paris that started this work.
The authors acknowledge use of the IBM Quantum devices for this work.
GCS, FJ, and PH acknowledge support by the Bundesministerium für Wirtschaft und Energie through the project ``EnerQuant” (Project- ID 03EI1025C). 
FJ acknowledges the DFG support through the Emmy-Noether grant (Project-ID 377616843). This work is 
supported by the DFG Collaborative Research Centre “SFB 1225 (ISOQUANT), by the Bundesministerium für Bildung und Forschung through the project ``HFAK" (Project- ID 13N15632).
PH acknowledges support by the ERC Starting Grant StrEnQTh (project ID 804305), Provincia Autonoma di Trento, and by Q@TN, the joint lab between University of Trento, FBK-Fondazione Bruno Kessler, INFN-National Institute for Nuclear Physics and CNR-National Research Council. 

%

\clearpage

\title{Supplemental Material: Squeezing and quantum approximate optimization}
\date{\today}
\maketitle
\def\theequation{S\arabic{equation}}
\def\thefigure{S\arabic{figure}}
In this Supplemental Material, we (i) discuss the details of the optimization method used to obtain the parameters $\{\boldsymbol\gamma,\boldsymbol\beta\}$, (ii) describe why squeezed states are entangled, (iii) define and connect multipartite entanglement to quantum Fisher information and squeezing, (iv)
discuss the practical advantages of using QAOA for metrology in different hardware architectures, (v) argue why squeezing can be a better benchmark than quantum volume in QAOA, (vi) explain the discontinuities observed in the benchmark in Fig.2(b), (vii) give details of the \emph{ibmq\_mumbai}, (viii) discuss the advantages of using multiple-layers of QAOA, (ix) explain why increasing the duration of $\hat{H}_C$ is not helpful compared to the alternating layers in QAOA, (x) discuss the creation of arbitrary Dicke states, and (xi) show how to initialize QAOA with squeezed states to lower the number of optimization parameter.
\section{Optimization method}
The 12 qubit example in the main text is run with the Qiskit QAOA Runtime program~\cite{Weidenfeller2022}.
To optimize the $\{\boldsymbol\gamma,\boldsymbol\beta\}$ we use the simultaneous perturbation stochastic approximation (SPSA) algorithm~\cite{spall1998overview} which simultaneously optimizes multiple parameters and can handle noisy environments.
We do not initialize the optimizer with values for the learning rate and a perturbation. 
Instead, we let SPSA calibrate itself in the first 25 iterations.
To obtain good solutions we allow SPSA a maximum of $500$ iterations and gather a total of $2^{15}$ shots per iteration.

\section{Entanglement from squeezing}
Measurements of collective spin observables can reveal entanglement. In particular, separable states satisfy~\cite{toth2007detection}
\begin{align}
    \langle \hat{L}_x^2 \rangle +\langle \hat{L}_y^2 \rangle \leq \frac{n}{2}\bigg(\frac{n}{2}+\frac{1}{2}\bigg) \Longleftrightarrow  \frac{n}{4}\leq \langle \lzop^2 \rangle = \text{Var}(\lzop)  \,.
\end{align}
The second implication is reached using the identity $\langle \hat{L}^2 \rangle = \langle \hat{L}_x^2 \rangle + \langle \hat{L}_y^2 \rangle+\langle \hat{L}_z^2 \rangle=\frac{n}{2}(\frac{n}{2}+1)$, and $\langle \lzop \rangle=0$ for our target states. 
Any squeezed state defined through Eq.~(\ref{eq: squeezing}) violates the relation above and is thus entangled. 
Moreover, $\langle \hat{L}_x^2 \rangle +\langle \hat{L}_y^2 \rangle$ reaches the maximum $\frac{n}{2} (\frac{n}{2}+1)$ in the Dicke state~\cite{toth2007detection}, which is the same as having $\text{Var}(\lzop)=0$. 
In Fig.~\ref{fig:entanglement}(a), we show how the obtained values of $\langle \lzop^2 \rangle$ are close to the minimum limit $0$, revealing the existence of significant entanglement.

\section{Multipartite entanglement, quantum Fisher information, and squeezing}

\begin{figure}[hbt!]
    \centering
    \includegraphics[width=1\columnwidth]{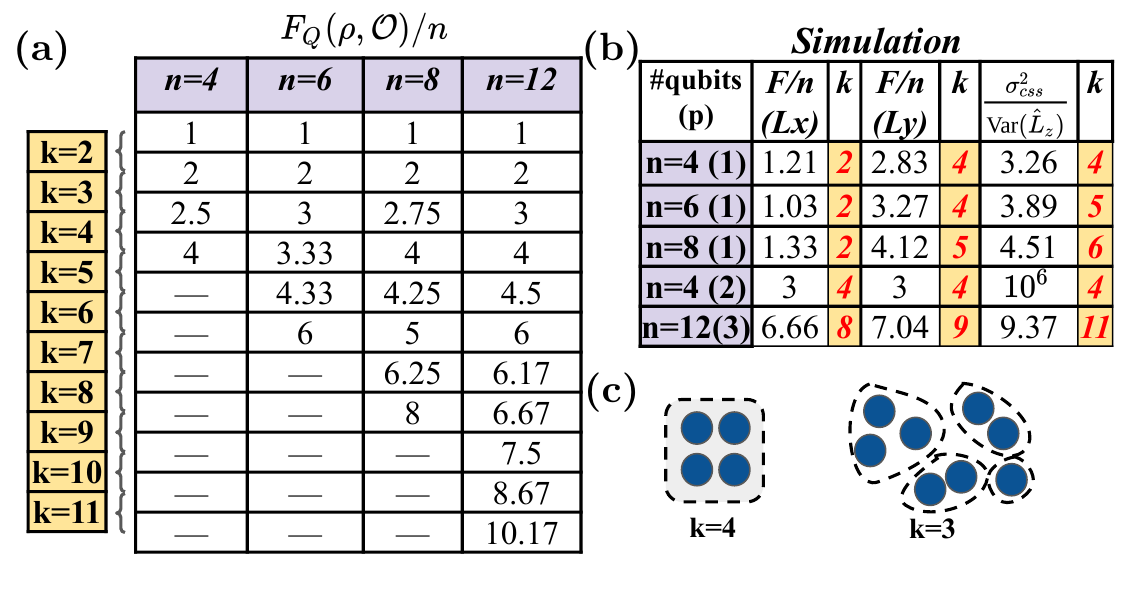}
    \caption{Multipartite entanglement from quantum Fisher information and number squeezing. (a) $F_Q$ witnessing $k$-partite entanglement for different $n$, (b) In the simulations, $F_Q$ obtained with $\hat{L}_y$ is larger than $\hat{L}_x$. The numbers of entangled particles ($k$) estimated from squeezing ($\sigma^2_{\mathrm{css}}/\text{Var}(\lzop)=10^{-\mathcal{S}/10})$ are close to the numbers obtained from $F_Q[\hat{L}_y]$ for most cases. In a proper Dicke state as obtained with $n=4 (p=2)$, the $\text{Var}(\lzop)$ becomes extremely small, leading to the very large value of $\sigma^2_{\mathrm{css}}/\text{Var}(\lzop)$ seen in the fourth row. (c) Illustrative examples of $k$-partite entanglement as entanglement depth.}
    \label{fig:entfishsqu}
\end{figure}

A pure state of $n$-qubits, written as a product $\ket{\psi}=\otimes_{j=1}^M \ket{\psi_j}$, is $k$-partite entangled when at least one state $\ket{\psi_j}$ contains non-factorizable $k$-qubits \cite{hyllus2012fisher,Apellaniz2015}. 
This definition is the same as the entanglement depth
~\cite{lucke2014detecting}, see Fig.~\ref{fig:entfishsqu}(c). 
A sufficient condition for $(k+1)$-partite entanglement stems from the quantum Fisher information $F_Q$: a state reaching $F_Q [\rho_n,\mathcal{O}]> (sk^2+r^2)$---where $s =\lfloor n/k \rfloor$ denotes the integer division of $n$ by $k$, and $r$ is the remainder---is at least $(k+1)$-partite entangled~\cite{hyllus2012fisher,toth2012multipartite}.

While $F_Q$ is becoming a useful witness for entanglement in quantum many-body systems~\cite{hauke2016measuring,smith2016many,wang2014quantum,yin2019,mathew2020experimental,laurell2021}, its origin is as a key figure of merit in quantum metrology, where $F_Q$ quantifies the distinguishability of a state $\rho$ from $\rho'=e^{-i\theta \mathcal{O}}\rho e^{i\theta \mathcal{O}}$, generated by the Hermitian operator $\mathcal{O}$ with infinitesimal $\theta$. 
Thus, a large $F_Q$ implies a high measurement precision for estimating the value of $\theta$~\cite{braunstein1994,pezze2014quantum}. 
For pure states $\psi$, the quantum Fisher information becomes simply $F_Q[\psi,\mathcal{O}]=4 \textrm{Var}(\mathcal{O})_\psi$~\cite{toth2014quantum}, whereas for mixed states it provides a lower bound on the variance. 

The target state of the QAOA for MaxCut on $\mathcal{G}_n$, the Dicke state, is invariant under a unitary evolution generated by $\lzop$ but is 
highly sensitive to rotations around the $x$- or $y$-axes of the collective Bloch sphere~\cite{Strobel2014}. Thus, to obtain a large $F_Q$, it is advantageous to choose $\mathcal{O}=\hat{L}_x$ or---even more so---$\mathcal{O}=\hat{L}_y$.
We report $F_Q$ in Fig.~\ref{fig:entfishsqu}(b) for both $\mathcal{O}=\hat{L}_x,\hat{L}_y$ and the resulting $k$-partite entanglement witnessed by it for the ideal simulations. 
In this ideal scenario of noiseless numerical simulations, the large values of $F_Q[\psi,\hat{L}_{x,y}]$ are directly related to the anti-squeezing of the final state along the equator of Bloch sphere. 

In the hardware, where the system is no longer in a pure state, it is considerably more challenging to directly access $F_Q$~\cite{hauke2016measuring,CostadeAlmeida2021}.
However, for Gaussian states, one can nevertheless use the empirical relation~\cite{Strobel2014}
\begin{align}
    F_Q/n[\hat{L}_y] \simeq   \sigma^2_{\mathrm{css}}/\text{Var}(\lzop)=10^{-\mathcal{S}/10}
\end{align}
between $F_Q$ and squeezing.
For the simulation, the estimated $k$ using this relation is close to the exact estimation from $F_Q$ in most of cases, see Fig.~\ref{fig:entfishsqu}(b), except for the depth-three QAOA, where the states are no longer Gaussian~\cite{gross2007non}.
Assuming that the above relation holds for depth-one QAOA, where the states are expected to be Gaussian, we obtain the estimates for $k$-partite entanglement in the hardware implementation reported in Fig.~\ref{fig:entanglement}(b) of the main-text.

\section{Metrology}

In this section, we study the time taken to measure a phase $\theta$ with the CSS verus QAOA-generated Dicke states within different hardware architectures (but without an attempt at comparing the different architectures to each other, which often have different aims and boundary conditions that are difficult to compare on an even footing).
The coherent state is easily prepared by a single rotation around the $y$-axis. It has no entanglement, and $m$ measurements of the phase $\theta$ have a variance bound by $\Delta^2\theta\geq 1/(m N)$.
By contrast, a QAOA prepared state with $F_Q$ above the $k+1$-partite entangled limit takes more time to prepare than the CSS, but it requires a smaller number of measurements to reach the same variance since $\Delta^2\theta\geq 1/(m k N)$ is lowered by a factor of $1/k$.

One may then wonder whether, in a practical application, the improved precision can offset the larger preparation time. 
Crucially, the optimization cost of QAOA can be ignored in these considerations since the optimal $\boldsymbol{\gamma}$ and $\boldsymbol{\beta}$ parameters are reusable across different measurements and experiments.
We therefore compute the duration of a single measurement repetition $t_\text{repet.}$, which is the sum of the duration of the gates in the circuit to prepare the state $t_\text{gates}$ and the readout time including the reset of the measurement apparatus $t_\text{rr}$, i.e., $t_\text{repet.}=t_\text{gates}+t_\text{rr}$.
The gate duration for the coherent spin state $t_\text{gate}^\text{CSS}$ is the duration of a single-qubit gate, while the QAOA protocol requires two-qubit gates, whose number can depend on the available universal gate set and the hardware connectivity.

The QAOA-generated states are advantageous when the time $t^\text{QAOA}=t_\text{repet.}^\text{QAOA}m_\text{QAOA}$ to achieve a certain precision is smaller than the time $t^\text{CSS}=t_\text{repet.}^\text{CSS}m_\text{CSS}$ to achieve the same precision with coherent states. 
If the QAOA-prepared state achieves $F_Q=kN$, we have for equal precision $m_\text{QAOA}k=m_\text{CSS}$, i.e., the QAOA-prepared states are advantageous if 
\begin{align} \label{eqn:qaoa_vs_css}
    t_\text{gates}^\text{QAOA}+t_\text{rr}<k\left(t_\text{gates}^\text{CSS}+t_\text{rr}\right).
\end{align}

\paragraph{Superconducting qubits:} The duration of a QAOA layer is impacted by the qubit connectivity.
Each QAOA layer on $N$ linearly connected superconducting qubits requires $3(N-2)$ layers of simultaneously executable CNOT gates which includes SWAP gates~\cite{Weidenfeller2022}.
Under the assumption that QAOA can create $k+1$-partite entanglement in $p=\log_2(k)$ layers~\cite{Farhi2020}, the duration $t_\text{gates}^\text{QAOA}=3(N-2)\log_2(k)t_\text{cx}$ with $t_\text{cx}$ the duration of a CNOT gate.
Here, we neglected the duration of single-qubit gates. 
With $k=N$, Eq.~(\ref{eqn:qaoa_vs_css}) yields
\begin{align}
    3(N-2)\log_2(N)t_\text{cx}+t_\text{rr}\lesssim N t_\text{rr}
\end{align}
which, for large $N$ and the durations in Tab.~\ref{tab:hardware}, amounts to
\begin{align}
    N \lesssim 2^{t_\text{rr}/(3t_\text{cx})}=2^{1000/3}. 
\end{align}
Although the linear layout of the hardware poses a limit onto when QAOA-generated states remain useful, this limit is extremely high. 
Assuming noisy hardware achieves only a finite $k$, one has 
\begin{align}
    (N-2)\log_2 N <(k-1)t_\text{rr}/(3t_\text{cx})\,. 
\end{align}
For $k=2$, e.g., the QAOA-generated states would remain advantageous up to about 60 qubits arranged in a linear chain, which lies at the size limit of current hardware. These numbers are conservative estimates that can be significantly increased by improved noise resilience and higher hardware connectivity.

\paragraph{Trapped-ion qubits:}  
Large multipartite entangled states of trapped-ions can be generated by a single application of the M\o lmer–S\o rensen gate (MS)~\cite{sorensen1999quantum,sackett2000experimental} where interaction strength among all qubit pairs is equal~\cite{lanyon2011universal}.
Therefore, if we neglect the duration of single-qubit gates, $t_\text{gates}^\text{QAOA}$ only depends on the number of QAOA layers $p=\log_2(k)$, and $t_\text{gates}^\text{QAOA}=\log_2(k) t_{ms}$ with $t_{ms}$ being the duration of a MS gate.
Following Tab.~\ref{tab:hardware}, QAOA-generated $k+1$-partite entangled states are therefore advantageous when
\begin{align}
    \log_2(k) t_{\textrm{ms}} \lesssim (k-1) t_{\textrm{rr}} \Rightarrow k^{\frac{1}{k-1}} \lesssim 2^{t_{\textrm{rr}}/t_{\textrm{rm}}}= 2^{25}.
\end{align}
Since $k^{1/{(k-1)}}$ is a decreasing function, QAOA generated states are always advantageous in trapped-ion setups.

\paragraph{Cold-atoms:}
We now consider cold-atoms in Bose-Einstein condensates which can, e.g., manipulate states with of the order of 400 atoms~\cite{Strobel2014}.
Following QAOA, we assume that $\log_2(k)$ layers of the one-axis-twisting Hamiltonian interleaved with $x$-rotations can generate $k+1$-partite entanglement.
The squeezed state is thus created in a time $t_\text{gates}^\text{QAOA}=\log_2(k)t_\text{OAT}$.
We neglect the duration of $x$ and $y$-rotations. Equation~(\ref{eqn:qaoa_vs_css}) implies $\log_2(k)t_\text{OAT}<(k-1) t_\text{rr}$, showing that given a Bose-condensed atom cloud of fixed size it is always favorable to create spin squeezed states for metrology since $t_\text{OAT}\ll t_\text{rr}$, see Tab.~\ref{tab:hardware}.

\begin{table}[htbp!]
    \centering
    \begin{tabular}{l r r r} \hline\hline
        Platform               & Single-qubit & Entanglement & Readout \& Reset $t_\text{rr}$ \\ \hline
        Transmons & $10$ ns~\cite{Werninghaus2020} & $100$ ns~\cite{Jurcevic2021} & $100~\mu$s~\cite{Wack2021}\\
        Trapped ions~\cite{pogorelov2021compact}          & $15~\mu$s & $200~\mu$s & $300~\mu$s $+ 5$ ms~\cite{schindler2013quantum} \\
        Cold atoms (BEC) & & $10$ ms~\cite{Strobel2014}& $1$~s\\ \hline\hline
    \end{tabular}
    \caption{
    Duration of key operations presented as orders of magnitude only.
    The entangling operation for the transmons, trapped-ions and cold atoms is the two-qubit CNOT gate, the M\o lmer–S\o rensen gate, and the one axis twisting Hamiltonian, respectively.
    }
    \label{tab:hardware}

\end{table}

\section{Quantum Volume and QAOA benchmarking}

\begin{figure}[htbp!]
    \centering
    \includegraphics[width=1.0\columnwidth, clip, trim=20 5 10 20]{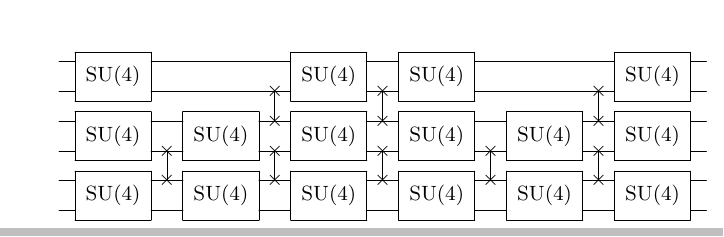}
    \caption{Example of a six-qubit quantum volume circuit as presented in Ref.~\cite{Jurcevic2021}, which shows the layers of ${\rm SU}(4)$ and SWAP gates.}
    \label{fig:qv_example}
\end{figure}

A processor with a Quantum Volume of $QV=2^n$ can reliably, as defined by the generation of heavy output bit-strings, execute circuits that apply $n$ layers of ${\rm SU}(4)$ gates on random permutations of $n$ qubits~\cite{Cross2019}.
When transpiled to a line of $n$ qubits, QV circuits result in $n$ layers of ${\rm SU}(4)$ gates that have at most $\left\lfloor \frac{n}{2} \right\rfloor$ individual ${\rm SU}(4)$ gates simultaneously executed on the qubits~\cite{Jurcevic2021}.
In between these ${\rm SU}(4)$ layers, there are at most $\left\lfloor \frac{n}{2} \right\rfloor$ SWAP gates, see Fig.~\ref{fig:qv_example}.
Furthermore, each ${\rm SU}(4)$ and SWAP gate require at most and exactly three CNOT gates, respectively~\cite{Vidal2004}. 
Under these conditions, the total number of CNOT gates is at most
\begin{align}
  3n \left\lfloor \frac{n}{2} \right\rfloor + 3(n-1) \left\lfloor \frac{n}{2} \right\rfloor\,,
\end{align}
which approaches $3n^2$ as $n$ becomes large.
By comparison, the cost operator of QAOA circuits of complete graphs transpiled to a line requires exactly $\frac{3}{2}n(n-1) - n + 1$ CNOT gates, approaching $3n^2/2$ for large $n$.
This suggests that a $2^n$ Quantum Volume is a good performance indicator for a depth $p=2$ QAOA on $n$ qubits.
Importantly, this comparison is only possible as long as the QAOA circuit is executed using the same error mitigation and transpilation methods as those employed to measure QV~\cite{Pelofske2022}.
However, QV fails to capture the depth dependency $p$ of QAOA.
The benchmark that we develop overcomes this limitation as the QAOA depth should be chosen such that the measured squeezing is maximum.
This also provides the maximum $p$ for which it makes sense to run QAOA on the benchmarked noisy hardware.

\begin{figure}[htbp!]
    \centering
    \includegraphics[width=\columnwidth,clip,trim=15 5 0 5]{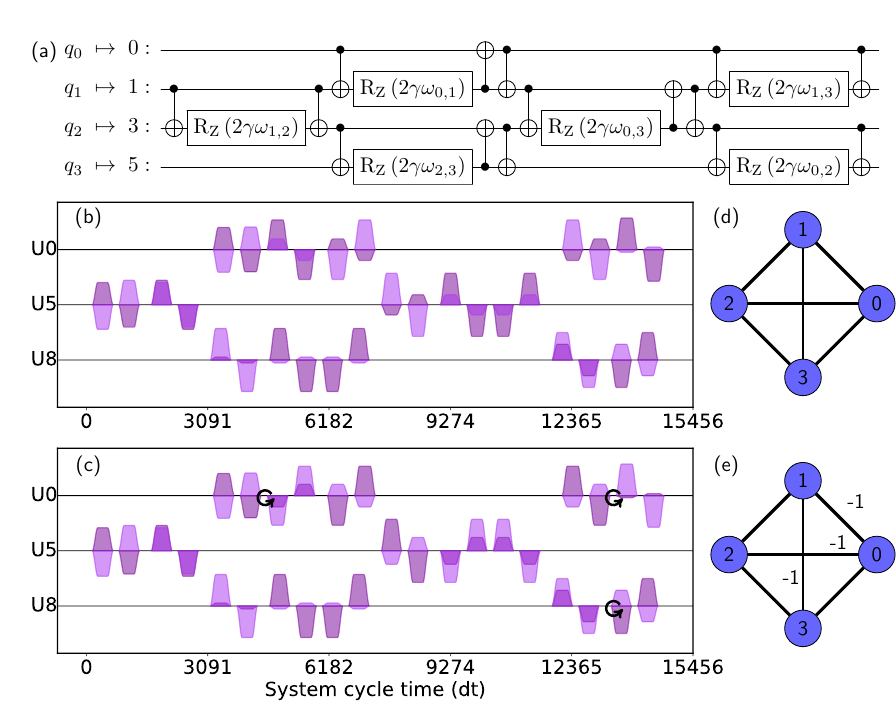}
    \caption{Cross-resonance pulse schedules of the squeezing circuit and an arbitrary QUBO.
      (a) Quantum circuit of a general four qubit fully connected cost operator $e^{-i\gamma\hat H_C}$ transpiled to qubits 0, 1, 3, and 5 of \emph{ibm\_lagos}. 
      (b) Pulse schedule of the cost operator used to generate the symmetric Dicke state, i.e., $\omega_{i,j}=1~\forall i, j$.
      (c) Pulse schedule of a MaxCut instance with edge weights $\omega_{0,1}=\omega_{0,2}=\omega_{1,3}=-1$ and $\omega_{0,3}=\omega_{1,2}=\omega_{2,3}=1$.
      The circular arrows show where the phase shifts differ from the pulse schedule in (b).
      (d) and (e) MaxCut graph corresponding to the pulse schedule in (b) and (c), respectively.
      The duration of a single sample of the arbitrary waveform generators is ${\rm d}t=0.222~{\rm ns}$.
      The light and dark pulses show the in-phase and quadrature of each complex amplitude pulse applied to control channels U0, U5, and U8 of \emph{ibm\_lagos}.}
    \label{fig:schedules}
\end{figure}

From a hardware perspective, the squeezing circuit, exemplified in Fig.~\ref{fig:schedules}(a), captures the complexity of the pulses that execute an arbitrary fully-connected QUBO.
Indeed, the difference between the pulse schedules only amounts to phase changes, indicated by circular arrows in Fig.~\ref{fig:schedules}(c).
The duration and magnitude of the cross-resonance pulses are identical, compare Fig.~\ref{fig:schedules}(b) and (c).
Therefore, much like Quantum Volume, the hardware benchmark based on squeezing captures effects such as limited qubit connectivity, unitary gate errors, decoherence, and cross-talk.
Furthermore, from a hardware perspective the squeezing circuit is also the hardest to implement since QUBOs that are not fully connected, i.e., $\exists~(i,j)~\vert~\Sigma_{i,j}=0$, require less pulses.

\section{Discontinuities in the QAOA hardware benchmark}
 
According to Eq.~(\ref{eq:Palpha}), the states in the domain $(\lceil m_- \rceil, \lfloor m_+ \rfloor)$ are included in $P_\alpha$, where $m_\pm(n,\alpha)=\frac{n}{2}\pm\frac{n}{2}\sqrt{1-\alpha}$.
Since $\lfloor m_+ \rfloor$ and $\lceil m_- \rceil$ must both be integers, the span of the domain $\lfloor m_+ \rfloor-\lceil m_- \rceil$ remains constant over a large $n$ range and changes abruptly when $\lfloor \frac{n}{2}\sqrt{1-\alpha} \rfloor \in \mathbb{Z} $ changes value. 
We denote the values of $n$ at which such changes occur as $n_\text{dis.}$, which correspond to the discrete jumps along the $n$-axis in Fig.~\ref{fig:colorplot}(b) of the main text.
For $\alpha=99.9\%$ and $n$ even, we obtain discontinuities in $P_{99.9\%}$ at $n_\text{dis.}=64,128,190,254$.

\section{Hardware details}

\begin{figure}[htbp!]
    \centering
    \includegraphics[width=\columnwidth]{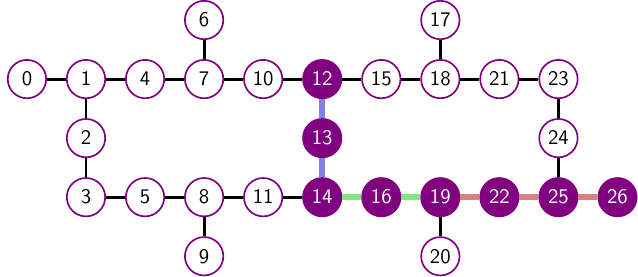}
    \caption{Coupling map of \emph{ibmq\_mumbai} with the qubits used shown in violet.
    The four, six, and eight qubit data were measured on the linearly connected qubits $\{19, 22, 25, 26\}$, $\{14, 16, 19, 22, 25, 26\}$, and $\{12, 13, 14, 16, 19, 22, 25, 26\}$, respectively, chosen based on the CNOT gate fidelity.
    }
    \label{fig:mumbai_map}
\end{figure}

The superconducting qubit data is gathered on the \emph{ibmq\_mumbai} system which has 27 fixed-frequency qubits connected through resonators; its coupling map is shown in Fig.~\ref{fig:mumbai_map}.
We chose a set of qubits that form a line with the smallest possible CNOT gate error.
Each circuit is measured with 4000 shots.
The properties of the device such as $T_1$ times and CNOT gate error are shown in Tab.~\ref{tab:mumbai}.

\begin{table}[htbp!]
    \centering
    \begin{tabular}{l r r | c l r} \hline\hline
        \multicolumn{3}{c|}{CNOT gate} & & & \\
        Qubit pair & $\quad$error (\%) & $\quad$duration (ns) & & Qubit & $\quad T_1~(\mu{\rm s})$ \\ \hline
        (12, 13) & 0.77 & 548 & & 12 & 166 \\ 
        (13, 14) & 1.26 & 320 & & 13 & 137 \\
        (14, 16) & 1.04 & 348 & & 14 & 174 \\
        (16, 19) & 0.77 & 747 & & 16 & 118 \\
        (19, 22) & 0.66 & 363 & & 19 & 227 \\
        (22, 25) & 0.58 & 484 & & 22 & 122 \\
        (25, 26) & 0.50 & 348 & & 25 & 194 \\ 
                 & 0.80$\pm$0.27 & 451$\pm$155 & & 26 & 103 \\ \hline\hline
    \end{tabular}
    \caption{
    Properties of the relevant CNOT gates as reported by \emph{ibmq\_mumbai} on the date of the circuit execution.
    The average $T_1$ of the selected qubits is $155\pm43~\mu{\rm s}$.
    }
    \label{tab:mumbai}
\end{table}

\section{Increasing the duration of \texorpdfstring{$\hat{H}_C$}{}}\label{app:onelayer}
Squeezing is generated by $\lzop^2$~\cite{Strobel2014}, which suggests that simply applying $\hat{H}_C \propto \lzop^2$ for a longer duration, corresponding to a larger coefficient $\gamma$ in the QAOA, may transform the coherent state to a squeezed state, after which we can use the mixer $\hat{H}_M$ to reveal the squeezing along $\lzop$ as in the main text. 
In this way, one layer of QAOA would suffice to create any squeezing which would also require fewer CNOT gates than when $p>1$. 
To test this hypothesis, we run depth-one QAOA using $\gamma=\gamma_1+\gamma_2+\gamma_3$ where the $\gamma_i$ are taken from Fig.~\ref{fig:stepQAOA} in the main text, as they contain the source of ``total" squeezing. 
The result is a fragmented Wigner distribution on the Bloch sphere without observable squeezing in any direction, see Fig.~\ref{fig:bad_wigner}(a).
Furthermore, no squeezing is detected along $z$ for any value of the tomography angle $\beta$, see Fig.~\ref{fig:bad_wigner}(b). 
This finding is in agreement with the known observation that over-squeezing can be detrimental for precision~\cite{StrobelThesis}, however, the states here do not wrap around points near poles because $\lzop^2$ and $\hat{L}_x$ are not applied simultaneously as in Ref.~\cite{StrobelThesis}.

\section{Advantages of multi-layer QAOA}
\begin{figure}
    \centering
    \includegraphics[width=\columnwidth]{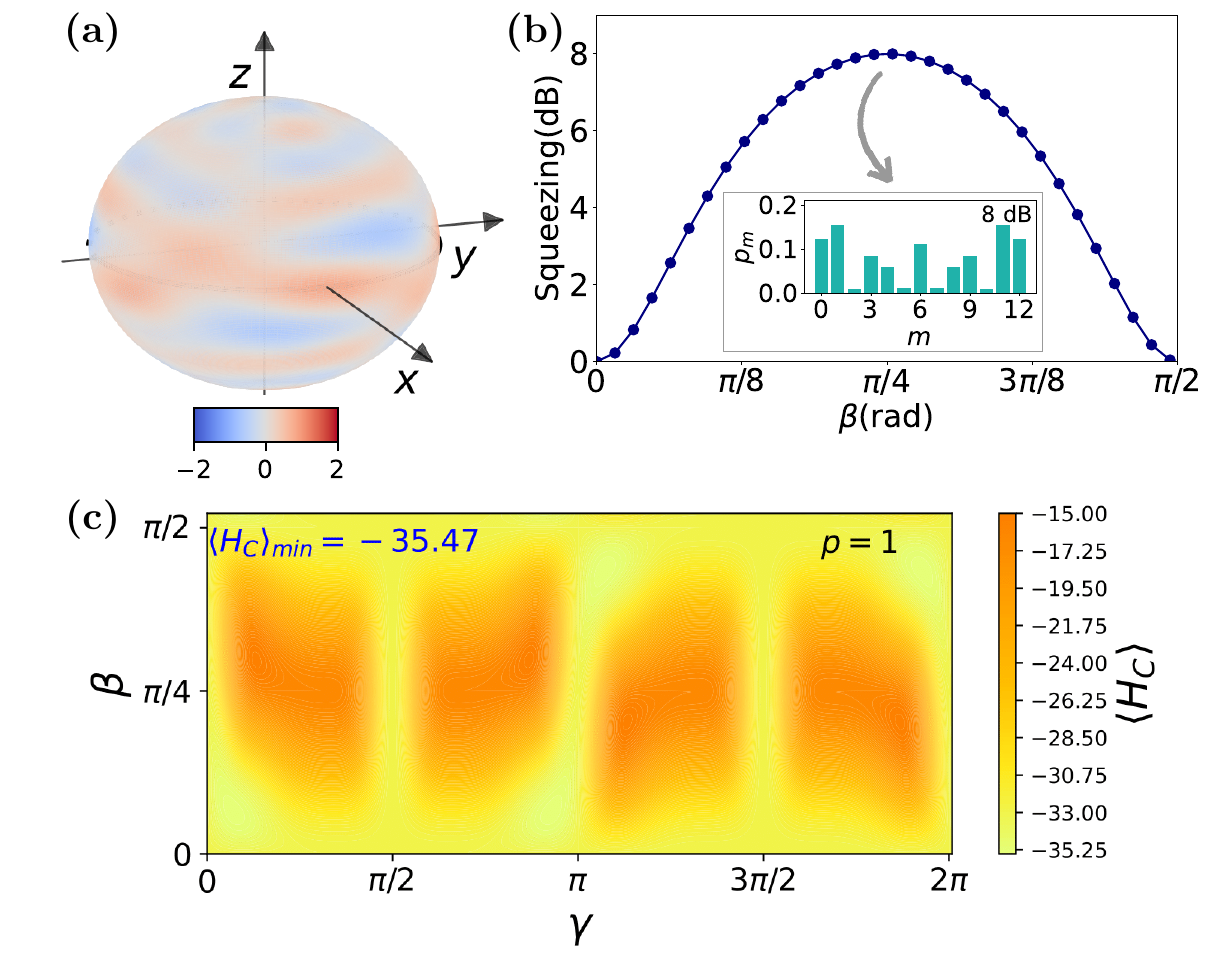}
    \caption{Illustration of the importance of alternating the cost-function and the mixer operator.
    (a) A fragmented Wigner distribution on the Bloch sphere is obtained when $\hat{H}_C$ is applied with $\gamma=\gamma_1+\gamma_2+\gamma_3$ from Fig.~\ref{fig:stepQAOA}. 
    (b) For the state in (a) no squeezing is observed at any $\beta$. 
    The inset shows the probability distribution at $\beta=\pi/4$, corresponding to $\mathcal{S}=8~{\rm dB}$, i.e. over-squeezing. 
    (c) The energy landscape of the depth-one QAOA reveals that the lowest energy it can reach is $-35.47$, which is inferior to $-35.68$ obtained in depth-three QAOA.}
    \label{fig:bad_wigner}
\end{figure}

One may object to the arguments in the preceding Appendix~\ref{app:onelayer} that the $\gamma$ we chose is sub-optimal. 
To address that, in Fig.~\ref{fig:bad_wigner}(c), we numerically map the energy landscape of depth-one QAOA in the $\{\gamma,\beta\}$ plane. 
The results reveal a minimum energy of $\langle \hat{H}_C \rangle_{\mathrm{min}}=-35.47$ which corresponds to $|\langle D^{12}_6|\psi\rangle|^2=98.53\%$. 
These results are inferior to those we obtain from the depth-three QAOA, i.e., $\langle \hat{H}_C \rangle_{\mathrm{min}}=-35.68$ and $|\langle D^{12}_6|\psi\rangle|^2=99.08\%$.
Alternating multiple layers of $\hat{H}_C$ and $\hat{H}_M$ is therefore advantageous over a single application of the one-axis-twisting operator.

To quantify the obtainable improvement as a function of the number of layers used, we can define a new performance metric $\Delta^n_p(\%)$, which compares the energy reduction over the initial ansatz obtained by $p$-layers of QAOA with the one achieved by the ideal target state.
In the $n=12$ case, the initial coherent state and the target Dicke state have $\langle \hat{H}_C \rangle=-33$, and $-36$, respectively. 
Thus, a depth-one QAOA (corresponding to the usual squeezing protocol) can reach a maximum $\Delta^{12}_1=2.47/3$. In contrast, the depth-three QAOA can reach $\Delta^{12}_3=2.68/3$, as shown in Fig.~\ref{fig:stepQAOA}.
Thus, according to this metric a depth-three QAOA is $0.21/3=7\%$ better than a depth-one QAOA.

\section{
Creating arbitrary Dicke states}
In this section, we show how to create arbitrary Dicke states (Eq.~\ref{dicke}) by minimizing a QUBO cost function with QAOA.
Let $\ket{x_{n-1}...x_0}$ be a basis state in which qubit $i$ is in state $x_i\in\{0, 1\}$.
Each basis state in $D_k^n$ satisfies the equation $\sum_{i=0}^{n-1} x_i=k$, which is a constraint on the binary variables $x_i$. 
We express this constraint as the QUBO problem
\begin{align}\label{eqn:min_constraint}
    \min_{x\in\{0, 1\}^n}\left(k-\sum_{i=0}^{n-1} x_i\right)^2.
\end{align}
The solution to this optimization problem is a superposition of all basis states with $k$ qubits in the excited state, i.e., $D_k^n$.
We apply the change of variables $x_i=(z_i+1)/2$ to rewrite $(k-\sum_i x_i)^2$ as 
\begin{align}\label{eqn:z_min_constraint}
    k^2 - kn +\frac{n}{4}(n+1)+\left(\frac{n}{2}-k\right) \sum_{i=0}^{n-1} z_i +\frac{1}{2} \sum_{i<j} z_i z_j.
\end{align}
After promoting each $z_i$ variable to a Pauli spin operator $\hat Z_i$, Eq.~(\ref{eqn:z_min_constraint}) yields a cost Hamiltonian to minimize
\begin{equation}\label{eqn:hc}
    \hat H_C=\left(\frac{n}{2}-k\right) \sum_{i=0}^{n-1} \hat Z_i +\frac{1}{2}\sum_{i<j} \hat Z_i \hat Z_j.
\end{equation}
When $k=n/2$, we recover the MaxCut problem on the symmetric graph.
For $k\neq n/2$, we have an extra term $(n/2-k)\sum\hat Z_i$ that biases the total spin towards $\langle \hat Z\rangle=k$.
The Hamiltonian in Eq.~(\ref{eqn:hc}) can therefore be used to generate the Dicke state $D_k^n$ with QAOA.

\begin{figure*}[htbp!]
    \centering
    \includegraphics[width=1\textwidth, clip, trim= 0 0 0 0]{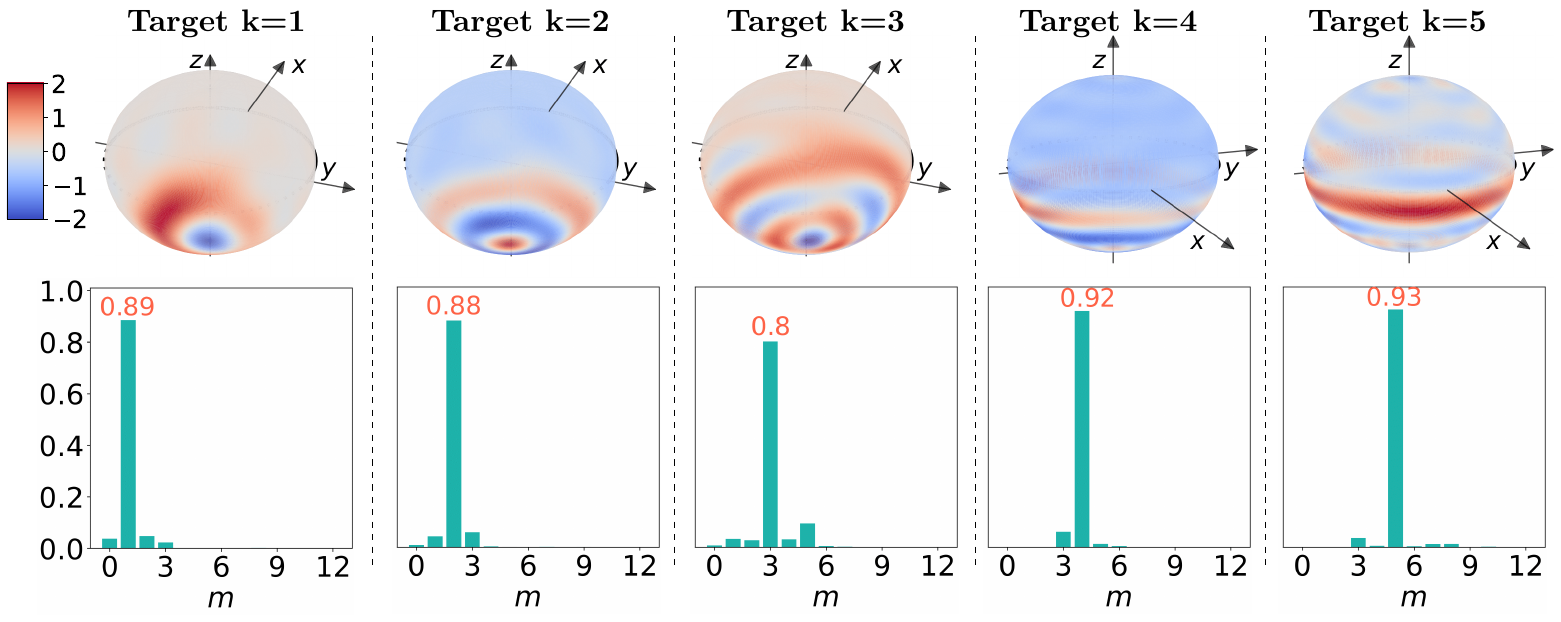}
    \caption{Metrologically useful arbitrary Dicke states generated by a depth-three QAOA by minimizing the cost Hamiltonian Eq.~(\ref{eqn:hc}).
    The top panels show the Wigner quasi-probability distribution on the Bloch spheres.
    The bottom panels show the corresponding histograms of the total spin operator $\langle \hat Z \rangle=m$.
    The orange numbers in each historgram show the overlap probability density $|\langle D_k^{12}|\psi\rangle|^2$ with the target Dicke states.}
    \label{fig:allk}
\end{figure*}

For n=12 qubits, we use the cost Hamiltonian in Eq.~(\ref{eqn:hc}) to simulate the generation of Dicke states with $k=1,2,3,4,5$.
With three QAOA layers, we obtain fidelities in excess of 80\%, see Fig.~\ref{fig:allk}.
The corresponding QAOA parameters $\boldsymbol{\gamma}$ and $\boldsymbol{\beta}$ are shown in Tab.~\ref{tab:arbitrary_dicke}.

\begin{table}[htbp!]
    \centering
    \begin{tabular} {c | c c| c c | c c }  \hline \hline   
        Num. spin up $k$ & $\gamma_1$ & $\beta_1$ & $\gamma_2$ & $\beta_2$ & 
        $\gamma_3$ & $\beta_3$ \\ \hline
        1 & 0.101   & 0.903   & 0.317 & 1.324   & 1.506  &  -0.155 \\ 
        2 & 0.093  & 1.106  & 0.427  &  1.409   & 1.457  &  -0.068  \\
        3 & 0.149  & 1.205 &  1.645  &  1.576  &  0.472 & -0.076   \\
        4 & 0.111   & 1.220 &  0.441  &  1.690 &  1.028  &  0.062 \\
        5 & 0.231 & 1.340  &  1.643   &  1.500   & 1.774  &  0.004  \\
         \hline\hline
    \end{tabular}
    \caption{
    The parameters $(\gamma_i,\beta_i)$ of an optimized depth-three QAOA circuit to create $k=1,2,3,4,5$ Dicke states.
    }
    \label{tab:arbitrary_dicke}
\end{table}

\section{Warm-start with a Dicke state}

In this section, we explore in how far the symmetric MaxCut problem that has a Dicke state as ground state can help solve also non-trivial asymmetric problems.
We show how such squeezed states increase the likelihood to sample good cuts on random Erdős–Rényi graphs.
Each edge $\omega_{i,j}$ of a graph is sampled from a Gaussian distribution $\mathcal{N}(\mu, \epsilon)$ and then rounded to one decimal place to increase the separation in the cut-values of the graph.
We compare standard QAOA with $p$ layers to a QAOA with $p_\text{s}+p$ layers in which the first $p_\text{s}$ layers have fixed parameters to produce a squeezed state.
Both methods, therefore, have $2p$ parameters that require optimization for each graph instance.
For the second approach, in addition, $2p_\text{s}$ parameters are optimized once with the symmetric MaxCut problem as target and are reused for different problem instances.
For each $n\in\{4,6,8,10,12\}$, we sample 100 graph instances from $\mathcal{N}(\mu, \epsilon)$ for which we chose $\mu=4$ and $\epsilon=0.5$ and optimize the cut-value for varying $p$.
The resulting energy normalized to the minimum energy and averaged over the 100 graph realizations is used to compare both methods.
To ensure that $p$ layers always produce a result that is at least as good as the one for $p-1$ layers, we bootstrap the optimization parameters.
The initial guess of the parameters for layer $p$ are based on the optimized parameters of layer $p-1$, i.e., $(\beta_1, \beta_2, ....., \beta_p, \gamma_1, \gamma_2,......, \gamma_p)_{initial}=(\beta^{\rm{opt}}_1, \beta^{\rm{opt}}_2, .....,\beta^{\rm{opt}}_{p-1}, 0, \gamma^{\rm{opt}}_1, \gamma^{\rm{opt}}_2,......,\gamma^{\rm{opt}}_{p-1}, 0)$.

\begin{figure*}
    \centering
    \includegraphics[width=1\textwidth]{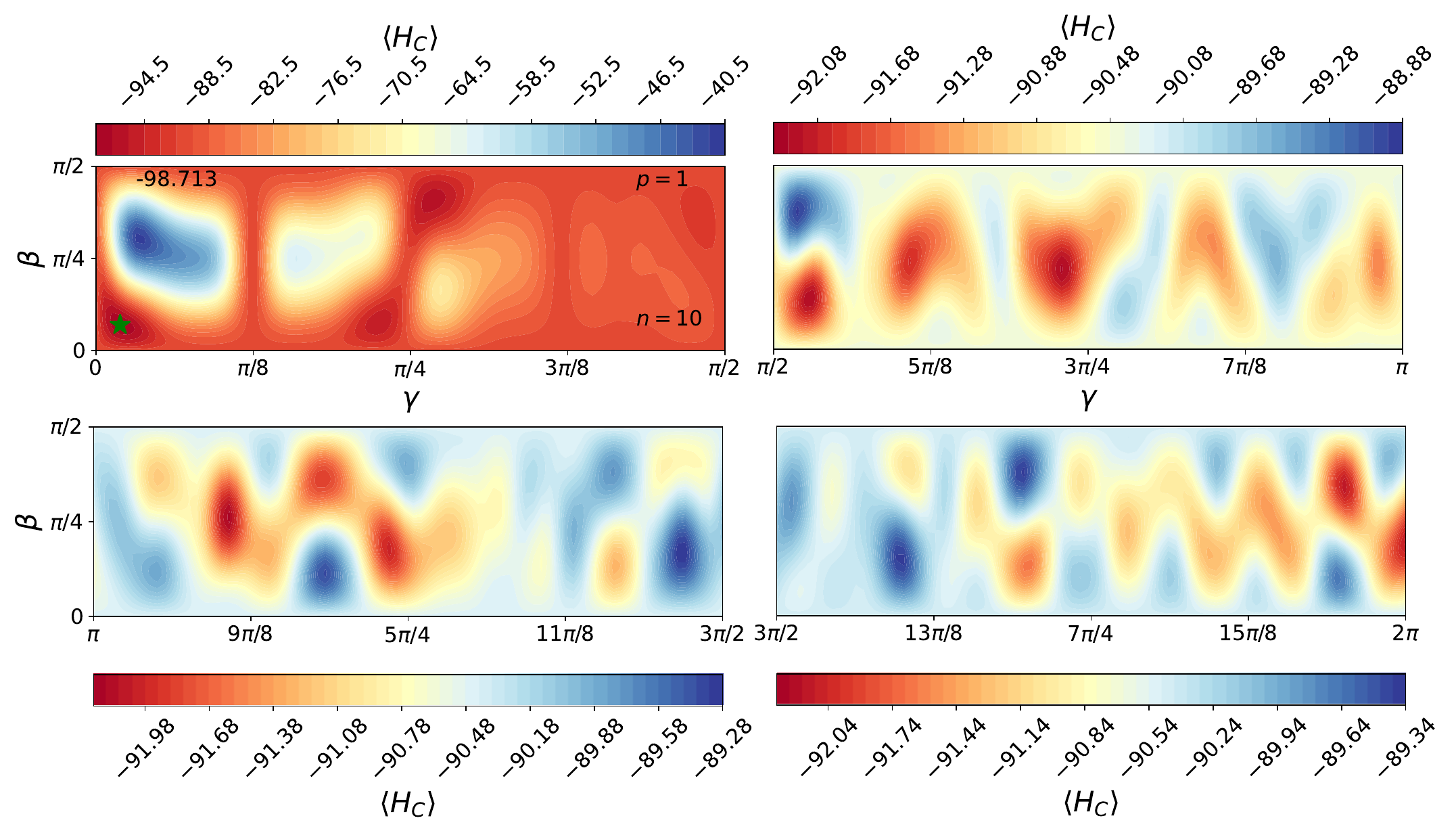}
    \caption{Complexity of the optimization landscape for a $n=10$ vertices graph with edge weights $[$3.1, 3.5, 3.9, 3.6, 4.6, 4.2, 4.8, 3.8, 4.1, 4.8, 5.0, 4.8, 4.0, 3.8, 3.2, 4.1, 4.2, 4.6, 4.3, 3.5, 3.9, 3.8, 3.2, 3.2, 4.7, 3.7, 4.1, 3.5, 4.1, 4.0, 4.2, 3.6, 4.4, 4.1, 3.5, 4.2, 3.7, 3.4, 4.4, 4.4, 3.6, 4.0, 4.3, 4.9, 4.1$]$ and QAOA depth $p=1$. 
    Out of the full landscape, we show the first $0$ to $2\pi$ portion of the $\gamma$ landscape in four subplots $\gamma \in [0,\pi/2],[\pi/2,\pi],[\pi,3\pi/2],[3\pi/2,2\pi]$ with different color scales to increase the contrast between the local minima and maxima.
    This reveals a large number of local minima.
    The small improvement in solution quality with increasing $p$ can therefore be attributed to the many local minima, created from the interference of the frequencies generated by the different edge weights.}
    \label{fig:warm_landscape}
\end{figure*}

\begin{figure}[htbp!]
    \centering
    \includegraphics[width=1\columnwidth]{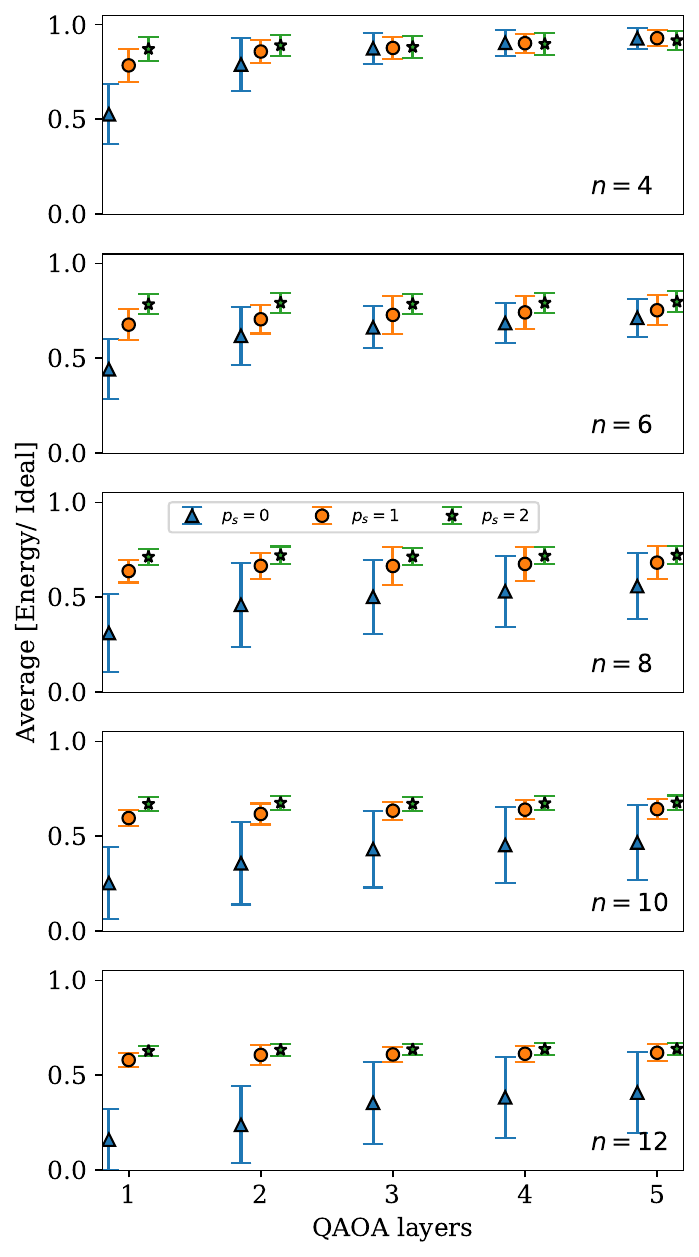}
    \caption{Advantage of QAOA initialized with squeezed states. 
    The blue triangles show standard QAOA initialized from a coherent spin state.
    The orange circles and green stars show QAOA initialized with a spin-squeezed state created with $p_s=1$ and $2$ QAOA layers, respectively.
    The $x$-axis is the QAOA depth after the initial state and the y-axis is the energy normalized to the ideal value.
    The markers and error bars indicate the average and variance of 100 graph instances drawn from $\mathcal{N}(4,0.5)$ with different sizes $n=4$ (top) to $n=12$ (bottom).
    Squeezed initial states boost the average energy of the QAOA optimized state as shown by the orange and green markers having an energy that is closer to the ideal energy than the blue markers.
    The energy increases only modestly as $p$ increases due to the complexity of the optimization landscape.
    }
    \label{fig:warm_start}
\end{figure}

QAOA initialized with squeezed states, shown as orange circles and green stars in Fig.~\ref{fig:warm_start}, significantly improves the average energy when compared to QAOA initialized from an equal superposition, shown as blue triangles in Fig.~\ref{fig:warm_start}.
We observe little improvement in solution quality with increasing $p$.
We attribute this to the complexity of the optimization landscape which has many local minima, even at depth-one, due to the interference of the frequencies generated by the different edge weights, see Fig.~\ref{fig:warm_landscape}.
In the four qubits case, the energy for $p\geq 3$ layers of both methods is comparable.
As the system size is increased, we observe a greater advantage for QAOA initialized with a squeezed state.
These results indicate that, when solving a family of problems, it may be advantageous to initialize QAOA with a state that corresponds to the average problem even when such a problem is trivial to solve.

\begin{thebibliography}{93}%
\makeatletter
\providecommand \@ifxundefined [1]{%
 \@ifx{#1\undefined}
}%
\providecommand \@ifnum [1]{%
 \ifnum #1\expandafter \@firstoftwo
 \else \expandafter \@secondoftwo
 \fi
}%
\providecommand \@ifx [1]{%
 \ifx #1\expandafter \@firstoftwo
 \else \expandafter \@secondoftwo
 \fi
}%
\providecommand \natexlab [1]{#1}%
\providecommand \enquote  [1]{``#1''}%
\providecommand \bibnamefont  [1]{#1}%
\providecommand \bibfnamefont [1]{#1}%
\providecommand \citenamefont [1]{#1}%
\providecommand \href@noop [0]{\@secondoftwo}%
\providecommand \href [0]{\begingroup \@sanitize@url \@href}%
\providecommand \@href[1]{\@@startlink{#1}\@@href}%
\providecommand \@@href[1]{\endgroup#1\@@endlink}%
\providecommand \@sanitize@url [0]{\catcode `\\12\catcode `\$12\catcode
  `\&12\catcode `\#12\catcode `\^12\catcode `\_12\catcode `\%12\relax}%
\providecommand \@@startlink[1]{}%
\providecommand \@@endlink[0]{}%
\providecommand \url  [0]{\begingroup\@sanitize@url \@url }%
\providecommand \@url [1]{\endgroup\@href {#1}{\urlprefix }}%
\providecommand \urlprefix  [0]{URL }%
\providecommand \Eprint [0]{\href }%
\providecommand \doibase [0]{http://dx.doi.org/}%
\providecommand \selectlanguage [0]{\@gobble}%
\providecommand \bibinfo  [0]{\@secondoftwo}%
\providecommand \bibfield  [0]{\@secondoftwo}%
\providecommand \translation [1]{[#1]}%
\providecommand \BibitemOpen [0]{}%
\providecommand \bibitemStop [0]{}%
\providecommand \bibitemNoStop [0]{.\EOS\space}%
\providecommand \EOS [0]{\spacefactor3000\relax}%
\providecommand \BibitemShut  [1]{\csname bibitem#1\endcsname}%
\let\auto@bib@innerbib\@empty
\bibitem [{\citenamefont {Farhi}\ \emph {et~al.}(2014)\citenamefont {Farhi},
  \citenamefont {Goldstone},\ and\ \citenamefont {Gutmann}}]{Farhi2014}%
  \BibitemOpen
  \bibfield  {author} {\bibinfo {author} {\bibfnamefont {Edward}\ \bibnamefont
  {Farhi}}, \bibinfo {author} {\bibfnamefont {Jeffrey}\ \bibnamefont
  {Goldstone}}, \ and\ \bibinfo {author} {\bibfnamefont {Sam}\ \bibnamefont
  {Gutmann}},\ }\href@noop {} {\enquote {\bibinfo {title} {A quantum
  approximate optimization algorithm},}\ } (\bibinfo {year} {2014}),\ \Eprint
  {http://arxiv.org/abs/1411.4028} {arXiv:1411.4028} \BibitemShut {NoStop}%
\bibitem [{\citenamefont {Torta}\ \emph {et~al.}(2021)\citenamefont {Torta},
  \citenamefont {Mbeng}, \citenamefont {Baldassi}, \citenamefont {Zecchina},\
  and\ \citenamefont {Santoro}}]{torta2021quantum}%
  \BibitemOpen
  \bibfield  {author} {\bibinfo {author} {\bibfnamefont {Pietro}\ \bibnamefont
  {Torta}}, \bibinfo {author} {\bibfnamefont {Glen~B.}\ \bibnamefont {Mbeng}},
  \bibinfo {author} {\bibfnamefont {Carlo}\ \bibnamefont {Baldassi}}, \bibinfo
  {author} {\bibfnamefont {Riccardo}\ \bibnamefont {Zecchina}}, \ and\ \bibinfo
  {author} {\bibfnamefont {Giuseppe~E.}\ \bibnamefont {Santoro}},\ }\bibfield
  {title} {\enquote {\bibinfo {title} {Quantum approximate optimization
  algorithm applied to the binary perceptron},}\ }\href
  {https://doi.org/10.48550/arXiv.2112.10219} {\bibfield  {journal} {\bibinfo
  {journal} {arXiv:2112.10219}\ } (\bibinfo {year} {2021})}\BibitemShut
  {NoStop}%
\bibitem [{\citenamefont {Headley}\ \emph {et~al.}(2020)\citenamefont
  {Headley}, \citenamefont {M{\"u}ller}, \citenamefont {Martin}, \citenamefont
  {Solano}, \citenamefont {Sanz},\ and\ \citenamefont
  {Wilhelm}}]{headley2020approximating}%
  \BibitemOpen
  \bibfield  {author} {\bibinfo {author} {\bibfnamefont {David}\ \bibnamefont
  {Headley}}, \bibinfo {author} {\bibfnamefont {Thorge}\ \bibnamefont
  {M{\"u}ller}}, \bibinfo {author} {\bibfnamefont {Ana}\ \bibnamefont
  {Martin}}, \bibinfo {author} {\bibfnamefont {Enrique}\ \bibnamefont
  {Solano}}, \bibinfo {author} {\bibfnamefont {Mikel}\ \bibnamefont {Sanz}}, \
  and\ \bibinfo {author} {\bibfnamefont {Frank~K.}\ \bibnamefont {Wilhelm}},\
  }\bibfield  {title} {\enquote {\bibinfo {title} {Approximating the quantum
  approximate optimisation algorithm},}\ }\href
  {https://doi.org/10.48550/arXiv.2002.12215} {\bibfield  {journal} {\bibinfo
  {journal} {arXiv:2002.12215}\ } (\bibinfo {year} {2020})}\BibitemShut
  {NoStop}%
\bibitem [{\citenamefont {Lucas}(2014)}]{Lucas2014}%
  \BibitemOpen
  \bibfield  {author} {\bibinfo {author} {\bibfnamefont {Andrew}\ \bibnamefont
  {Lucas}},\ }\bibfield  {title} {\enquote {\bibinfo {title} {Ising
  formulations of many {NP} problems},}\ }\href
  {https://doi.org/10.3389/fphy.2014.00005} {\bibfield  {journal} {\bibinfo
  {journal} {Front. Phys.}\ }\textbf {\bibinfo {volume} {2}},\ \bibinfo {pages}
  {5} (\bibinfo {year} {2014})}\BibitemShut {NoStop}%
\bibitem [{\citenamefont {Liang}\ \emph {et~al.}(2020)\citenamefont {Liang},
  \citenamefont {Li},\ and\ \citenamefont {Leichenauer}}]{Liang2020}%
  \BibitemOpen
  \bibfield  {author} {\bibinfo {author} {\bibfnamefont {Daniel}\ \bibnamefont
  {Liang}}, \bibinfo {author} {\bibfnamefont {Li}~\bibnamefont {Li}}, \ and\
  \bibinfo {author} {\bibfnamefont {Stefan}\ \bibnamefont {Leichenauer}},\
  }\bibfield  {title} {\enquote {\bibinfo {title} {Investigating quantum
  approximate optimization algorithms under bang-bang protocols},}\ }\href
  {\doibase 10.1103/PhysRevResearch.2.033402} {\bibfield  {journal} {\bibinfo
  {journal} {Phys. Rev. Research}\ }\textbf {\bibinfo {volume} {2}},\ \bibinfo
  {pages} {033402} (\bibinfo {year} {2020})}\BibitemShut {NoStop}%
\bibitem [{\citenamefont {Lee}\ \emph {et~al.}(2021)\citenamefont {Lee},
  \citenamefont {Magann}, \citenamefont {Rabitz},\ and\ \citenamefont
  {Arenz}}]{Lee2021}%
  \BibitemOpen
  \bibfield  {author} {\bibinfo {author} {\bibfnamefont {Juneseo}\ \bibnamefont
  {Lee}}, \bibinfo {author} {\bibfnamefont {Alicia~B.}\ \bibnamefont {Magann}},
  \bibinfo {author} {\bibfnamefont {Herschel~A.}\ \bibnamefont {Rabitz}}, \
  and\ \bibinfo {author} {\bibfnamefont {Christian}\ \bibnamefont {Arenz}},\
  }\bibfield  {title} {\enquote {\bibinfo {title} {Progress toward favorable
  landscapes in quantum combinatorial optimization},}\ }\href {\doibase
  10.1103/PhysRevA.104.032401} {\bibfield  {journal} {\bibinfo  {journal}
  {Phys. Rev. A}\ }\textbf {\bibinfo {volume} {104}},\ \bibinfo {pages}
  {032401} (\bibinfo {year} {2021})}\BibitemShut {NoStop}%
\bibitem [{\citenamefont {Streif}\ and\ \citenamefont
  {Leib}(2019)}]{Streif2019}%
  \BibitemOpen
  \bibfield  {author} {\bibinfo {author} {\bibfnamefont {Michael}\ \bibnamefont
  {Streif}}\ and\ \bibinfo {author} {\bibfnamefont {Martin}\ \bibnamefont
  {Leib}},\ }\href@noop {} {\enquote {\bibinfo {title} {Comparison of {QAOA}
  with quantum and simulated annealing},}\ } (\bibinfo {year} {2019}),\ \Eprint
  {http://arxiv.org/abs/1901.01903} {arXiv:1901.01903} \BibitemShut {NoStop}%
\bibitem [{\citenamefont {Wurtz}\ and\ \citenamefont {Love}(2022)}]{Wurtz2022}%
  \BibitemOpen
  \bibfield  {author} {\bibinfo {author} {\bibfnamefont {Jonathan}\
  \bibnamefont {Wurtz}}\ and\ \bibinfo {author} {\bibfnamefont {Peter~J.}\
  \bibnamefont {Love}},\ }\bibfield  {title} {\enquote {\bibinfo {title}
  {Counterdiabaticity and the quantum approximate optimization algorithm},}\
  }\href {\doibase 10.22331/q-2022-01-27-635} {\bibfield  {journal} {\bibinfo
  {journal} {Quantum}\ }\textbf {\bibinfo {volume} {6}},\ \bibinfo {pages}
  {635} (\bibinfo {year} {2022})}\BibitemShut {NoStop}%
\bibitem [{\citenamefont {Pezz\'e}\ and\ \citenamefont
  {Smerzi}(2009)}]{Pezze2009}%
  \BibitemOpen
  \bibfield  {author} {\bibinfo {author} {\bibfnamefont {Luca}\ \bibnamefont
  {Pezz\'e}}\ and\ \bibinfo {author} {\bibfnamefont {Augusto}\ \bibnamefont
  {Smerzi}},\ }\bibfield  {title} {\enquote {\bibinfo {title} {Entanglement,
  nonlinear dynamics, and the {Heisenberg} limit},}\ }\href {\doibase
  10.1103/PhysRevLett.102.100401} {\bibfield  {journal} {\bibinfo  {journal}
  {Phys. Rev. Lett.}\ }\textbf {\bibinfo {volume} {102}},\ \bibinfo {pages}
  {100401} (\bibinfo {year} {2009})}\BibitemShut {NoStop}%
\bibitem [{\citenamefont {Pezz\`e}\ \emph {et~al.}(2018)\citenamefont
  {Pezz\`e}, \citenamefont {Smerzi}, \citenamefont {Oberthaler}, \citenamefont
  {Schmied},\ and\ \citenamefont {Treutlein}}]{Pezze2018}%
  \BibitemOpen
  \bibfield  {author} {\bibinfo {author} {\bibfnamefont {Luca}\ \bibnamefont
  {Pezz\`e}}, \bibinfo {author} {\bibfnamefont {Augusto}\ \bibnamefont
  {Smerzi}}, \bibinfo {author} {\bibfnamefont {Markus~K.}\ \bibnamefont
  {Oberthaler}}, \bibinfo {author} {\bibfnamefont {Roman}\ \bibnamefont
  {Schmied}}, \ and\ \bibinfo {author} {\bibfnamefont {Philipp}\ \bibnamefont
  {Treutlein}},\ }\bibfield  {title} {\enquote {\bibinfo {title} {Quantum
  metrology with nonclassical states of atomic ensembles},}\ }\href {\doibase
  10.1103/RevModPhys.90.035005} {\bibfield  {journal} {\bibinfo  {journal}
  {Rev. Mod. Phys.}\ }\textbf {\bibinfo {volume} {90}},\ \bibinfo {pages}
  {035005} (\bibinfo {year} {2018})}\BibitemShut {NoStop}%
\bibitem [{\citenamefont {Degen}\ \emph {et~al.}(2017)\citenamefont {Degen},
  \citenamefont {Reinhard},\ and\ \citenamefont
  {Cappellaro}}]{degen2017quantum}%
  \BibitemOpen
  \bibfield  {author} {\bibinfo {author} {\bibfnamefont {Christian~L.}\
  \bibnamefont {Degen}}, \bibinfo {author} {\bibfnamefont {Friedemann}\
  \bibnamefont {Reinhard}}, \ and\ \bibinfo {author} {\bibfnamefont {Paola}\
  \bibnamefont {Cappellaro}},\ }\bibfield  {title} {\enquote {\bibinfo {title}
  {Quantum sensing},}\ }\href {\doibase 10.1103/RevModPhys.89.035002}
  {\bibfield  {journal} {\bibinfo  {journal} {Rev. Mod. Phys.}\ }\textbf
  {\bibinfo {volume} {89}},\ \bibinfo {pages} {035002} (\bibinfo {year}
  {2017})}\BibitemShut {NoStop}%
\bibitem [{\citenamefont {Gross}\ \emph {et~al.}(2010)\citenamefont {Gross},
  \citenamefont {Zibold}, \citenamefont {Nicklas}, \citenamefont {Est{\`e}ve},\
  and\ \citenamefont {Oberthaler}}]{Gross2010}%
  \BibitemOpen
  \bibfield  {author} {\bibinfo {author} {\bibfnamefont {Christian}\
  \bibnamefont {Gross}}, \bibinfo {author} {\bibfnamefont {Tilman}\
  \bibnamefont {Zibold}}, \bibinfo {author} {\bibfnamefont {Euler}\
  \bibnamefont {Nicklas}}, \bibinfo {author} {\bibfnamefont {Jerome}\
  \bibnamefont {Est{\`e}ve}}, \ and\ \bibinfo {author} {\bibfnamefont
  {Markus~K.}\ \bibnamefont {Oberthaler}},\ }\bibfield  {title} {\enquote
  {\bibinfo {title} {Nonlinear atom interferometer surpasses classical
  precision limit},}\ }\href {\doibase 10.1038/nature08919} {\bibfield
  {journal} {\bibinfo  {journal} {Nature}\ }\textbf {\bibinfo {volume} {464}},\
  \bibinfo {pages} {1165--1169} (\bibinfo {year} {2010})}\BibitemShut {NoStop}%
\bibitem [{\citenamefont {Sewell}\ \emph {et~al.}(2012)\citenamefont {Sewell},
  \citenamefont {Koschorreck}, \citenamefont {Napolitano}, \citenamefont
  {Dubost}, \citenamefont {Behbood},\ and\ \citenamefont
  {Mitchell}}]{Sewell2012}%
  \BibitemOpen
  \bibfield  {author} {\bibinfo {author} {\bibfnamefont {Robert~J.}\
  \bibnamefont {Sewell}}, \bibinfo {author} {\bibfnamefont {Marco}\
  \bibnamefont {Koschorreck}}, \bibinfo {author} {\bibfnamefont {Mario}\
  \bibnamefont {Napolitano}}, \bibinfo {author} {\bibfnamefont {Brice}\
  \bibnamefont {Dubost}}, \bibinfo {author} {\bibfnamefont {Naeimeh}\
  \bibnamefont {Behbood}}, \ and\ \bibinfo {author} {\bibfnamefont {Morgan~W.}\
  \bibnamefont {Mitchell}},\ }\bibfield  {title} {\enquote {\bibinfo {title}
  {Magnetic sensitivity beyond the projection noise limit by spin squeezing},}\
  }\href {\doibase 10.1103/PhysRevLett.109.253605} {\bibfield  {journal}
  {\bibinfo  {journal} {Phys. Rev. Lett.}\ }\textbf {\bibinfo {volume} {109}},\
  \bibinfo {pages} {253605} (\bibinfo {year} {2012})}\BibitemShut {NoStop}%
\bibitem [{\citenamefont {Barsotti}\ \emph {et~al.}(2018)\citenamefont
  {Barsotti}, \citenamefont {Harms},\ and\ \citenamefont
  {Schnabel}}]{barsotti2018squeezed}%
  \BibitemOpen
  \bibfield  {author} {\bibinfo {author} {\bibfnamefont {Lisa}\ \bibnamefont
  {Barsotti}}, \bibinfo {author} {\bibfnamefont {Jan}\ \bibnamefont {Harms}}, \
  and\ \bibinfo {author} {\bibfnamefont {Roman}\ \bibnamefont {Schnabel}},\
  }\bibfield  {title} {\enquote {\bibinfo {title} {Squeezed vacuum states of
  light for gravitational wave detectors},}\ }\href {\doibase
  10.1088/1361-6633/aab906} {\bibfield  {journal} {\bibinfo  {journal} {Rep.
  Prog. Phys.}\ }\textbf {\bibinfo {volume} {82}},\ \bibinfo {pages} {016905}
  (\bibinfo {year} {2018})}\BibitemShut {NoStop}%
\bibitem [{\citenamefont {Dicke}(1954)}]{Dicke1954}%
  \BibitemOpen
  \bibfield  {author} {\bibinfo {author} {\bibfnamefont {Robert~H.}\
  \bibnamefont {Dicke}},\ }\bibfield  {title} {\enquote {\bibinfo {title}
  {Coherence in spontaneous radiation processes},}\ }\href {\doibase
  10.1103/PhysRev.93.99} {\bibfield  {journal} {\bibinfo  {journal} {Phys.
  Rev.}\ }\textbf {\bibinfo {volume} {93}},\ \bibinfo {pages} {99--110}
  (\bibinfo {year} {1954})}\BibitemShut {NoStop}%
\bibitem [{\citenamefont {Giovannetti}\ \emph {et~al.}(2006)\citenamefont
  {Giovannetti}, \citenamefont {Lloyd},\ and\ \citenamefont
  {Maccone}}]{Giovannetti2006}%
  \BibitemOpen
  \bibfield  {author} {\bibinfo {author} {\bibfnamefont {Vittorio}\
  \bibnamefont {Giovannetti}}, \bibinfo {author} {\bibfnamefont {Seth}\
  \bibnamefont {Lloyd}}, \ and\ \bibinfo {author} {\bibfnamefont {Lorenzo}\
  \bibnamefont {Maccone}},\ }\bibfield  {title} {\enquote {\bibinfo {title}
  {Quantum metrology},}\ }\href {\doibase 10.1103/PhysRevLett.96.010401}
  {\bibfield  {journal} {\bibinfo  {journal} {Phys. Rev. Lett.}\ }\textbf
  {\bibinfo {volume} {96}},\ \bibinfo {pages} {010401} (\bibinfo {year}
  {2006})}\BibitemShut {NoStop}%
\bibitem [{\citenamefont {Omran}\ \emph {et~al.}(2019)\citenamefont {Omran},
  \citenamefont {Levine}, \citenamefont {Keesling}, \citenamefont {Semeghini},
  \citenamefont {Wang}, \citenamefont {Ebadi}, \citenamefont {Bernien},
  \citenamefont {Zibrov}, \citenamefont {Pichler}, \citenamefont {Choi},\ and\
  \citenamefont {\emph{et al.}}}]{Omran2019}%
  \BibitemOpen
  \bibfield  {author} {\bibinfo {author} {\bibfnamefont {Ahmed}\ \bibnamefont
  {Omran}}, \bibinfo {author} {\bibfnamefont {Harry}\ \bibnamefont {Levine}},
  \bibinfo {author} {\bibfnamefont {Alexander}\ \bibnamefont {Keesling}},
  \bibinfo {author} {\bibfnamefont {Giulia}\ \bibnamefont {Semeghini}},
  \bibinfo {author} {\bibfnamefont {Tout~T.}\ \bibnamefont {Wang}}, \bibinfo
  {author} {\bibfnamefont {Sepehr}\ \bibnamefont {Ebadi}}, \bibinfo {author}
  {\bibfnamefont {Hannes}\ \bibnamefont {Bernien}}, \bibinfo {author}
  {\bibfnamefont {Alexander~S.}\ \bibnamefont {Zibrov}}, \bibinfo {author}
  {\bibfnamefont {Hannes}\ \bibnamefont {Pichler}}, \bibinfo {author}
  {\bibfnamefont {Soonwon}\ \bibnamefont {Choi}}, \ and\ \bibinfo {author}
  {\bibnamefont {\emph{et al.}}},\ }\bibfield  {title} {\enquote {\bibinfo
  {title} {Generation and manipulation of {Schrödinger} cat states in
  {Rydberg} atom arrays},}\ }\href {\doibase 10.1126/science.aax9743}
  {\bibfield  {journal} {\bibinfo  {journal} {Science}\ }\textbf {\bibinfo
  {volume} {365}},\ \bibinfo {pages} {570--574} (\bibinfo {year}
  {2019})}\BibitemShut {NoStop}%
\bibitem [{\citenamefont {Marciniak}\ \emph {et~al.}(2022)\citenamefont
  {Marciniak}, \citenamefont {Feldker}, \citenamefont {Pogorelov},
  \citenamefont {Kaubruegger}, \citenamefont {Vasilyev}, \citenamefont {van
  Bijnen}, \citenamefont {Schindler}, \citenamefont {Zoller}, \citenamefont
  {Blatt},\ and\ \citenamefont {Monz}}]{Marciniak2022}%
  \BibitemOpen
  \bibfield  {author} {\bibinfo {author} {\bibfnamefont {Christian~D.}\
  \bibnamefont {Marciniak}}, \bibinfo {author} {\bibfnamefont {Thomas}\
  \bibnamefont {Feldker}}, \bibinfo {author} {\bibfnamefont {Ivan}\
  \bibnamefont {Pogorelov}}, \bibinfo {author} {\bibfnamefont {Raphael}\
  \bibnamefont {Kaubruegger}}, \bibinfo {author} {\bibfnamefont {Denis~V.}\
  \bibnamefont {Vasilyev}}, \bibinfo {author} {\bibfnamefont {Rick}\
  \bibnamefont {van Bijnen}}, \bibinfo {author} {\bibfnamefont {Philipp}\
  \bibnamefont {Schindler}}, \bibinfo {author} {\bibfnamefont {Peter}\
  \bibnamefont {Zoller}}, \bibinfo {author} {\bibfnamefont {Rainer}\
  \bibnamefont {Blatt}}, \ and\ \bibinfo {author} {\bibfnamefont {Thomas}\
  \bibnamefont {Monz}},\ }\bibfield  {title} {\enquote {\bibinfo {title}
  {Optimal metrology with programmable quantum sensors},}\ }\href {\doibase
  10.1038/s41586-022-04435-4} {\bibfield  {journal} {\bibinfo  {journal}
  {Nature}\ }\textbf {\bibinfo {volume} {603}},\ \bibinfo {pages} {604--609}
  (\bibinfo {year} {2022})}\BibitemShut {NoStop}%
\bibitem [{\citenamefont {Arrazola}\ \emph {et~al.}(2021)\citenamefont
  {Arrazola}, \citenamefont {Bergholm}, \citenamefont {Br{\'a}dler},
  \citenamefont {Bromley}, \citenamefont {Collins}, \citenamefont {Dhand},
  \citenamefont {Fumagalli}, \citenamefont {Gerrits}, \citenamefont {Goussev},
  \citenamefont {Helt},\ and\ \citenamefont {\emph{et
  al.}}}]{arrazola2021quantum}%
  \BibitemOpen
  \bibfield  {author} {\bibinfo {author} {\bibfnamefont {Juan~M.}\ \bibnamefont
  {Arrazola}}, \bibinfo {author} {\bibfnamefont {Ville}\ \bibnamefont
  {Bergholm}}, \bibinfo {author} {\bibfnamefont {Kamil}\ \bibnamefont
  {Br{\'a}dler}}, \bibinfo {author} {\bibfnamefont {Thomas~R.}\ \bibnamefont
  {Bromley}}, \bibinfo {author} {\bibfnamefont {Matt~J.}\ \bibnamefont
  {Collins}}, \bibinfo {author} {\bibfnamefont {Ish}\ \bibnamefont {Dhand}},
  \bibinfo {author} {\bibfnamefont {Alberto}\ \bibnamefont {Fumagalli}},
  \bibinfo {author} {\bibfnamefont {Thomas}\ \bibnamefont {Gerrits}}, \bibinfo
  {author} {\bibfnamefont {Andrey}\ \bibnamefont {Goussev}}, \bibinfo {author}
  {\bibfnamefont {Lukas~G.}\ \bibnamefont {Helt}}, \ and\ \bibinfo {author}
  {\bibnamefont {\emph{et al.}}},\ }\bibfield  {title} {\enquote {\bibinfo
  {title} {Quantum circuits with many photons on a programmable nanophotonic
  chip},}\ }\href {\doibase 10.1038/s41586-021-03202-1} {\bibfield  {journal}
  {\bibinfo  {journal} {Nature}\ }\textbf {\bibinfo {volume} {591}},\ \bibinfo
  {pages} {54--60} (\bibinfo {year} {2021})}\BibitemShut {NoStop}%
\bibitem [{\citenamefont {Gross}(2007)}]{gross2007non}%
  \BibitemOpen
  \bibfield  {author} {\bibinfo {author} {\bibfnamefont {David}\ \bibnamefont
  {Gross}},\ }\bibfield  {title} {\enquote {\bibinfo {title} {Non-negative
  wigner functions in prime dimensions},}\ }\href {\doibase
  https://doi.org/10.1007/s00340-006-2510-9} {\bibfield  {journal} {\bibinfo
  {journal} {Appl. Phys. B}\ }\textbf {\bibinfo {volume} {86}},\ \bibinfo
  {pages} {367--370} (\bibinfo {year} {2007})}\BibitemShut {NoStop}%
\bibitem [{\citenamefont {Esteve}\ \emph {et~al.}(2008)\citenamefont {Esteve},
  \citenamefont {Gross}, \citenamefont {Weller}, \citenamefont {Giovanazzi},\
  and\ \citenamefont {Oberthaler}}]{esteve2008squeezing}%
  \BibitemOpen
  \bibfield  {author} {\bibinfo {author} {\bibfnamefont {Jerome}\ \bibnamefont
  {Esteve}}, \bibinfo {author} {\bibfnamefont {Christian}\ \bibnamefont
  {Gross}}, \bibinfo {author} {\bibfnamefont {Andreas}\ \bibnamefont {Weller}},
  \bibinfo {author} {\bibfnamefont {Stefano}\ \bibnamefont {Giovanazzi}}, \
  and\ \bibinfo {author} {\bibfnamefont {Markus~K.}\ \bibnamefont
  {Oberthaler}},\ }\bibfield  {title} {\enquote {\bibinfo {title} {Squeezing
  and entanglement in a {Bose--Einstein} condensate},}\ }\href {\doibase
  10.1038/nature07332} {\bibfield  {journal} {\bibinfo  {journal} {Nature}\
  }\textbf {\bibinfo {volume} {455}},\ \bibinfo {pages} {1216--1219} (\bibinfo
  {year} {2008})}\BibitemShut {NoStop}%
\bibitem [{\citenamefont {Purdy}\ \emph {et~al.}(2013)\citenamefont {Purdy},
  \citenamefont {Yu}, \citenamefont {Peterson}, \citenamefont {Kampel},\ and\
  \citenamefont {Regal}}]{purdy2013strong}%
  \BibitemOpen
  \bibfield  {author} {\bibinfo {author} {\bibfnamefont {Thomas~P.}\
  \bibnamefont {Purdy}}, \bibinfo {author} {\bibfnamefont {Pen-Li}\
  \bibnamefont {Yu}}, \bibinfo {author} {\bibfnamefont {Robert~W.}\
  \bibnamefont {Peterson}}, \bibinfo {author} {\bibfnamefont {Nir~S.}\
  \bibnamefont {Kampel}}, \ and\ \bibinfo {author} {\bibfnamefont {Cindy~A.}\
  \bibnamefont {Regal}},\ }\bibfield  {title} {\enquote {\bibinfo {title}
  {Strong optomechanical squeezing of light},}\ }\href {\doibase
  10.1103/PhysRevX.3.031012} {\bibfield  {journal} {\bibinfo  {journal} {Phys.
  Rev. X}\ }\textbf {\bibinfo {volume} {3}},\ \bibinfo {pages} {031012}
  (\bibinfo {year} {2013})}\BibitemShut {NoStop}%
\bibitem [{\citenamefont {Strobel}\ \emph {et~al.}(2014)\citenamefont
  {Strobel}, \citenamefont {Muessel}, \citenamefont {Linnemann}, \citenamefont
  {Zibold}, \citenamefont {Hume}, \citenamefont {Pezzè}, \citenamefont
  {Smerzi},\ and\ \citenamefont {Oberthaler}}]{Strobel2014}%
  \BibitemOpen
  \bibfield  {author} {\bibinfo {author} {\bibfnamefont {Helmut}\ \bibnamefont
  {Strobel}}, \bibinfo {author} {\bibfnamefont {Wolfgang}\ \bibnamefont
  {Muessel}}, \bibinfo {author} {\bibfnamefont {Daniel}\ \bibnamefont
  {Linnemann}}, \bibinfo {author} {\bibfnamefont {Tilman}\ \bibnamefont
  {Zibold}}, \bibinfo {author} {\bibfnamefont {David~B.}\ \bibnamefont {Hume}},
  \bibinfo {author} {\bibfnamefont {Luca}\ \bibnamefont {Pezzè}}, \bibinfo
  {author} {\bibfnamefont {Augusto}\ \bibnamefont {Smerzi}}, \ and\ \bibinfo
  {author} {\bibfnamefont {Markus~K.}\ \bibnamefont {Oberthaler}},\ }\bibfield
  {title} {\enquote {\bibinfo {title} {Fisher information and entanglement of
  non-gaussian spin states},}\ }\href {\doibase 10.1126/science.1250147}
  {\bibfield  {journal} {\bibinfo  {journal} {Science}\ }\textbf {\bibinfo
  {volume} {345}},\ \bibinfo {pages} {424–427} (\bibinfo {year}
  {2014})}\BibitemShut {NoStop}%
\bibitem [{\citenamefont {Muessel}\ \emph {et~al.}(2015)\citenamefont
  {Muessel}, \citenamefont {Strobel}, \citenamefont {Linnemann}, \citenamefont
  {Zibold}, \citenamefont {Juli\'a-D\'{\i}az},\ and\ \citenamefont
  {Oberthaler}}]{muessel2015twist}%
  \BibitemOpen
  \bibfield  {author} {\bibinfo {author} {\bibfnamefont {W.}~\bibnamefont
  {Muessel}}, \bibinfo {author} {\bibfnamefont {H.}~\bibnamefont {Strobel}},
  \bibinfo {author} {\bibfnamefont {D.}~\bibnamefont {Linnemann}}, \bibinfo
  {author} {\bibfnamefont {T.}~\bibnamefont {Zibold}}, \bibinfo {author}
  {\bibfnamefont {B.}~\bibnamefont {Juli\'a-D\'{\i}az}}, \ and\ \bibinfo
  {author} {\bibfnamefont {M.~K.}\ \bibnamefont {Oberthaler}},\ }\bibfield
  {title} {\enquote {\bibinfo {title} {Twist-and-turn spin squeezing in
  bose-einstein condensates},}\ }\href {\doibase 10.1103/PhysRevA.92.023603}
  {\bibfield  {journal} {\bibinfo  {journal} {Phys. Rev. A}\ }\textbf {\bibinfo
  {volume} {92}},\ \bibinfo {pages} {023603} (\bibinfo {year}
  {2015})}\BibitemShut {NoStop}%
\bibitem [{\citenamefont {Xu}\ \emph {et~al.}(2022)\citenamefont {Xu},
  \citenamefont {Zhang}, \citenamefont {Sun}, \citenamefont {Li}, \citenamefont
  {Song}, \citenamefont {Xiang}, \citenamefont {Huang}, \citenamefont {Li},
  \citenamefont {Shi}, \citenamefont {Chen}, \citenamefont {Song},
  \citenamefont {Zheng}, \citenamefont {Nori}, \citenamefont {Wang},\ and\
  \citenamefont {Fan}}]{Xu2022}%
  \BibitemOpen
  \bibfield  {author} {\bibinfo {author} {\bibfnamefont {Kai}\ \bibnamefont
  {Xu}}, \bibinfo {author} {\bibfnamefont {Yu-Ran}\ \bibnamefont {Zhang}},
  \bibinfo {author} {\bibfnamefont {Zheng-Hang}\ \bibnamefont {Sun}}, \bibinfo
  {author} {\bibfnamefont {Hekang}\ \bibnamefont {Li}}, \bibinfo {author}
  {\bibfnamefont {Pengtao}\ \bibnamefont {Song}}, \bibinfo {author}
  {\bibfnamefont {Zhongcheng}\ \bibnamefont {Xiang}}, \bibinfo {author}
  {\bibfnamefont {Kaixuan}\ \bibnamefont {Huang}}, \bibinfo {author}
  {\bibfnamefont {Hao}\ \bibnamefont {Li}}, \bibinfo {author} {\bibfnamefont
  {Yun-Hao}\ \bibnamefont {Shi}}, \bibinfo {author} {\bibfnamefont {Chi-Tong}\
  \bibnamefont {Chen}}, \bibinfo {author} {\bibfnamefont {Xiaohui}\
  \bibnamefont {Song}}, \bibinfo {author} {\bibfnamefont {Dongning}\
  \bibnamefont {Zheng}}, \bibinfo {author} {\bibfnamefont {Franco}\
  \bibnamefont {Nori}}, \bibinfo {author} {\bibfnamefont {H.}~\bibnamefont
  {Wang}}, \ and\ \bibinfo {author} {\bibfnamefont {Heng}\ \bibnamefont
  {Fan}},\ }\bibfield  {title} {\enquote {\bibinfo {title} {Metrological
  characterization of non-gaussian entangled states of superconducting
  qubits},}\ }\href {\doibase 10.1103/PhysRevLett.128.150501} {\bibfield
  {journal} {\bibinfo  {journal} {Phys. Rev. Lett.}\ }\textbf {\bibinfo
  {volume} {128}},\ \bibinfo {pages} {150501} (\bibinfo {year}
  {2022})}\BibitemShut {NoStop}%
\bibitem [{\citenamefont {S{\o}rensen}\ \emph {et~al.}(2001)\citenamefont
  {S{\o}rensen}, \citenamefont {Duan}, \citenamefont {Cirac},\ and\
  \citenamefont {Zoller}}]{sorensen2001many}%
  \BibitemOpen
  \bibfield  {author} {\bibinfo {author} {\bibfnamefont {Anders}\ \bibnamefont
  {S{\o}rensen}}, \bibinfo {author} {\bibfnamefont {L.-M.}\ \bibnamefont
  {Duan}}, \bibinfo {author} {\bibfnamefont {Juan~Ignacio}\ \bibnamefont
  {Cirac}}, \ and\ \bibinfo {author} {\bibfnamefont {Peter}\ \bibnamefont
  {Zoller}},\ }\bibfield  {title} {\enquote {\bibinfo {title} {Many-particle
  entanglement with {Bose--Einstein} condensates},}\ }\href
  {https://www.nature.com/articles/35051038} {\bibfield  {journal} {\bibinfo
  {journal} {Nature}\ }\textbf {\bibinfo {volume} {409}},\ \bibinfo {pages}
  {63--66} (\bibinfo {year} {2001})}\BibitemShut {NoStop}%
\bibitem [{\citenamefont {Korbicz}\ \emph {et~al.}(2005)\citenamefont
  {Korbicz}, \citenamefont {Cirac},\ and\ \citenamefont
  {Lewenstein}}]{Korbicz2005}%
  \BibitemOpen
  \bibfield  {author} {\bibinfo {author} {\bibfnamefont {Jaroslaw~K.}\
  \bibnamefont {Korbicz}}, \bibinfo {author} {\bibfnamefont {Ignacio~J.}\
  \bibnamefont {Cirac}}, \ and\ \bibinfo {author} {\bibfnamefont {Maciej}\
  \bibnamefont {Lewenstein}},\ }\bibfield  {title} {\enquote {\bibinfo {title}
  {Spin squeezing inequalities and entanglement of $n$ qubit states},}\ }\href
  {\doibase 10.1103/PhysRevLett.95.120502} {\bibfield  {journal} {\bibinfo
  {journal} {Phys. Rev. Lett.}\ }\textbf {\bibinfo {volume} {95}},\ \bibinfo
  {pages} {120502} (\bibinfo {year} {2005})}\BibitemShut {NoStop}%
\bibitem [{\citenamefont {Korbicz}\ \emph {et~al.}(2006)\citenamefont
  {Korbicz}, \citenamefont {G\"uhne}, \citenamefont {Lewenstein}, \citenamefont
  {H\"affner}, \citenamefont {Roos},\ and\ \citenamefont
  {Blatt}}]{korbicz2006}%
  \BibitemOpen
  \bibfield  {author} {\bibinfo {author} {\bibfnamefont {Jaroslaw~K.}\
  \bibnamefont {Korbicz}}, \bibinfo {author} {\bibfnamefont {Otfried}\
  \bibnamefont {G\"uhne}}, \bibinfo {author} {\bibfnamefont {Maciej}\
  \bibnamefont {Lewenstein}}, \bibinfo {author} {\bibfnamefont {Hartmut}\
  \bibnamefont {H\"affner}}, \bibinfo {author} {\bibfnamefont {Christian~F.}\
  \bibnamefont {Roos}}, \ and\ \bibinfo {author} {\bibfnamefont {Rainer}\
  \bibnamefont {Blatt}},\ }\bibfield  {title} {\enquote {\bibinfo {title}
  {Generalized spin-squeezing inequalities in $n$-qubit systems: Theory and
  experiment},}\ }\href {\doibase 10.1103/PhysRevA.74.052319} {\bibfield
  {journal} {\bibinfo  {journal} {Phys. Rev. A}\ }\textbf {\bibinfo {volume}
  {74}},\ \bibinfo {pages} {052319} (\bibinfo {year} {2006})}\BibitemShut
  {NoStop}%
\bibitem [{\citenamefont {G{\"u}hne}\ and\ \citenamefont
  {T{\'o}th}(2009)}]{guhne2009entanglement}%
  \BibitemOpen
  \bibfield  {author} {\bibinfo {author} {\bibfnamefont {Otfried}\ \bibnamefont
  {G{\"u}hne}}\ and\ \bibinfo {author} {\bibfnamefont {G{\'e}za}\ \bibnamefont
  {T{\'o}th}},\ }\bibfield  {title} {\enquote {\bibinfo {title} {Entanglement
  detection},}\ }\href {https://doi.org/10.1016/j.physrep.2009.02.004}
  {\bibfield  {journal} {\bibinfo  {journal} {Phys. Rep.}\ }\textbf {\bibinfo
  {volume} {474}},\ \bibinfo {pages} {1--75} (\bibinfo {year}
  {2009})}\BibitemShut {NoStop}%
\bibitem [{\citenamefont {T{\'o}th}(2007)}]{toth2007detection}%
  \BibitemOpen
  \bibfield  {author} {\bibinfo {author} {\bibfnamefont {G{\'e}za}\
  \bibnamefont {T{\'o}th}},\ }\bibfield  {title} {\enquote {\bibinfo {title}
  {Detection of multipartite entanglement in the vicinity of symmetric dicke
  states},}\ }\href {https://doi.org/10.1364/JOSAB.24.000275} {\bibfield
  {journal} {\bibinfo  {journal} {J. Opt. Soc. Am. B}\ }\textbf {\bibinfo
  {volume} {24}},\ \bibinfo {pages} {275--282} (\bibinfo {year}
  {2007})}\BibitemShut {NoStop}%
\bibitem [{\citenamefont {Kitagawa}\ and\ \citenamefont
  {Ueda}(1993)}]{kitagawa1993squeezed}%
  \BibitemOpen
  \bibfield  {author} {\bibinfo {author} {\bibfnamefont {Masahiro}\
  \bibnamefont {Kitagawa}}\ and\ \bibinfo {author} {\bibfnamefont {Masahito}\
  \bibnamefont {Ueda}},\ }\bibfield  {title} {\enquote {\bibinfo {title}
  {Squeezed spin states},}\ }\href {\doibase 10.1103/PhysRevA.47.5138}
  {\bibfield  {journal} {\bibinfo  {journal} {Phys. Rev. A}\ }\textbf {\bibinfo
  {volume} {47}},\ \bibinfo {pages} {5138--5143} (\bibinfo {year}
  {1993})}\BibitemShut {NoStop}%
\bibitem [{\citenamefont {Wang}\ \emph {et~al.}(2001)\citenamefont {Wang},
  \citenamefont {S\o{}ndberg~S\o{}rensen},\ and\ \citenamefont
  {M\o{}lmer}}]{Wang2001}%
  \BibitemOpen
  \bibfield  {author} {\bibinfo {author} {\bibfnamefont {Xiaoguang}\
  \bibnamefont {Wang}}, \bibinfo {author} {\bibfnamefont {Anders}\ \bibnamefont
  {S\o{}ndberg~S\o{}rensen}}, \ and\ \bibinfo {author} {\bibfnamefont {Klaus}\
  \bibnamefont {M\o{}lmer}},\ }\bibfield  {title} {\enquote {\bibinfo {title}
  {Spin squeezing in the {Ising} model},}\ }\href {\doibase
  10.1103/PhysRevA.64.053815} {\bibfield  {journal} {\bibinfo  {journal} {Phys.
  Rev. A}\ }\textbf {\bibinfo {volume} {64}},\ \bibinfo {pages} {053815}
  (\bibinfo {year} {2001})}\BibitemShut {NoStop}%
\bibitem [{\citenamefont {Ma}\ \emph {et~al.}(2011)\citenamefont {Ma},
  \citenamefont {Wang}, \citenamefont {Sun},\ and\ \citenamefont
  {Nori}}]{ma2011quantum}%
  \BibitemOpen
  \bibfield  {author} {\bibinfo {author} {\bibfnamefont {Jian}\ \bibnamefont
  {Ma}}, \bibinfo {author} {\bibfnamefont {Xiaoguang}\ \bibnamefont {Wang}},
  \bibinfo {author} {\bibfnamefont {Chang-Pu}\ \bibnamefont {Sun}}, \ and\
  \bibinfo {author} {\bibfnamefont {Franco}\ \bibnamefont {Nori}},\ }\bibfield
  {title} {\enquote {\bibinfo {title} {Quantum spin squeezing},}\ }\href
  {https://doi.org/10.1016/j.physrep.2011.08.003} {\bibfield  {journal}
  {\bibinfo  {journal} {Phys. Rep.}\ }\textbf {\bibinfo {volume} {509}},\
  \bibinfo {pages} {89--165} (\bibinfo {year} {2011})}\BibitemShut {NoStop}%
\bibitem [{Not()}]{NoteXmarked}%
  \BibitemOpen
  \href@noop {} {}\bibinfo {note} {See supplemental material where we (i)
  discuss the details of the optimization method used to obtain the parameters
  $\{\boldsymbol\gamma,\boldsymbol\beta\}$, (ii) describe why squeezed states
  are entangled, (iii) define and connect multipartite entanglement to quantum
  Fisher information and squeezing, (iv) discuss the practical advantages of
  using QAOA for metrology in different hardware architectures, (v) argue why
  squeezing can be a better benchmark than quantum volume in QAOA, (vi) explain
  the discontinuities observed in the benchmark in Fig.2(b), (vii) give details
  of the \emph{ibmq\_mumbai}, (viii) discuss the advantages of using
  multiple-layers of QAOA, (ix) explain why increasing the duration of
  $\hat{H}_C$ is not helpful compared to the alternating layers in QAOA, (x)
  discuss the creation of arbitrary Dicke states, (xi) show how to initialize
  QAOA with squeezed states to lower the number of optimization parameter, and
  contains references
  \cite{Weidenfeller2022,spall1998overview,toth2007detection,hyllus2012fisher,Apellaniz2015,lucke2014detecting,toth2012multipartite,hauke2016measuring,smith2016many,wang2014quantum,yin2019,mathew2020experimental,laurell2021,braunstein1994,pezze2014quantum,toth2014quantum,Strobel2014,CostadeAlmeida2021,gross2007non,Cross2019,Jurcevic2021,Vidal2004,Pelofske2022,StrobelThesis,Farhi2020,sorensen1999quantum,lanyon2011universal,Werninghaus2020,Wack2021,pogorelov2021compact,schindler2013quantum,sackett2000experimental}}\BibitemShut
  {NoStop}%
\bibitem [{\citenamefont {Hyllus}\ \emph {et~al.}(2012)\citenamefont {Hyllus},
  \citenamefont {Laskowski}, \citenamefont {Krischek}, \citenamefont
  {Schwemmer}, \citenamefont {Wieczorek}, \citenamefont {Weinfurter},
  \citenamefont {Pezz{\'e}},\ and\ \citenamefont {Smerzi}}]{hyllus2012fisher}%
  \BibitemOpen
  \bibfield  {author} {\bibinfo {author} {\bibfnamefont {Philipp}\ \bibnamefont
  {Hyllus}}, \bibinfo {author} {\bibfnamefont {Wies{\l}aw}\ \bibnamefont
  {Laskowski}}, \bibinfo {author} {\bibfnamefont {Roland}\ \bibnamefont
  {Krischek}}, \bibinfo {author} {\bibfnamefont {Christian}\ \bibnamefont
  {Schwemmer}}, \bibinfo {author} {\bibfnamefont {Witlef}\ \bibnamefont
  {Wieczorek}}, \bibinfo {author} {\bibfnamefont {Harald}\ \bibnamefont
  {Weinfurter}}, \bibinfo {author} {\bibfnamefont {Luca}\ \bibnamefont
  {Pezz{\'e}}}, \ and\ \bibinfo {author} {\bibfnamefont {Augusto}\ \bibnamefont
  {Smerzi}},\ }\bibfield  {title} {\enquote {\bibinfo {title} {Fisher
  information and multiparticle entanglement},}\ }\href {\doibase
  10.1103/PhysRevA.85.022321} {\bibfield  {journal} {\bibinfo  {journal} {Phys.
  Rev. A}\ }\textbf {\bibinfo {volume} {85}},\ \bibinfo {pages} {022321}
  (\bibinfo {year} {2012})}\BibitemShut {NoStop}%
\bibitem [{\citenamefont {T{\'o}th}(2012)}]{toth2012multipartite}%
  \BibitemOpen
  \bibfield  {author} {\bibinfo {author} {\bibfnamefont {G{\'e}za}\
  \bibnamefont {T{\'o}th}},\ }\bibfield  {title} {\enquote {\bibinfo {title}
  {Multipartite entanglement and high-precision metrology},}\ }\href {\doibase
  10.1103/PhysRevA.85.022322} {\bibfield  {journal} {\bibinfo  {journal} {Phys.
  Rev. A}\ }\textbf {\bibinfo {volume} {85}},\ \bibinfo {pages} {022322}
  (\bibinfo {year} {2012})}\BibitemShut {NoStop}%
\bibitem [{\citenamefont {Magesan}\ \emph {et~al.}(2011)\citenamefont
  {Magesan}, \citenamefont {Gambetta},\ and\ \citenamefont
  {Emerson}}]{Magesan2011}%
  \BibitemOpen
  \bibfield  {author} {\bibinfo {author} {\bibfnamefont {Easwar}\ \bibnamefont
  {Magesan}}, \bibinfo {author} {\bibfnamefont {Jay~M.}\ \bibnamefont
  {Gambetta}}, \ and\ \bibinfo {author} {\bibfnamefont {Joseph}\ \bibnamefont
  {Emerson}},\ }\bibfield  {title} {\enquote {\bibinfo {title} {Scalable and
  robust randomized benchmarking of quantum processes},}\ }\href {\doibase
  10.1103/PhysRevLett.106.180504} {\bibfield  {journal} {\bibinfo  {journal}
  {Phys. Rev. Lett.}\ }\textbf {\bibinfo {volume} {106}},\ \bibinfo {pages}
  {180504} (\bibinfo {year} {2011})}\BibitemShut {NoStop}%
\bibitem [{\citenamefont {Magesan}\ \emph {et~al.}(2012)\citenamefont
  {Magesan}, \citenamefont {Gambetta}, \citenamefont {Johnson}, \citenamefont
  {Ryan}, \citenamefont {Chow}, \citenamefont {Merkel}, \citenamefont
  {da~Silva}, \citenamefont {Keefe}, \citenamefont {Rothwell},\ and\
  \citenamefont {\emph{et al.}}}]{Magesan2012b}%
  \BibitemOpen
  \bibfield  {author} {\bibinfo {author} {\bibfnamefont {Easwar}\ \bibnamefont
  {Magesan}}, \bibinfo {author} {\bibfnamefont {Jay~M.}\ \bibnamefont
  {Gambetta}}, \bibinfo {author} {\bibfnamefont {Blake~R.}\ \bibnamefont
  {Johnson}}, \bibinfo {author} {\bibfnamefont {Colm~A.}\ \bibnamefont {Ryan}},
  \bibinfo {author} {\bibfnamefont {Jerry~M.}\ \bibnamefont {Chow}}, \bibinfo
  {author} {\bibfnamefont {Seth~T.}\ \bibnamefont {Merkel}}, \bibinfo {author}
  {\bibfnamefont {Marcus~P.}\ \bibnamefont {da~Silva}}, \bibinfo {author}
  {\bibfnamefont {George~A.}\ \bibnamefont {Keefe}}, \bibinfo {author}
  {\bibfnamefont {Mary~B.}\ \bibnamefont {Rothwell}}, \ and\ \bibinfo {author}
  {\bibnamefont {\emph{et al.}}},\ }\bibfield  {title} {\enquote {\bibinfo
  {title} {Efficient measurement of quantum gate error by interleaved
  randomized benchmarking},}\ }\href {\doibase 10.1103/PhysRevLett.109.080505}
  {\bibfield  {journal} {\bibinfo  {journal} {Phys. Rev. Lett.}\ }\textbf
  {\bibinfo {volume} {109}},\ \bibinfo {pages} {080505} (\bibinfo {year}
  {2012})}\BibitemShut {NoStop}%
\bibitem [{\citenamefont {C\'orcoles}\ \emph {et~al.}(2013)\citenamefont
  {C\'orcoles}, \citenamefont {Gambetta}, \citenamefont {Chow}, \citenamefont
  {Smolin}, \citenamefont {Ware}, \citenamefont {Strand}, \citenamefont
  {Plourde},\ and\ \citenamefont {Steffen}}]{Corcoles2013}%
  \BibitemOpen
  \bibfield  {author} {\bibinfo {author} {\bibfnamefont {Antonio~D.}\
  \bibnamefont {C\'orcoles}}, \bibinfo {author} {\bibfnamefont {Jay~M.}\
  \bibnamefont {Gambetta}}, \bibinfo {author} {\bibfnamefont {Jerry~M.}\
  \bibnamefont {Chow}}, \bibinfo {author} {\bibfnamefont {John~A.}\
  \bibnamefont {Smolin}}, \bibinfo {author} {\bibfnamefont {Matthew}\
  \bibnamefont {Ware}}, \bibinfo {author} {\bibfnamefont {Joel}\ \bibnamefont
  {Strand}}, \bibinfo {author} {\bibfnamefont {Britton L.~T.}\ \bibnamefont
  {Plourde}}, \ and\ \bibinfo {author} {\bibfnamefont {Matthias}\ \bibnamefont
  {Steffen}},\ }\bibfield  {title} {\enquote {\bibinfo {title} {Process
  verification of two-qubit quantum gates by randomized benchmarking},}\ }\href
  {\doibase 10.1103/PhysRevA.87.030301} {\bibfield  {journal} {\bibinfo
  {journal} {Phys. Rev. A}\ }\textbf {\bibinfo {volume} {87}},\ \bibinfo
  {pages} {030301} (\bibinfo {year} {2013})}\BibitemShut {NoStop}%
\bibitem [{\citenamefont {O'Brien}\ \emph {et~al.}(2004)\citenamefont
  {O'Brien}, \citenamefont {Pryde}, \citenamefont {Gilchrist}, \citenamefont
  {James}, \citenamefont {Langford}, \citenamefont {Ralph},\ and\ \citenamefont
  {White}}]{OBrien2004}%
  \BibitemOpen
  \bibfield  {author} {\bibinfo {author} {\bibfnamefont {Jeremy~L.}\
  \bibnamefont {O'Brien}}, \bibinfo {author} {\bibfnamefont {Geoff~J.}\
  \bibnamefont {Pryde}}, \bibinfo {author} {\bibfnamefont {Alexei}\
  \bibnamefont {Gilchrist}}, \bibinfo {author} {\bibfnamefont {Daniel F.~V.}\
  \bibnamefont {James}}, \bibinfo {author} {\bibfnamefont {Nathan~K.}\
  \bibnamefont {Langford}}, \bibinfo {author} {\bibfnamefont {Timothy~C.}\
  \bibnamefont {Ralph}}, \ and\ \bibinfo {author} {\bibfnamefont {Andrew~G.}\
  \bibnamefont {White}},\ }\bibfield  {title} {\enquote {\bibinfo {title}
  {Quantum process tomography of a controlled-{NOT} gate},}\ }\href {\doibase
  10.1103/PhysRevLett.93.080502} {\bibfield  {journal} {\bibinfo  {journal}
  {Phys. Rev. Lett.}\ }\textbf {\bibinfo {volume} {93}},\ \bibinfo {pages}
  {080502} (\bibinfo {year} {2004})}\BibitemShut {NoStop}%
\bibitem [{\citenamefont {Bialczak}\ \emph {et~al.}(2010)\citenamefont
  {Bialczak}, \citenamefont {Ansmann}, \citenamefont {Hofheinz}, \citenamefont
  {Lucero}, \citenamefont {Neeley}, \citenamefont {O'Connell}, \citenamefont
  {Sank}, \citenamefont {Wang}, \citenamefont {Wenner}, \citenamefont
  {Steffen},\ and\ \citenamefont {\emph{et al.}}}]{Bialczak2010}%
  \BibitemOpen
  \bibfield  {author} {\bibinfo {author} {\bibfnamefont {Radoslaw~C.}\
  \bibnamefont {Bialczak}}, \bibinfo {author} {\bibfnamefont {Markus}\
  \bibnamefont {Ansmann}}, \bibinfo {author} {\bibfnamefont {Max}\ \bibnamefont
  {Hofheinz}}, \bibinfo {author} {\bibfnamefont {Erik}\ \bibnamefont {Lucero}},
  \bibinfo {author} {\bibfnamefont {Matthew}\ \bibnamefont {Neeley}}, \bibinfo
  {author} {\bibfnamefont {Aaron~D.}\ \bibnamefont {O'Connell}}, \bibinfo
  {author} {\bibfnamefont {Daniel}\ \bibnamefont {Sank}}, \bibinfo {author}
  {\bibfnamefont {Haohua}\ \bibnamefont {Wang}}, \bibinfo {author}
  {\bibfnamefont {James}\ \bibnamefont {Wenner}}, \bibinfo {author}
  {\bibfnamefont {Matthias}\ \bibnamefont {Steffen}}, \ and\ \bibinfo {author}
  {\bibnamefont {\emph{et al.}}},\ }\bibfield  {title} {\enquote {\bibinfo
  {title} {Quantum process tomography of a universal entangling gate
  implemented with josephson phase qubits},}\ }\href {\doibase
  10.1038/nphys1639} {\bibfield  {journal} {\bibinfo  {journal} {Nat. Phys.}\
  }\textbf {\bibinfo {volume} {6}},\ \bibinfo {pages} {409--413} (\bibinfo
  {year} {2010})}\BibitemShut {NoStop}%
\bibitem [{\citenamefont {Cross}\ \emph {et~al.}(2019)\citenamefont {Cross},
  \citenamefont {Bishop}, \citenamefont {Sheldon}, \citenamefont {Nation},\
  and\ \citenamefont {Gambetta}}]{Cross2019}%
  \BibitemOpen
  \bibfield  {author} {\bibinfo {author} {\bibfnamefont {Andrew~W.}\
  \bibnamefont {Cross}}, \bibinfo {author} {\bibfnamefont {Lev~S.}\
  \bibnamefont {Bishop}}, \bibinfo {author} {\bibfnamefont {Sarah}\
  \bibnamefont {Sheldon}}, \bibinfo {author} {\bibfnamefont {Paul~D.}\
  \bibnamefont {Nation}}, \ and\ \bibinfo {author} {\bibfnamefont {Jay~M.}\
  \bibnamefont {Gambetta}},\ }\bibfield  {title} {\enquote {\bibinfo {title}
  {Validating quantum computers using randomized model circuits},}\ }\href
  {\doibase 10.1103/PhysRevA.100.032328} {\bibfield  {journal} {\bibinfo
  {journal} {Phys. Rev. A}\ }\textbf {\bibinfo {volume} {100}},\ \bibinfo
  {pages} {032328} (\bibinfo {year} {2019})}\BibitemShut {NoStop}%
\bibitem [{\citenamefont {Jurcevic}\ \emph {et~al.}(2021)\citenamefont
  {Jurcevic}, \citenamefont {Javadi-Abhari}, \citenamefont {Bishop},
  \citenamefont {Lauer}, \citenamefont {Bogorin}, \citenamefont {Brink},
  \citenamefont {Capelluto}, \citenamefont {G{\"u}nl{\"u}k}, \citenamefont
  {Itoko}, \citenamefont {Kanazawa},\ and\ \citenamefont {\emph{et
  al.}}}]{Jurcevic2021}%
  \BibitemOpen
  \bibfield  {author} {\bibinfo {author} {\bibfnamefont {Petar}\ \bibnamefont
  {Jurcevic}}, \bibinfo {author} {\bibfnamefont {Ali}\ \bibnamefont
  {Javadi-Abhari}}, \bibinfo {author} {\bibfnamefont {Lev~S.}\ \bibnamefont
  {Bishop}}, \bibinfo {author} {\bibfnamefont {Isaac}\ \bibnamefont {Lauer}},
  \bibinfo {author} {\bibfnamefont {Daniela~F.}\ \bibnamefont {Bogorin}},
  \bibinfo {author} {\bibfnamefont {Markus}\ \bibnamefont {Brink}}, \bibinfo
  {author} {\bibfnamefont {Lauren}\ \bibnamefont {Capelluto}}, \bibinfo
  {author} {\bibfnamefont {Oktay}\ \bibnamefont {G{\"u}nl{\"u}k}}, \bibinfo
  {author} {\bibfnamefont {Toshinari}\ \bibnamefont {Itoko}}, \bibinfo {author}
  {\bibfnamefont {Naoki}\ \bibnamefont {Kanazawa}}, \ and\ \bibinfo {author}
  {\bibnamefont {\emph{et al.}}},\ }\bibfield  {title} {\enquote {\bibinfo
  {title} {Demonstration of quantum volume 64 on a superconducting quantum
  computing system},}\ }\href {\doibase 10.1088/2058-9565/abe519} {\bibfield
  {journal} {\bibinfo  {journal} {Quantum Sci. Technol.}\ }\textbf {\bibinfo
  {volume} {6}},\ \bibinfo {pages} {025020} (\bibinfo {year}
  {2021})}\BibitemShut {NoStop}%
\bibitem [{\citenamefont {Pelofske}\ \emph {et~al.}(2022)\citenamefont
  {Pelofske}, \citenamefont {Bärtschi},\ and\ \citenamefont
  {Eidenbenz}}]{Pelofske2022}%
  \BibitemOpen
  \bibfield  {author} {\bibinfo {author} {\bibfnamefont {Elijah}\ \bibnamefont
  {Pelofske}}, \bibinfo {author} {\bibfnamefont {Andreas}\ \bibnamefont
  {Bärtschi}}, \ and\ \bibinfo {author} {\bibfnamefont {Stephan}\ \bibnamefont
  {Eidenbenz}},\ }\href@noop {} {\enquote {\bibinfo {title} {Quantum volume in
  practice: What users can expect from {NISQ} devices},}\ } (\bibinfo {year}
  {2022}),\ \Eprint {http://arxiv.org/abs/2203.03816} {arXiv:2203.03816}
  \BibitemShut {NoStop}%
\bibitem [{\citenamefont {Arute}\ \emph {et~al.}(2020)\citenamefont {Arute},
  \citenamefont {Arya}, \citenamefont {Babbush}, \citenamefont {Bacon},
  \citenamefont {Bardin}, \citenamefont {Barends}, \citenamefont {Boixo},
  \citenamefont {Broughton}, \citenamefont {Buckley}, \citenamefont {Buell},\
  and\ \citenamefont {\emph{et al.}}}]{Arute2020}%
  \BibitemOpen
  \bibfield  {author} {\bibinfo {author} {\bibfnamefont {Frank}\ \bibnamefont
  {Arute}}, \bibinfo {author} {\bibfnamefont {Kunal}\ \bibnamefont {Arya}},
  \bibinfo {author} {\bibfnamefont {Ryan}\ \bibnamefont {Babbush}}, \bibinfo
  {author} {\bibfnamefont {Dave}\ \bibnamefont {Bacon}}, \bibinfo {author}
  {\bibfnamefont {Joseph~C.}\ \bibnamefont {Bardin}}, \bibinfo {author}
  {\bibfnamefont {Rami}\ \bibnamefont {Barends}}, \bibinfo {author}
  {\bibfnamefont {Sergio}\ \bibnamefont {Boixo}}, \bibinfo {author}
  {\bibfnamefont {Michael}\ \bibnamefont {Broughton}}, \bibinfo {author}
  {\bibfnamefont {Bob~B.}\ \bibnamefont {Buckley}}, \bibinfo {author}
  {\bibfnamefont {David~A.}\ \bibnamefont {Buell}}, \ and\ \bibinfo {author}
  {\bibnamefont {\emph{et al.}}},\ }\bibfield  {title} {\enquote {\bibinfo
  {title} {{Hartree-Fock} on a superconducting qubit quantum computer},}\
  }\href {\doibase 10.1126/science.abb9811} {\bibfield  {journal} {\bibinfo
  {journal} {Science}\ }\textbf {\bibinfo {volume} {369}},\ \bibinfo {pages}
  {1084--1089} (\bibinfo {year} {2020})}\BibitemShut {NoStop}%
\bibitem [{\citenamefont {Benedetti}\ \emph {et~al.}(2019)\citenamefont
  {Benedetti}, \citenamefont {Garcia-Pintos}, \citenamefont {Perdomo},
  \citenamefont {Leyton-Ortega}, \citenamefont {Nam},\ and\ \citenamefont
  {Perdomo-Ortiz}}]{Benedetti2019}%
  \BibitemOpen
  \bibfield  {author} {\bibinfo {author} {\bibfnamefont {Marcello}\
  \bibnamefont {Benedetti}}, \bibinfo {author} {\bibfnamefont {Delfina}\
  \bibnamefont {Garcia-Pintos}}, \bibinfo {author} {\bibfnamefont {Oscar}\
  \bibnamefont {Perdomo}}, \bibinfo {author} {\bibfnamefont {Vicente}\
  \bibnamefont {Leyton-Ortega}}, \bibinfo {author} {\bibfnamefont {Yunseong}\
  \bibnamefont {Nam}}, \ and\ \bibinfo {author} {\bibfnamefont {Alejandro}\
  \bibnamefont {Perdomo-Ortiz}},\ }\bibfield  {title} {\enquote {\bibinfo
  {title} {A generative modeling approach for benchmarking and training shallow
  quantum circuits},}\ }\href {\doibase 10.1038/s41534-019-0157-8} {\bibfield
  {journal} {\bibinfo  {journal} {Npj Quantum Inf.}\ }\textbf {\bibinfo
  {volume} {5}},\ \bibinfo {pages} {45} (\bibinfo {year} {2019})}\BibitemShut
  {NoStop}%
\bibitem [{\citenamefont {Dallaire-Demers}\ and\ \citenamefont
  {Killoran}(2018)}]{DallaireDemers2018}%
  \BibitemOpen
  \bibfield  {author} {\bibinfo {author} {\bibfnamefont {Pierre-Luc}\
  \bibnamefont {Dallaire-Demers}}\ and\ \bibinfo {author} {\bibfnamefont
  {Nathan}\ \bibnamefont {Killoran}},\ }\bibfield  {title} {\enquote {\bibinfo
  {title} {Quantum generative adversarial networks},}\ }\href {\doibase
  10.1103/PhysRevA.98.012324} {\bibfield  {journal} {\bibinfo  {journal} {Phys.
  Rev. A}\ }\textbf {\bibinfo {volume} {98}},\ \bibinfo {pages} {012324}
  (\bibinfo {year} {2018})}\BibitemShut {NoStop}%
\bibitem [{\citenamefont {Karamlou}\ \emph {et~al.}(2021)\citenamefont
  {Karamlou}, \citenamefont {Simon}, \citenamefont {Katabarwa}, \citenamefont
  {Scholten}, \citenamefont {Peropadre},\ and\ \citenamefont
  {Cao}}]{Karamlou2021}%
  \BibitemOpen
  \bibfield  {author} {\bibinfo {author} {\bibfnamefont {Amir~H.}\ \bibnamefont
  {Karamlou}}, \bibinfo {author} {\bibfnamefont {William~A.}\ \bibnamefont
  {Simon}}, \bibinfo {author} {\bibfnamefont {Amara}\ \bibnamefont
  {Katabarwa}}, \bibinfo {author} {\bibfnamefont {Travis~L.}\ \bibnamefont
  {Scholten}}, \bibinfo {author} {\bibfnamefont {Borja}\ \bibnamefont
  {Peropadre}}, \ and\ \bibinfo {author} {\bibfnamefont {Yudong}\ \bibnamefont
  {Cao}},\ }\bibfield  {title} {\enquote {\bibinfo {title} {Analyzing the
  performance of variational quantum factoring on a superconducting quantum
  processor},}\ }\href {\doibase 10.1038/s41534-021-00478-z} {\bibfield
  {journal} {\bibinfo  {journal} {Npj Quantum Inf.}\ }\textbf {\bibinfo
  {volume} {7}},\ \bibinfo {pages} {156} (\bibinfo {year} {2021})}\BibitemShut
  {NoStop}%
\bibitem [{\citenamefont {Dallaire-Demers}\ \emph {et~al.}(2020)\citenamefont
  {Dallaire-Demers}, \citenamefont {Stechly}, \citenamefont {Gonthier},
  \citenamefont {Bashige}, \citenamefont {Romero},\ and\ \citenamefont
  {Cao}}]{DallaireDemers2020}%
  \BibitemOpen
  \bibfield  {author} {\bibinfo {author} {\bibfnamefont {Pierre-Luc}\
  \bibnamefont {Dallaire-Demers}}, \bibinfo {author} {\bibfnamefont {Michal}\
  \bibnamefont {Stechly}}, \bibinfo {author} {\bibfnamefont {Jerome~F.}\
  \bibnamefont {Gonthier}}, \bibinfo {author} {\bibfnamefont
  {Ntwali~Toussaint}\ \bibnamefont {Bashige}}, \bibinfo {author} {\bibfnamefont
  {Jonathan}\ \bibnamefont {Romero}}, \ and\ \bibinfo {author} {\bibfnamefont
  {Yudong}\ \bibnamefont {Cao}},\ }\href@noop {} {\enquote {\bibinfo {title}
  {An application benchmark for fermionic quantum simulations},}\ } (\bibinfo
  {year} {2020}),\ \Eprint {http://arxiv.org/abs/2003.01862} {arXiv:2003.01862}
  \BibitemShut {NoStop}%
\bibitem [{\citenamefont {Schmoll}\ and\ \citenamefont
  {Or\'us}(2017)}]{Schmoll2017}%
  \BibitemOpen
  \bibfield  {author} {\bibinfo {author} {\bibfnamefont {Philipp}\ \bibnamefont
  {Schmoll}}\ and\ \bibinfo {author} {\bibfnamefont {Rom\'an}\ \bibnamefont
  {Or\'us}},\ }\bibfield  {title} {\enquote {\bibinfo {title} {Kitaev honeycomb
  tensor networks: Exact unitary circuits and applications},}\ }\href {\doibase
  10.1103/PhysRevB.95.045112} {\bibfield  {journal} {\bibinfo  {journal} {Phys.
  Rev. B}\ }\textbf {\bibinfo {volume} {95}},\ \bibinfo {pages} {045112}
  (\bibinfo {year} {2017})}\BibitemShut {NoStop}%
\bibitem [{\citenamefont {Fran{\c{c}}a}\ and\ \citenamefont
  {Garc{\'i}a-Patr{\'o}n}(2021)}]{Franca2021}%
  \BibitemOpen
  \bibfield  {author} {\bibinfo {author} {\bibfnamefont {Daniel~S.}\
  \bibnamefont {Fran{\c{c}}a}}\ and\ \bibinfo {author} {\bibfnamefont {Raul}\
  \bibnamefont {Garc{\'i}a-Patr{\'o}n}},\ }\bibfield  {title} {\enquote
  {\bibinfo {title} {Limitations of optimization algorithms on noisy quantum
  devices},}\ }\href {\doibase 10.1038/s41567-021-01356-3} {\bibfield
  {journal} {\bibinfo  {journal} {Nat. Phys.}\ }\textbf {\bibinfo {volume}
  {17}},\ \bibinfo {pages} {1221--1227} (\bibinfo {year} {2021})}\BibitemShut
  {NoStop}%
\bibitem [{\citenamefont {Weidenfeller}\ \emph {et~al.}(2022)\citenamefont
  {Weidenfeller}, \citenamefont {Valor}, \citenamefont {Gacon}, \citenamefont
  {Tornow}, \citenamefont {Bello}, \citenamefont {Woerner},\ and\ \citenamefont
  {Egger}}]{Weidenfeller2022}%
  \BibitemOpen
  \bibfield  {author} {\bibinfo {author} {\bibfnamefont {Johannes}\
  \bibnamefont {Weidenfeller}}, \bibinfo {author} {\bibfnamefont {Lucia~C.}\
  \bibnamefont {Valor}}, \bibinfo {author} {\bibfnamefont {Julien}\
  \bibnamefont {Gacon}}, \bibinfo {author} {\bibfnamefont {Caroline}\
  \bibnamefont {Tornow}}, \bibinfo {author} {\bibfnamefont {Luciano}\
  \bibnamefont {Bello}}, \bibinfo {author} {\bibfnamefont {Stefan}\
  \bibnamefont {Woerner}}, \ and\ \bibinfo {author} {\bibfnamefont {Daniel~J.}\
  \bibnamefont {Egger}},\ }\href@noop {} {\enquote {\bibinfo {title} {Scaling
  of the quantum approximate optimization algorithm on superconducting qubit
  based hardware},}\ } (\bibinfo {year} {2022}),\ \Eprint
  {http://arxiv.org/abs/2202.03459} {arXiv:2202.03459} \BibitemShut {NoStop}%
\bibitem [{\citenamefont {McKay}\ \emph {et~al.}(2017)\citenamefont {McKay},
  \citenamefont {Wood}, \citenamefont {Sheldon}, \citenamefont {Chow},\ and\
  \citenamefont {Gambetta}}]{McKay2017}%
  \BibitemOpen
  \bibfield  {author} {\bibinfo {author} {\bibfnamefont {David~C.}\
  \bibnamefont {McKay}}, \bibinfo {author} {\bibfnamefont {Christopher~J.}\
  \bibnamefont {Wood}}, \bibinfo {author} {\bibfnamefont {Sarah}\ \bibnamefont
  {Sheldon}}, \bibinfo {author} {\bibfnamefont {Jerry~M.}\ \bibnamefont
  {Chow}}, \ and\ \bibinfo {author} {\bibfnamefont {Jay~M.}\ \bibnamefont
  {Gambetta}},\ }\bibfield  {title} {\enquote {\bibinfo {title} {Efficient $z$
  gates for quantum computing},}\ }\href {\doibase 10.1103/PhysRevA.96.022330}
  {\bibfield  {journal} {\bibinfo  {journal} {Phys. Rev. A}\ }\textbf {\bibinfo
  {volume} {96}},\ \bibinfo {pages} {022330} (\bibinfo {year}
  {2017})}\BibitemShut {NoStop}%
\bibitem [{\citenamefont {Krantz}\ \emph {et~al.}(2019)\citenamefont {Krantz},
  \citenamefont {Kjaergaard}, \citenamefont {Yan}, \citenamefont {Orlando},
  \citenamefont {Gustavsson},\ and\ \citenamefont {Oliver}}]{Krantz2019}%
  \BibitemOpen
  \bibfield  {author} {\bibinfo {author} {\bibfnamefont {Philip}\ \bibnamefont
  {Krantz}}, \bibinfo {author} {\bibfnamefont {Morten}\ \bibnamefont
  {Kjaergaard}}, \bibinfo {author} {\bibfnamefont {Fei}\ \bibnamefont {Yan}},
  \bibinfo {author} {\bibfnamefont {Terry~P.}\ \bibnamefont {Orlando}},
  \bibinfo {author} {\bibfnamefont {Simon}\ \bibnamefont {Gustavsson}}, \ and\
  \bibinfo {author} {\bibfnamefont {William~D.}\ \bibnamefont {Oliver}},\
  }\bibfield  {title} {\enquote {\bibinfo {title} {A quantum engineer's guide
  to superconducting qubits},}\ }\href {\doibase 10.1063/1.5089550} {\bibfield
  {journal} {\bibinfo  {journal} {Appl. Phys. Rev.}\ }\textbf {\bibinfo
  {volume} {6}},\ \bibinfo {pages} {021318} (\bibinfo {year}
  {2019})}\BibitemShut {NoStop}%
\bibitem [{\citenamefont {Anis}\ \emph {et~al.}(2021)\citenamefont {Anis},
  \citenamefont {Abby-Mitchell}, \citenamefont {Abraham}, \citenamefont
  {Offei}, \citenamefont {Agarwal}, \citenamefont {Agliardi}, \citenamefont
  {Aharoni}, \citenamefont {Akhalwaya}, \citenamefont {Aleksandrowicz},
  \citenamefont {Alexander},\ and\ \citenamefont {\emph{et al.}}}]{Qiskit}%
  \BibitemOpen
  \bibfield  {author} {\bibinfo {author} {\bibfnamefont {MD~Sajid}\
  \bibnamefont {Anis}}, \bibinfo {author} {\bibnamefont {Abby-Mitchell}},
  \bibinfo {author} {\bibfnamefont {H{\'e}ctor}\ \bibnamefont {Abraham}},
  \bibinfo {author} {\bibfnamefont {Adu}\ \bibnamefont {Offei}}, \bibinfo
  {author} {\bibfnamefont {Rochisha}\ \bibnamefont {Agarwal}}, \bibinfo
  {author} {\bibfnamefont {Gabriele}\ \bibnamefont {Agliardi}}, \bibinfo
  {author} {\bibfnamefont {Merav}\ \bibnamefont {Aharoni}}, \bibinfo {author}
  {\bibfnamefont {Ismail~Yunus}\ \bibnamefont {Akhalwaya}}, \bibinfo {author}
  {\bibfnamefont {Gadi}\ \bibnamefont {Aleksandrowicz}}, \bibinfo {author}
  {\bibfnamefont {Thomas}\ \bibnamefont {Alexander}}, \ and\ \bibinfo {author}
  {\bibnamefont {\emph{et al.}}},\ }\href {\doibase 10.5281/zenodo.2573505}
  {\enquote {\bibinfo {title} {Qiskit: An open-source framework for quantum
  computing},}\ } (\bibinfo {year} {2021})\BibitemShut {NoStop}%
\bibitem [{\citenamefont {Jin}\ \emph {et~al.}(2021)\citenamefont {Jin},
  \citenamefont {Fong}, \citenamefont {Chen}, \citenamefont {Hayes},
  \citenamefont {Zhang}, \citenamefont {Zhang}, \citenamefont {Hua},
  \citenamefont {Zheng},\ and\ \citenamefont {Zhang}}]{Jin2021}%
  \BibitemOpen
  \bibfield  {author} {\bibinfo {author} {\bibfnamefont {Yuwei}\ \bibnamefont
  {Jin}}, \bibinfo {author} {\bibfnamefont {Lucent}\ \bibnamefont {Fong}},
  \bibinfo {author} {\bibfnamefont {Yanhao}\ \bibnamefont {Chen}}, \bibinfo
  {author} {\bibfnamefont {Ari~B.}\ \bibnamefont {Hayes}}, \bibinfo {author}
  {\bibfnamefont {Shuo}\ \bibnamefont {Zhang}}, \bibinfo {author}
  {\bibfnamefont {Chi}\ \bibnamefont {Zhang}}, \bibinfo {author} {\bibfnamefont
  {Fei}\ \bibnamefont {Hua}}, \bibinfo {author} {\bibnamefont {Zheng}}, \ and\
  \bibinfo {author} {\bibnamefont {Zhang}},\ }\href@noop {} {\enquote {\bibinfo
  {title} {A structured method for compilation of {QAOA} circuits in quantum
  computing},}\ } (\bibinfo {year} {2021}),\ \Eprint
  {http://arxiv.org/abs/2112.06143} {arXiv:2112.06143} \BibitemShut {NoStop}%
\bibitem [{\citenamefont {Sheldon}\ \emph {et~al.}(2016)\citenamefont
  {Sheldon}, \citenamefont {Magesan}, \citenamefont {Chow},\ and\ \citenamefont
  {Gambetta}}]{Sheldon2016}%
  \BibitemOpen
  \bibfield  {author} {\bibinfo {author} {\bibfnamefont {Sarah}\ \bibnamefont
  {Sheldon}}, \bibinfo {author} {\bibfnamefont {Easwar}\ \bibnamefont
  {Magesan}}, \bibinfo {author} {\bibfnamefont {Jerry~M.}\ \bibnamefont
  {Chow}}, \ and\ \bibinfo {author} {\bibfnamefont {Jay~M.}\ \bibnamefont
  {Gambetta}},\ }\bibfield  {title} {\enquote {\bibinfo {title} {Procedure for
  systematically tuning up cross-talk in the cross-resonance gate},}\ }\href
  {\doibase 10.1103/PhysRevA.93.060302} {\bibfield  {journal} {\bibinfo
  {journal} {Phys. Rev. A}\ }\textbf {\bibinfo {volume} {93}},\ \bibinfo
  {pages} {060302} (\bibinfo {year} {2016})}\BibitemShut {NoStop}%
\bibitem [{\citenamefont {Sundaresan}\ \emph {et~al.}(2020)\citenamefont
  {Sundaresan}, \citenamefont {Lauer}, \citenamefont {Pritchett}, \citenamefont
  {Magesan}, \citenamefont {Jurcevic},\ and\ \citenamefont
  {Gambetta}}]{Sundaresan2020}%
  \BibitemOpen
  \bibfield  {author} {\bibinfo {author} {\bibfnamefont {Neereja}\ \bibnamefont
  {Sundaresan}}, \bibinfo {author} {\bibfnamefont {Isaac}\ \bibnamefont
  {Lauer}}, \bibinfo {author} {\bibfnamefont {Emily}\ \bibnamefont
  {Pritchett}}, \bibinfo {author} {\bibfnamefont {Easwar}\ \bibnamefont
  {Magesan}}, \bibinfo {author} {\bibfnamefont {Petar}\ \bibnamefont
  {Jurcevic}}, \ and\ \bibinfo {author} {\bibfnamefont {Jay~M.}\ \bibnamefont
  {Gambetta}},\ }\bibfield  {title} {\enquote {\bibinfo {title} {Reducing
  unitary and spectator errors in cross resonance with optimized rotary
  echoes},}\ }\href {\doibase 10.1103/PRXQuantum.1.020318} {\bibfield
  {journal} {\bibinfo  {journal} {PRX Quantum}\ }\textbf {\bibinfo {volume}
  {1}},\ \bibinfo {pages} {020318} (\bibinfo {year} {2020})}\BibitemShut
  {NoStop}%
\bibitem [{\citenamefont {Earnest}\ \emph {et~al.}(2021)\citenamefont
  {Earnest}, \citenamefont {Tornow},\ and\ \citenamefont
  {Egger}}]{Earnest2021}%
  \BibitemOpen
  \bibfield  {author} {\bibinfo {author} {\bibfnamefont {Nathan}\ \bibnamefont
  {Earnest}}, \bibinfo {author} {\bibfnamefont {Caroline}\ \bibnamefont
  {Tornow}}, \ and\ \bibinfo {author} {\bibfnamefont {Daniel~J.}\ \bibnamefont
  {Egger}},\ }\bibfield  {title} {\enquote {\bibinfo {title} {Pulse-efficient
  circuit transpilation for quantum applications on cross-resonance-based
  hardware},}\ }\href {\doibase 10.1103/PhysRevResearch.3.043088} {\bibfield
  {journal} {\bibinfo  {journal} {Phys. Rev. Research}\ }\textbf {\bibinfo
  {volume} {3}},\ \bibinfo {pages} {043088} (\bibinfo {year}
  {2021})}\BibitemShut {NoStop}%
\bibitem [{\citenamefont {Alexander}\ \emph {et~al.}(2020)\citenamefont
  {Alexander}, \citenamefont {Kanazawa}, \citenamefont {Egger}, \citenamefont
  {Capelluto}, \citenamefont {Wood}, \citenamefont {Javadi-Abhari},\ and\
  \citenamefont {McKay}}]{Alexander2020}%
  \BibitemOpen
  \bibfield  {author} {\bibinfo {author} {\bibfnamefont {Thomas}\ \bibnamefont
  {Alexander}}, \bibinfo {author} {\bibfnamefont {Naoki}\ \bibnamefont
  {Kanazawa}}, \bibinfo {author} {\bibfnamefont {Daniel~J.}\ \bibnamefont
  {Egger}}, \bibinfo {author} {\bibfnamefont {Lauren}\ \bibnamefont
  {Capelluto}}, \bibinfo {author} {\bibfnamefont {Christopher~J.}\ \bibnamefont
  {Wood}}, \bibinfo {author} {\bibfnamefont {Ali}\ \bibnamefont
  {Javadi-Abhari}}, \ and\ \bibinfo {author} {\bibfnamefont {David~C.}\
  \bibnamefont {McKay}},\ }\bibfield  {title} {\enquote {\bibinfo {title}
  {Qiskit pulse: programming quantum computers through the cloud with
  pulses},}\ }\href {\doibase 10.1088/2058-9565/aba404} {\bibfield  {journal}
  {\bibinfo  {journal} {Quantum Sci. Technol.}\ }\textbf {\bibinfo {volume}
  {5}},\ \bibinfo {pages} {044006} (\bibinfo {year} {2020})}\BibitemShut
  {NoStop}%
\bibitem [{\citenamefont {Bravyi}\ \emph {et~al.}(2021)\citenamefont {Bravyi},
  \citenamefont {Sheldon}, \citenamefont {Kandala}, \citenamefont {Mckay},\
  and\ \citenamefont {Gambetta}}]{Bravyi2020}%
  \BibitemOpen
  \bibfield  {author} {\bibinfo {author} {\bibfnamefont {Sergey}\ \bibnamefont
  {Bravyi}}, \bibinfo {author} {\bibfnamefont {Sarah}\ \bibnamefont {Sheldon}},
  \bibinfo {author} {\bibfnamefont {Abhinav}\ \bibnamefont {Kandala}}, \bibinfo
  {author} {\bibfnamefont {David~C.}\ \bibnamefont {Mckay}}, \ and\ \bibinfo
  {author} {\bibfnamefont {Jay~M.}\ \bibnamefont {Gambetta}},\ }\bibfield
  {title} {\enquote {\bibinfo {title} {Mitigating measurement errors in
  multiqubit experiments},}\ }\href {\doibase 10.1103/PhysRevA.103.042605}
  {\bibfield  {journal} {\bibinfo  {journal} {Phys. Rev. A}\ }\textbf {\bibinfo
  {volume} {103}},\ \bibinfo {pages} {042605} (\bibinfo {year}
  {2021})}\BibitemShut {NoStop}%
\bibitem [{\citenamefont {Barron}\ and\ \citenamefont
  {Wood}(2020)}]{Barron2020a}%
  \BibitemOpen
  \bibfield  {author} {\bibinfo {author} {\bibfnamefont {George~S.}\
  \bibnamefont {Barron}}\ and\ \bibinfo {author} {\bibfnamefont
  {Christopher~J.}\ \bibnamefont {Wood}},\ }\href@noop {} {\enquote {\bibinfo
  {title} {Measurement error mitigation for variational quantum algorithms},}\
  } (\bibinfo {year} {2020}),\ \Eprint {http://arxiv.org/abs/2010.08520}
  {arXiv:2010.08520} \BibitemShut {NoStop}%
\bibitem [{\citenamefont {Sackett}\ \emph {et~al.}(2000)\citenamefont
  {Sackett}, \citenamefont {Kielpinski}, \citenamefont {King}, \citenamefont
  {Langer}, \citenamefont {Meyer}, \citenamefont {Myatt}, \citenamefont {Rowe},
  \citenamefont {Turchette}, \citenamefont {Itano}, \citenamefont {Wineland}
  \emph {et~al.}}]{sackett2000experimental}%
  \BibitemOpen
  \bibfield  {author} {\bibinfo {author} {\bibfnamefont {Cass~A}\ \bibnamefont
  {Sackett}}, \bibinfo {author} {\bibfnamefont {David}\ \bibnamefont
  {Kielpinski}}, \bibinfo {author} {\bibfnamefont {Brian~E}\ \bibnamefont
  {King}}, \bibinfo {author} {\bibfnamefont {Christopher}\ \bibnamefont
  {Langer}}, \bibinfo {author} {\bibfnamefont {Volker}\ \bibnamefont {Meyer}},
  \bibinfo {author} {\bibfnamefont {Christopher~J}\ \bibnamefont {Myatt}},
  \bibinfo {author} {\bibfnamefont {M}~\bibnamefont {Rowe}}, \bibinfo {author}
  {\bibfnamefont {QA}~\bibnamefont {Turchette}}, \bibinfo {author}
  {\bibfnamefont {Wayne~M}\ \bibnamefont {Itano}}, \bibinfo {author}
  {\bibfnamefont {David~J}\ \bibnamefont {Wineland}},  \emph {et~al.},\
  }\bibfield  {title} {\enquote {\bibinfo {title} {Experimental entanglement of
  four particles},}\ }\href {https://www.nature.com/articles/35005011}
  {\bibfield  {journal} {\bibinfo  {journal} {Nature}\ }\textbf {\bibinfo
  {volume} {404}},\ \bibinfo {pages} {256--259} (\bibinfo {year}
  {2000})}\BibitemShut {NoStop}%
\bibitem [{\citenamefont {Meyer}\ \emph {et~al.}(2001)\citenamefont {Meyer},
  \citenamefont {Rowe}, \citenamefont {Kielpinski}, \citenamefont {Sackett},
  \citenamefont {Itano}, \citenamefont {Monroe},\ and\ \citenamefont
  {Wineland}}]{meyer2001experimental}%
  \BibitemOpen
  \bibfield  {author} {\bibinfo {author} {\bibfnamefont {V.}~\bibnamefont
  {Meyer}}, \bibinfo {author} {\bibfnamefont {M.~A.}\ \bibnamefont {Rowe}},
  \bibinfo {author} {\bibfnamefont {D.}~\bibnamefont {Kielpinski}}, \bibinfo
  {author} {\bibfnamefont {C.~A.}\ \bibnamefont {Sackett}}, \bibinfo {author}
  {\bibfnamefont {W.~M.}\ \bibnamefont {Itano}}, \bibinfo {author}
  {\bibfnamefont {C.}~\bibnamefont {Monroe}}, \ and\ \bibinfo {author}
  {\bibfnamefont {D.~J.}\ \bibnamefont {Wineland}},\ }\bibfield  {title}
  {\enquote {\bibinfo {title} {Experimental demonstration of
  entanglement-enhanced rotation angle estimation using trapped ions},}\ }\href
  {\doibase 10.1103/PhysRevLett.86.5870} {\bibfield  {journal} {\bibinfo
  {journal} {Phys. Rev. Lett.}\ }\textbf {\bibinfo {volume} {86}},\ \bibinfo
  {pages} {5870--5873} (\bibinfo {year} {2001})}\BibitemShut {NoStop}%
\bibitem [{\citenamefont {Leibfried}\ \emph {et~al.}(2003)\citenamefont
  {Leibfried}, \citenamefont {DeMarco}, \citenamefont {Meyer}, \citenamefont
  {Lucas}, \citenamefont {Barrett}, \citenamefont {Britton}, \citenamefont
  {Itano}, \citenamefont {Jelenkovi{\'c}}, \citenamefont {Langer},
  \citenamefont {Rosenband} \emph {et~al.}}]{leibfried2003experimental}%
  \BibitemOpen
  \bibfield  {author} {\bibinfo {author} {\bibfnamefont {Dietrich}\
  \bibnamefont {Leibfried}}, \bibinfo {author} {\bibfnamefont {Brian}\
  \bibnamefont {DeMarco}}, \bibinfo {author} {\bibfnamefont {Volker}\
  \bibnamefont {Meyer}}, \bibinfo {author} {\bibfnamefont {David}\ \bibnamefont
  {Lucas}}, \bibinfo {author} {\bibfnamefont {Murray}\ \bibnamefont {Barrett}},
  \bibinfo {author} {\bibfnamefont {Joe}\ \bibnamefont {Britton}}, \bibinfo
  {author} {\bibfnamefont {Wayne~M}\ \bibnamefont {Itano}}, \bibinfo {author}
  {\bibfnamefont {B}~\bibnamefont {Jelenkovi{\'c}}}, \bibinfo {author}
  {\bibfnamefont {Chris}\ \bibnamefont {Langer}}, \bibinfo {author}
  {\bibfnamefont {Till}\ \bibnamefont {Rosenband}},  \emph {et~al.},\
  }\bibfield  {title} {\enquote {\bibinfo {title} {Experimental demonstration
  of a robust, high-fidelity geometric two ion-qubit phase gate},}\ }\href
  {https://www.nature.com/articles/nature01492#citeas} {\bibfield  {journal}
  {\bibinfo  {journal} {Nature}\ }\textbf {\bibinfo {volume} {422}},\ \bibinfo
  {pages} {412--415} (\bibinfo {year} {2003})}\BibitemShut {NoStop}%
\bibitem [{\citenamefont {Leibfried}\ \emph {et~al.}(2004)\citenamefont
  {Leibfried}, \citenamefont {Barrett}, \citenamefont {Schaetz}, \citenamefont
  {Britton}, \citenamefont {Chiaverini}, \citenamefont {Itano}, \citenamefont
  {Jost}, \citenamefont {Langer},\ and\ \citenamefont
  {Wineland}}]{leibfried2004toward}%
  \BibitemOpen
  \bibfield  {author} {\bibinfo {author} {\bibfnamefont {Dietrich}\
  \bibnamefont {Leibfried}}, \bibinfo {author} {\bibfnamefont {Murray~D}\
  \bibnamefont {Barrett}}, \bibinfo {author} {\bibfnamefont {T}~\bibnamefont
  {Schaetz}}, \bibinfo {author} {\bibfnamefont {Joseph}\ \bibnamefont
  {Britton}}, \bibinfo {author} {\bibfnamefont {J}~\bibnamefont {Chiaverini}},
  \bibinfo {author} {\bibfnamefont {Wayne~M}\ \bibnamefont {Itano}}, \bibinfo
  {author} {\bibfnamefont {John~D}\ \bibnamefont {Jost}}, \bibinfo {author}
  {\bibfnamefont {Christopher}\ \bibnamefont {Langer}}, \ and\ \bibinfo
  {author} {\bibfnamefont {David~J}\ \bibnamefont {Wineland}},\ }\bibfield
  {title} {\enquote {\bibinfo {title} {Toward heisenberg-limited spectroscopy
  with multiparticle entangled states},}\ }\href
  {https://www.science.org/doi/10.1126/science.1097576} {\bibfield  {journal}
  {\bibinfo  {journal} {Science}\ }\textbf {\bibinfo {volume} {304}},\ \bibinfo
  {pages} {1476--1478} (\bibinfo {year} {2004})}\BibitemShut {NoStop}%
\bibitem [{\citenamefont {Leibfried}\ \emph {et~al.}(2005)\citenamefont
  {Leibfried}, \citenamefont {Knill}, \citenamefont {Seidelin}, \citenamefont
  {Britton}, \citenamefont {Blakestad}, \citenamefont {Chiaverini},
  \citenamefont {Hume}, \citenamefont {Itano}, \citenamefont {Jost},
  \citenamefont {Langer} \emph {et~al.}}]{leibfried2005creation}%
  \BibitemOpen
  \bibfield  {author} {\bibinfo {author} {\bibfnamefont {Dietrich}\
  \bibnamefont {Leibfried}}, \bibinfo {author} {\bibfnamefont {Emanuel}\
  \bibnamefont {Knill}}, \bibinfo {author} {\bibfnamefont {Signe}\ \bibnamefont
  {Seidelin}}, \bibinfo {author} {\bibfnamefont {Joe}\ \bibnamefont {Britton}},
  \bibinfo {author} {\bibfnamefont {R~Brad}\ \bibnamefont {Blakestad}},
  \bibinfo {author} {\bibfnamefont {John}\ \bibnamefont {Chiaverini}}, \bibinfo
  {author} {\bibfnamefont {David~B}\ \bibnamefont {Hume}}, \bibinfo {author}
  {\bibfnamefont {Wayne~M}\ \bibnamefont {Itano}}, \bibinfo {author}
  {\bibfnamefont {John~D}\ \bibnamefont {Jost}}, \bibinfo {author}
  {\bibfnamefont {Christopher}\ \bibnamefont {Langer}},  \emph {et~al.},\
  }\bibfield  {title} {\enquote {\bibinfo {title} {Creation of a six-atom
  ‘schr{\"o}dinger cat’state},}\ }\href
  {https://www.nature.com/articles/nature04251#content} {\bibfield  {journal}
  {\bibinfo  {journal} {Nature}\ }\textbf {\bibinfo {volume} {438}},\ \bibinfo
  {pages} {639--642} (\bibinfo {year} {2005})}\BibitemShut {NoStop}%
\bibitem [{\citenamefont {Monz}\ \emph {et~al.}(2011)\citenamefont {Monz},
  \citenamefont {Schindler}, \citenamefont {Barreiro}, \citenamefont {Chwalla},
  \citenamefont {Nigg}, \citenamefont {Coish}, \citenamefont {Harlander},
  \citenamefont {H\"ansel}, \citenamefont {Hennrich},\ and\ \citenamefont
  {Blatt}}]{monz2011}%
  \BibitemOpen
  \bibfield  {author} {\bibinfo {author} {\bibfnamefont {Thomas}\ \bibnamefont
  {Monz}}, \bibinfo {author} {\bibfnamefont {Philipp}\ \bibnamefont
  {Schindler}}, \bibinfo {author} {\bibfnamefont {Julio~T.}\ \bibnamefont
  {Barreiro}}, \bibinfo {author} {\bibfnamefont {Michael}\ \bibnamefont
  {Chwalla}}, \bibinfo {author} {\bibfnamefont {Daniel}\ \bibnamefont {Nigg}},
  \bibinfo {author} {\bibfnamefont {William~A.}\ \bibnamefont {Coish}},
  \bibinfo {author} {\bibfnamefont {Maximilian}\ \bibnamefont {Harlander}},
  \bibinfo {author} {\bibfnamefont {Wolfgang}\ \bibnamefont {H\"ansel}},
  \bibinfo {author} {\bibfnamefont {Markus}\ \bibnamefont {Hennrich}}, \ and\
  \bibinfo {author} {\bibfnamefont {Rainer}\ \bibnamefont {Blatt}},\ }\bibfield
   {title} {\enquote {\bibinfo {title} {14-qubit entanglement: Creation and
  coherence},}\ }\href {\doibase 10.1103/PhysRevLett.106.130506} {\bibfield
  {journal} {\bibinfo  {journal} {Phys. Rev. Lett.}\ }\textbf {\bibinfo
  {volume} {106}},\ \bibinfo {pages} {130506} (\bibinfo {year}
  {2011})}\BibitemShut {NoStop}%
\bibitem [{\citenamefont {Kaubruegger}\ \emph {et~al.}(2019)\citenamefont
  {Kaubruegger}, \citenamefont {Silvi}, \citenamefont {Kokail}, \citenamefont
  {van Bijnen}, \citenamefont {Rey}, \citenamefont {Ye}, \citenamefont
  {Kaufman},\ and\ \citenamefont {Zoller}}]{kaubruegger2019variational}%
  \BibitemOpen
  \bibfield  {author} {\bibinfo {author} {\bibfnamefont {Raphael}\ \bibnamefont
  {Kaubruegger}}, \bibinfo {author} {\bibfnamefont {Pietro}\ \bibnamefont
  {Silvi}}, \bibinfo {author} {\bibfnamefont {Christian}\ \bibnamefont
  {Kokail}}, \bibinfo {author} {\bibfnamefont {Rick}\ \bibnamefont {van
  Bijnen}}, \bibinfo {author} {\bibfnamefont {Ana~Maria}\ \bibnamefont {Rey}},
  \bibinfo {author} {\bibfnamefont {Jun}\ \bibnamefont {Ye}}, \bibinfo {author}
  {\bibfnamefont {Adam~M.}\ \bibnamefont {Kaufman}}, \ and\ \bibinfo {author}
  {\bibfnamefont {Peter}\ \bibnamefont {Zoller}},\ }\bibfield  {title}
  {\enquote {\bibinfo {title} {Variational spin-squeezing algorithms on
  programmable quantum sensors},}\ }\href {\doibase
  10.1103/PhysRevLett.123.260505} {\bibfield  {journal} {\bibinfo  {journal}
  {Phys. Rev. Lett.}\ }\textbf {\bibinfo {volume} {123}},\ \bibinfo {pages}
  {260505} (\bibinfo {year} {2019})}\BibitemShut {NoStop}%
\bibitem [{\citenamefont {Koczor}\ \emph {et~al.}(2020)\citenamefont {Koczor},
  \citenamefont {Endo}, \citenamefont {Jones}, \citenamefont {Matsuzaki},\ and\
  \citenamefont {Benjamin}}]{koczor2020variational}%
  \BibitemOpen
  \bibfield  {author} {\bibinfo {author} {\bibfnamefont {B{\'a}lint}\
  \bibnamefont {Koczor}}, \bibinfo {author} {\bibfnamefont {Suguru}\
  \bibnamefont {Endo}}, \bibinfo {author} {\bibfnamefont {Tyson}\ \bibnamefont
  {Jones}}, \bibinfo {author} {\bibfnamefont {Yuichiro}\ \bibnamefont
  {Matsuzaki}}, \ and\ \bibinfo {author} {\bibfnamefont {Simon~C}\ \bibnamefont
  {Benjamin}},\ }\bibfield  {title} {\enquote {\bibinfo {title}
  {Variational-state quantum metrology},}\ }\href
  {https://iopscience.iop.org/article/10.1088/1367-2630/ab965e/meta} {\bibfield
   {journal} {\bibinfo  {journal} {New Journal of Physics}\ }\textbf {\bibinfo
  {volume} {22}},\ \bibinfo {pages} {083038} (\bibinfo {year}
  {2020})}\BibitemShut {NoStop}%
\bibitem [{\citenamefont {Meyer}\ \emph {et~al.}(2021)\citenamefont {Meyer},
  \citenamefont {Borregaard},\ and\ \citenamefont
  {Eisert}}]{meyer2021variational}%
  \BibitemOpen
  \bibfield  {author} {\bibinfo {author} {\bibfnamefont {Johannes~Jakob}\
  \bibnamefont {Meyer}}, \bibinfo {author} {\bibfnamefont {Johannes}\
  \bibnamefont {Borregaard}}, \ and\ \bibinfo {author} {\bibfnamefont {Jens}\
  \bibnamefont {Eisert}},\ }\bibfield  {title} {\enquote {\bibinfo {title} {A
  variational toolbox for quantum multi-parameter estimation},}\ }\href
  {https://www.nature.com/articles/s41534-021-00425-y} {\bibfield  {journal}
  {\bibinfo  {journal} {npj Quantum Information}\ }\textbf {\bibinfo {volume}
  {7}},\ \bibinfo {pages} {1--5} (\bibinfo {year} {2021})}\BibitemShut
  {NoStop}%
\bibitem [{\citenamefont {Spall}(1998)}]{spall1998overview}%
  \BibitemOpen
  \bibfield  {author} {\bibinfo {author} {\bibfnamefont {James~C.}\
  \bibnamefont {Spall}},\ }\bibfield  {title} {\enquote {\bibinfo {title} {An
  overview of the simultaneous perturbation method for efficient
  optimization},}\ }\href
  {https://www.jhuapl.edu/spsa/PDF-SPSA/Spall_An_Overview.PDF} {\bibfield
  {journal} {\bibinfo  {journal} {{Johns} {Hopkins} {APL} technical digest}\
  }\textbf {\bibinfo {volume} {19}},\ \bibinfo {pages} {482--492} (\bibinfo
  {year} {1998})}\BibitemShut {NoStop}%
\bibitem [{\citenamefont {Apellaniz}\ \emph {et~al.}(2015)\citenamefont
  {Apellaniz}, \citenamefont {Lücke}, \citenamefont {Peise}, \citenamefont
  {Klempt},\ and\ \citenamefont {T{\'{o}}th}}]{Apellaniz2015}%
  \BibitemOpen
  \bibfield  {author} {\bibinfo {author} {\bibfnamefont {Iagoba}\ \bibnamefont
  {Apellaniz}}, \bibinfo {author} {\bibfnamefont {Bernd}\ \bibnamefont
  {Lücke}}, \bibinfo {author} {\bibfnamefont {Jan}\ \bibnamefont {Peise}},
  \bibinfo {author} {\bibfnamefont {Carsten}\ \bibnamefont {Klempt}}, \ and\
  \bibinfo {author} {\bibfnamefont {G{\'{e}}za}\ \bibnamefont {T{\'{o}}th}},\
  }\bibfield  {title} {\enquote {\bibinfo {title} {Detecting metrologically
  useful entanglement in the vicinity of {Dicke} states},}\ }\href {\doibase
  10.1088/1367-2630/17/8/083027} {\bibfield  {journal} {\bibinfo  {journal}
  {New J. Phys.}\ }\textbf {\bibinfo {volume} {17}},\ \bibinfo {pages} {083027}
  (\bibinfo {year} {2015})}\BibitemShut {NoStop}%
\bibitem [{\citenamefont {L{\"u}cke}\ \emph {et~al.}(2014)\citenamefont
  {L{\"u}cke}, \citenamefont {Peise}, \citenamefont {Vitagliano}, \citenamefont
  {Arlt}, \citenamefont {Santos}, \citenamefont {T{\'o}th},\ and\ \citenamefont
  {Klempt}}]{lucke2014detecting}%
  \BibitemOpen
  \bibfield  {author} {\bibinfo {author} {\bibfnamefont {Bernd}\ \bibnamefont
  {L{\"u}cke}}, \bibinfo {author} {\bibfnamefont {Jan}\ \bibnamefont {Peise}},
  \bibinfo {author} {\bibfnamefont {Giuseppe}\ \bibnamefont {Vitagliano}},
  \bibinfo {author} {\bibfnamefont {Jan}\ \bibnamefont {Arlt}}, \bibinfo
  {author} {\bibfnamefont {Luis}\ \bibnamefont {Santos}}, \bibinfo {author}
  {\bibfnamefont {G{\'e}za}\ \bibnamefont {T{\'o}th}}, \ and\ \bibinfo {author}
  {\bibfnamefont {Carsten}\ \bibnamefont {Klempt}},\ }\bibfield  {title}
  {\enquote {\bibinfo {title} {Detecting multiparticle entanglement of {Dicke}
  states},}\ }\href {\doibase 10.1103/PhysRevLett.112.155304} {\bibfield
  {journal} {\bibinfo  {journal} {Phys. Rev. Lett.}\ }\textbf {\bibinfo
  {volume} {112}},\ \bibinfo {pages} {155304} (\bibinfo {year}
  {2014})}\BibitemShut {NoStop}%
\bibitem [{\citenamefont {Hauke}\ \emph {et~al.}(2016)\citenamefont {Hauke},
  \citenamefont {Heyl}, \citenamefont {Tagliacozzo},\ and\ \citenamefont
  {Zoller}}]{hauke2016measuring}%
  \BibitemOpen
  \bibfield  {author} {\bibinfo {author} {\bibfnamefont {Philipp}\ \bibnamefont
  {Hauke}}, \bibinfo {author} {\bibfnamefont {Markus}\ \bibnamefont {Heyl}},
  \bibinfo {author} {\bibfnamefont {Luca}\ \bibnamefont {Tagliacozzo}}, \ and\
  \bibinfo {author} {\bibfnamefont {Peter}\ \bibnamefont {Zoller}},\ }\bibfield
   {title} {\enquote {\bibinfo {title} {Measuring multipartite entanglement
  through dynamic susceptibilities},}\ }\href
  {https://doi.org/10.1038/nphys3700} {\bibfield  {journal} {\bibinfo
  {journal} {Nat. Phys.}\ }\textbf {\bibinfo {volume} {12}},\ \bibinfo {pages}
  {778--782} (\bibinfo {year} {2016})}\BibitemShut {NoStop}%
\bibitem [{\citenamefont {Smith}\ \emph {et~al.}(2016)\citenamefont {Smith},
  \citenamefont {Lee}, \citenamefont {Richerme}, \citenamefont {Neyenhuis},
  \citenamefont {Hess}, \citenamefont {Hauke}, \citenamefont {Heyl},
  \citenamefont {Huse},\ and\ \citenamefont {Monroe}}]{smith2016many}%
  \BibitemOpen
  \bibfield  {author} {\bibinfo {author} {\bibfnamefont {Jacob}\ \bibnamefont
  {Smith}}, \bibinfo {author} {\bibfnamefont {Aaron}\ \bibnamefont {Lee}},
  \bibinfo {author} {\bibfnamefont {Philip}\ \bibnamefont {Richerme}}, \bibinfo
  {author} {\bibfnamefont {Brian}\ \bibnamefont {Neyenhuis}}, \bibinfo {author}
  {\bibfnamefont {Paul~W.}\ \bibnamefont {Hess}}, \bibinfo {author}
  {\bibfnamefont {Philipp}\ \bibnamefont {Hauke}}, \bibinfo {author}
  {\bibfnamefont {Markus}\ \bibnamefont {Heyl}}, \bibinfo {author}
  {\bibfnamefont {David~A.}\ \bibnamefont {Huse}}, \ and\ \bibinfo {author}
  {\bibfnamefont {Christopher}\ \bibnamefont {Monroe}},\ }\bibfield  {title}
  {\enquote {\bibinfo {title} {Many-body localization in a quantum simulator
  with programmable random disorder},}\ }\href {\doibase
  https://doi.org/10.1038/nphys3783} {\bibfield  {journal} {\bibinfo  {journal}
  {Nat. Phys.}\ }\textbf {\bibinfo {volume} {12}},\ \bibinfo {pages} {907--911}
  (\bibinfo {year} {2016})}\BibitemShut {NoStop}%
\bibitem [{\citenamefont {Wang}\ \emph {et~al.}(2014)\citenamefont {Wang},
  \citenamefont {Wu}, \citenamefont {Yang}, \citenamefont {Jin}, \citenamefont
  {Lambert},\ and\ \citenamefont {Nori}}]{wang2014quantum}%
  \BibitemOpen
  \bibfield  {author} {\bibinfo {author} {\bibfnamefont {Teng-Long}\
  \bibnamefont {Wang}}, \bibinfo {author} {\bibfnamefont {Ling-Na}\
  \bibnamefont {Wu}}, \bibinfo {author} {\bibfnamefont {Wen}\ \bibnamefont
  {Yang}}, \bibinfo {author} {\bibfnamefont {Guang-Ri}\ \bibnamefont {Jin}},
  \bibinfo {author} {\bibfnamefont {Neill}\ \bibnamefont {Lambert}}, \ and\
  \bibinfo {author} {\bibfnamefont {Franco}\ \bibnamefont {Nori}},\ }\bibfield
  {title} {\enquote {\bibinfo {title} {Quantum fisher information as a
  signature of the superradiant quantum phase transition},}\ }\href
  {https://doi.org/10.1088/1367-2630/16/6/063039} {\bibfield  {journal}
  {\bibinfo  {journal} {New Journal of Physics}\ }\textbf {\bibinfo {volume}
  {16}},\ \bibinfo {pages} {063039} (\bibinfo {year} {2014})}\BibitemShut
  {NoStop}%
\bibitem [{\citenamefont {Yin}\ \emph {et~al.}(2019)\citenamefont {Yin},
  \citenamefont {Song}, \citenamefont {Zhang},\ and\ \citenamefont
  {Liu}}]{yin2019}%
  \BibitemOpen
  \bibfield  {author} {\bibinfo {author} {\bibfnamefont {Shaoying}\
  \bibnamefont {Yin}}, \bibinfo {author} {\bibfnamefont {Jie}\ \bibnamefont
  {Song}}, \bibinfo {author} {\bibfnamefont {Yujun}\ \bibnamefont {Zhang}}, \
  and\ \bibinfo {author} {\bibfnamefont {Shutian}\ \bibnamefont {Liu}},\
  }\bibfield  {title} {\enquote {\bibinfo {title} {Quantum fisher information
  in quantum critical systems with topological characterization},}\ }\href
  {\doibase 10.1103/PhysRevB.100.184417} {\bibfield  {journal} {\bibinfo
  {journal} {Phys. Rev. B}\ }\textbf {\bibinfo {volume} {100}},\ \bibinfo
  {pages} {184417} (\bibinfo {year} {2019})}\BibitemShut {NoStop}%
\bibitem [{\citenamefont {Mathew}\ \emph {et~al.}(2020)\citenamefont {Mathew},
  \citenamefont {Silva}, \citenamefont {Jain}, \citenamefont {Mohan},
  \citenamefont {Adroja}, \citenamefont {Sakai}, \citenamefont {Tomy},
  \citenamefont {Banerjee}, \citenamefont {Goreti}, \citenamefont {N.},
  \citenamefont {Singh},\ and\ \citenamefont
  {Jaiswal-Nagar}}]{mathew2020experimental}%
  \BibitemOpen
  \bibfield  {author} {\bibinfo {author} {\bibfnamefont {George}\ \bibnamefont
  {Mathew}}, \bibinfo {author} {\bibfnamefont {Saulo L.~L.}\ \bibnamefont
  {Silva}}, \bibinfo {author} {\bibfnamefont {Anil}\ \bibnamefont {Jain}},
  \bibinfo {author} {\bibfnamefont {Arya}\ \bibnamefont {Mohan}}, \bibinfo
  {author} {\bibfnamefont {Devashi~T.}\ \bibnamefont {Adroja}}, \bibinfo
  {author} {\bibfnamefont {V.~G.}\ \bibnamefont {Sakai}}, \bibinfo {author}
  {\bibfnamefont {C.~V.}\ \bibnamefont {Tomy}}, \bibinfo {author}
  {\bibfnamefont {Alok}\ \bibnamefont {Banerjee}}, \bibinfo {author}
  {\bibfnamefont {Rajendar}\ \bibnamefont {Goreti}}, \bibinfo {author}
  {\bibfnamefont {Aswathi~V.}\ \bibnamefont {N.}}, \bibinfo {author}
  {\bibfnamefont {Ranjit}\ \bibnamefont {Singh}}, \ and\ \bibinfo {author}
  {\bibfnamefont {D.}~\bibnamefont {Jaiswal-Nagar}},\ }\bibfield  {title}
  {\enquote {\bibinfo {title} {Experimental realization of multipartite
  entanglement via quantum {Fisher} information in a uniform antiferromagnetic
  quantum spin chain},}\ }\href {\doibase 10.1103/PhysRevResearch.2.043329}
  {\bibfield  {journal} {\bibinfo  {journal} {Phys. Rev. Research}\ }\textbf
  {\bibinfo {volume} {2}},\ \bibinfo {pages} {043329} (\bibinfo {year}
  {2020})}\BibitemShut {NoStop}%
\bibitem [{\citenamefont {Laurell}\ \emph {et~al.}(2021)\citenamefont
  {Laurell}, \citenamefont {Scheie}, \citenamefont {Mukherjee}, \citenamefont
  {Koza}, \citenamefont {Enderle}, \citenamefont {Tylczynski}, \citenamefont
  {Okamoto}, \citenamefont {Coldea}, \citenamefont {Tennant},\ and\
  \citenamefont {Alvarez}}]{laurell2021}%
  \BibitemOpen
  \bibfield  {author} {\bibinfo {author} {\bibfnamefont {Pontus}\ \bibnamefont
  {Laurell}}, \bibinfo {author} {\bibfnamefont {Allen}\ \bibnamefont {Scheie}},
  \bibinfo {author} {\bibfnamefont {Chiron~J.}\ \bibnamefont {Mukherjee}},
  \bibinfo {author} {\bibfnamefont {Michael~M.}\ \bibnamefont {Koza}}, \bibinfo
  {author} {\bibfnamefont {Mechtild}\ \bibnamefont {Enderle}}, \bibinfo
  {author} {\bibfnamefont {Zbigniew}\ \bibnamefont {Tylczynski}}, \bibinfo
  {author} {\bibfnamefont {Satoshi}\ \bibnamefont {Okamoto}}, \bibinfo {author}
  {\bibfnamefont {Radu}\ \bibnamefont {Coldea}}, \bibinfo {author}
  {\bibfnamefont {D.~Alan}\ \bibnamefont {Tennant}}, \ and\ \bibinfo {author}
  {\bibfnamefont {Gonzalo}\ \bibnamefont {Alvarez}},\ }\bibfield  {title}
  {\enquote {\bibinfo {title} {Quantifying and controlling entanglement in the
  quantum magnet {${\text{Cs}}_{2}{\text{CoCl}}_{4}$}},}\ }\href {\doibase
  10.1103/PhysRevLett.127.037201} {\bibfield  {journal} {\bibinfo  {journal}
  {Phys. Rev. Lett.}\ }\textbf {\bibinfo {volume} {127}},\ \bibinfo {pages}
  {037201} (\bibinfo {year} {2021})}\BibitemShut {NoStop}%
\bibitem [{\citenamefont {Braunstein}\ and\ \citenamefont
  {Caves}(1994)}]{braunstein1994}%
  \BibitemOpen
  \bibfield  {author} {\bibinfo {author} {\bibfnamefont {Samuel~L.}\
  \bibnamefont {Braunstein}}\ and\ \bibinfo {author} {\bibfnamefont
  {Carlton~M.}\ \bibnamefont {Caves}},\ }\bibfield  {title} {\enquote {\bibinfo
  {title} {Statistical distance and the geometry of quantum states},}\ }\href
  {\doibase 10.1103/PhysRevLett.72.3439} {\bibfield  {journal} {\bibinfo
  {journal} {Phys. Rev. Lett.}\ }\textbf {\bibinfo {volume} {72}},\ \bibinfo
  {pages} {3439--3443} (\bibinfo {year} {1994})}\BibitemShut {NoStop}%
\bibitem [{\citenamefont {Pezze}\ and\ \citenamefont
  {Smerzi}(2014)}]{pezze2014quantum}%
  \BibitemOpen
  \bibfield  {author} {\bibinfo {author} {\bibfnamefont {Luca}\ \bibnamefont
  {Pezze}}\ and\ \bibinfo {author} {\bibfnamefont {Augusto}\ \bibnamefont
  {Smerzi}},\ }\bibfield  {title} {\enquote {\bibinfo {title} {Quantum theory
  of phase estimation},}\ }\href {https://doi.org/10.48550/arXiv.1411.5164}
  {\bibfield  {journal} {\bibinfo  {journal} {arXiv:1411.5164}\ } (\bibinfo
  {year} {2014})}\BibitemShut {NoStop}%
\bibitem [{\citenamefont {T{\'o}th}\ and\ \citenamefont
  {Apellaniz}(2014)}]{toth2014quantum}%
  \BibitemOpen
  \bibfield  {author} {\bibinfo {author} {\bibfnamefont {G{\'e}za}\
  \bibnamefont {T{\'o}th}}\ and\ \bibinfo {author} {\bibfnamefont {Iagoba}\
  \bibnamefont {Apellaniz}},\ }\bibfield  {title} {\enquote {\bibinfo {title}
  {Quantum metrology from a quantum information science perspective},}\ }\href
  {https://doi.org/10.1088/1751-8113/47/42/424006} {\bibfield  {journal}
  {\bibinfo  {journal} {J. Phys. A: Math. Theor.}\ }\textbf {\bibinfo {volume}
  {47}},\ \bibinfo {pages} {424006} (\bibinfo {year} {2014})}\BibitemShut
  {NoStop}%
\bibitem [{\citenamefont {Costa~de Almeida}\ and\ \citenamefont
  {Hauke}(2021)}]{CostadeAlmeida2021}%
  \BibitemOpen
  \bibfield  {author} {\bibinfo {author} {\bibfnamefont {Ricardo}\ \bibnamefont
  {Costa~de Almeida}}\ and\ \bibinfo {author} {\bibfnamefont {Philipp}\
  \bibnamefont {Hauke}},\ }\bibfield  {title} {\enquote {\bibinfo {title} {From
  entanglement certification with quench dynamics to multipartite entanglement
  of interacting fermions},}\ }\href {\doibase
  10.1103/PhysRevResearch.3.L032051} {\bibfield  {journal} {\bibinfo  {journal}
  {Phys. Rev. Research}\ }\textbf {\bibinfo {volume} {3}},\ \bibinfo {pages}
  {L032051} (\bibinfo {year} {2021})}\BibitemShut {NoStop}%
\bibitem [{\citenamefont {Vidal}\ and\ \citenamefont
  {Dawson}(2004)}]{Vidal2004}%
  \BibitemOpen
  \bibfield  {author} {\bibinfo {author} {\bibfnamefont {Guifre}\ \bibnamefont
  {Vidal}}\ and\ \bibinfo {author} {\bibfnamefont {Christopher~M.}\
  \bibnamefont {Dawson}},\ }\bibfield  {title} {\enquote {\bibinfo {title}
  {Universal quantum circuit for two-qubit transformations with three
  controlled-{NOT} gates},}\ }\href {\doibase 10.1103/PhysRevA.69.010301}
  {\bibfield  {journal} {\bibinfo  {journal} {Phys. Rev. A}\ }\textbf {\bibinfo
  {volume} {69}},\ \bibinfo {pages} {010301} (\bibinfo {year}
  {2004})}\BibitemShut {NoStop}%
\bibitem [{\citenamefont {Strobel}(2016)}]{StrobelThesis}%
  \BibitemOpen
  \bibfield  {author} {\bibinfo {author} {\bibfnamefont {Helmut}\ \bibnamefont
  {Strobel}},\ }\emph {\bibinfo {title} {Fisher Information and entanglement of
  non-Gaussian spin states}},\ \href {\doibase
  https://doi.org/10.11588/heidok.00020251} {Ph.D. thesis},\ \bibinfo  {school}
  {Kirchhoff Institute for Physics, Universit\"at Heidelberg} (\bibinfo {year}
  {2016})\BibitemShut {NoStop}%
\bibitem [{\citenamefont {Farhi}\ \emph {et~al.}(2020)\citenamefont {Farhi},
  \citenamefont {Gamarnik},\ and\ \citenamefont {Gutmann}}]{Farhi2020}%
  \BibitemOpen
  \bibfield  {author} {\bibinfo {author} {\bibfnamefont {Edward}\ \bibnamefont
  {Farhi}}, \bibinfo {author} {\bibfnamefont {David}\ \bibnamefont {Gamarnik}},
  \ and\ \bibinfo {author} {\bibfnamefont {Sam}\ \bibnamefont {Gutmann}},\
  }\href {\doibase 10.48550/ARXIV.2004.09002} {\enquote {\bibinfo {title} {The
  quantum approximate optimization algorithm needs to see the whole graph: A
  typical case},}\ } (\bibinfo {year} {2020})\BibitemShut {NoStop}%
\bibitem [{\citenamefont {S\o{}rensen}\ and\ \citenamefont
  {M\o{}lmer}(1999)}]{sorensen1999quantum}%
  \BibitemOpen
  \bibfield  {author} {\bibinfo {author} {\bibfnamefont {Anders}\ \bibnamefont
  {S\o{}rensen}}\ and\ \bibinfo {author} {\bibfnamefont {Klaus}\ \bibnamefont
  {M\o{}lmer}},\ }\bibfield  {title} {\enquote {\bibinfo {title} {Quantum
  computation with ions in thermal motion},}\ }\href {\doibase
  10.1103/PhysRevLett.82.1971} {\bibfield  {journal} {\bibinfo  {journal}
  {Phys. Rev. Lett.}\ }\textbf {\bibinfo {volume} {82}},\ \bibinfo {pages}
  {1971--1974} (\bibinfo {year} {1999})}\BibitemShut {NoStop}%
\bibitem [{\citenamefont {Lanyon}\ \emph {et~al.}(2011)\citenamefont {Lanyon},
  \citenamefont {Hempel}, \citenamefont {Nigg}, \citenamefont {M{\"u}ller},
  \citenamefont {Gerritsma}, \citenamefont {Z{\"a}hringer}, \citenamefont
  {Schindler}, \citenamefont {Barreiro}, \citenamefont {Rambach}, \citenamefont
  {Kirchmair} \emph {et~al.}}]{lanyon2011universal}%
  \BibitemOpen
  \bibfield  {author} {\bibinfo {author} {\bibfnamefont {Ben~P}\ \bibnamefont
  {Lanyon}}, \bibinfo {author} {\bibfnamefont {Cornelius}\ \bibnamefont
  {Hempel}}, \bibinfo {author} {\bibfnamefont {Daniel}\ \bibnamefont {Nigg}},
  \bibinfo {author} {\bibfnamefont {Markus}\ \bibnamefont {M{\"u}ller}},
  \bibinfo {author} {\bibfnamefont {Rene}\ \bibnamefont {Gerritsma}}, \bibinfo
  {author} {\bibfnamefont {F}~\bibnamefont {Z{\"a}hringer}}, \bibinfo {author}
  {\bibfnamefont {Philipp}\ \bibnamefont {Schindler}}, \bibinfo {author}
  {\bibfnamefont {Julio~T}\ \bibnamefont {Barreiro}}, \bibinfo {author}
  {\bibfnamefont {Markus}\ \bibnamefont {Rambach}}, \bibinfo {author}
  {\bibfnamefont {Gerhard}\ \bibnamefont {Kirchmair}},  \emph {et~al.},\
  }\bibfield  {title} {\enquote {\bibinfo {title} {Universal digital quantum
  simulation with trapped ions},}\ }\href {\doibase DOI:
  10.1126/science.1208001} {\bibfield  {journal} {\bibinfo  {journal}
  {Science}\ }\textbf {\bibinfo {volume} {334}},\ \bibinfo {pages} {57--61}
  (\bibinfo {year} {2011})}\BibitemShut {NoStop}%
\bibitem [{\citenamefont {Werninghaus}\ \emph {et~al.}(2021)\citenamefont
  {Werninghaus}, \citenamefont {Egger}, \citenamefont {Roy}, \citenamefont
  {Machnes}, \citenamefont {Wilhelm},\ and\ \citenamefont
  {Filipp}}]{Werninghaus2020}%
  \BibitemOpen
  \bibfield  {author} {\bibinfo {author} {\bibfnamefont {Max}\ \bibnamefont
  {Werninghaus}}, \bibinfo {author} {\bibfnamefont {Daniel~J.}\ \bibnamefont
  {Egger}}, \bibinfo {author} {\bibfnamefont {Federico}\ \bibnamefont {Roy}},
  \bibinfo {author} {\bibfnamefont {Shai}\ \bibnamefont {Machnes}}, \bibinfo
  {author} {\bibfnamefont {Frank~K.}\ \bibnamefont {Wilhelm}}, \ and\ \bibinfo
  {author} {\bibfnamefont {Stefan}\ \bibnamefont {Filipp}},\ }\bibfield
  {title} {\enquote {\bibinfo {title} {Leakage reduction in fast
  superconducting qubit gates via optimal control},}\ }\href
  {https://doi.org/10.1038/s41534-020-00346-2} {\bibfield  {journal} {\bibinfo
  {journal} {npj Quantum Inf.}\ }\textbf {\bibinfo {volume} {7}} (\bibinfo
  {year} {2021})}\BibitemShut {NoStop}%
\bibitem [{\citenamefont {Wack}\ \emph {et~al.}(2021)\citenamefont {Wack},
  \citenamefont {Paik}, \citenamefont {Javadi-Abhari}, \citenamefont
  {Jurcevic}, \citenamefont {Faro}, \citenamefont {Gambetta},\ and\
  \citenamefont {Johnson}}]{Wack2021}%
  \BibitemOpen
  \bibfield  {author} {\bibinfo {author} {\bibfnamefont {Andrew}\ \bibnamefont
  {Wack}}, \bibinfo {author} {\bibfnamefont {Hanhee}\ \bibnamefont {Paik}},
  \bibinfo {author} {\bibfnamefont {Ali}\ \bibnamefont {Javadi-Abhari}},
  \bibinfo {author} {\bibfnamefont {Petar}\ \bibnamefont {Jurcevic}}, \bibinfo
  {author} {\bibfnamefont {Ismael}\ \bibnamefont {Faro}}, \bibinfo {author}
  {\bibfnamefont {Jay~M.}\ \bibnamefont {Gambetta}}, \ and\ \bibinfo {author}
  {\bibfnamefont {Blake~R.}\ \bibnamefont {Johnson}},\ }\href@noop {} {\enquote
  {\bibinfo {title} {Quality, speed, and scale: three key attributes to measure
  the performance of near-term quantum computers},}\ } (\bibinfo {year}
  {2021}),\ \Eprint {http://arxiv.org/abs/2110.14108} {arXiv:2110.14108
  [quant-ph]} \BibitemShut {NoStop}%
\bibitem [{\citenamefont {Pogorelov}\ \emph {et~al.}(2021)\citenamefont
  {Pogorelov}, \citenamefont {Feldker}, \citenamefont {Marciniak},
  \citenamefont {Postler}, \citenamefont {Jacob}, \citenamefont
  {Krieglsteiner}, \citenamefont {Podlesnic}, \citenamefont {Meth},
  \citenamefont {Negnevitsky}, \citenamefont {Stadler}, \citenamefont
  {H\"ofer}, \citenamefont {W\"achter}, \citenamefont {Lakhmanskiy},
  \citenamefont {Blatt}, \citenamefont {Schindler},\ and\ \citenamefont
  {Monz}}]{pogorelov2021compact}%
  \BibitemOpen
  \bibfield  {author} {\bibinfo {author} {\bibfnamefont {I.}~\bibnamefont
  {Pogorelov}}, \bibinfo {author} {\bibfnamefont {T.}~\bibnamefont {Feldker}},
  \bibinfo {author} {\bibfnamefont {Ch.~D.}\ \bibnamefont {Marciniak}},
  \bibinfo {author} {\bibfnamefont {L.}~\bibnamefont {Postler}}, \bibinfo
  {author} {\bibfnamefont {G.}~\bibnamefont {Jacob}}, \bibinfo {author}
  {\bibfnamefont {O.}~\bibnamefont {Krieglsteiner}}, \bibinfo {author}
  {\bibfnamefont {V.}~\bibnamefont {Podlesnic}}, \bibinfo {author}
  {\bibfnamefont {M.}~\bibnamefont {Meth}}, \bibinfo {author} {\bibfnamefont
  {V.}~\bibnamefont {Negnevitsky}}, \bibinfo {author} {\bibfnamefont
  {M.}~\bibnamefont {Stadler}}, \bibinfo {author} {\bibfnamefont
  {B.}~\bibnamefont {H\"ofer}}, \bibinfo {author} {\bibfnamefont
  {C.}~\bibnamefont {W\"achter}}, \bibinfo {author} {\bibfnamefont
  {K.}~\bibnamefont {Lakhmanskiy}}, \bibinfo {author} {\bibfnamefont
  {R.}~\bibnamefont {Blatt}}, \bibinfo {author} {\bibfnamefont
  {P.}~\bibnamefont {Schindler}}, \ and\ \bibinfo {author} {\bibfnamefont
  {T.}~\bibnamefont {Monz}},\ }\bibfield  {title} {\enquote {\bibinfo {title}
  {Compact ion-trap quantum computing demonstrator},}\ }\href {\doibase
  10.1103/PRXQuantum.2.020343} {\bibfield  {journal} {\bibinfo  {journal} {PRX
  Quantum}\ }\textbf {\bibinfo {volume} {2}},\ \bibinfo {pages} {020343}
  (\bibinfo {year} {2021})}\BibitemShut {NoStop}%
\bibitem [{\citenamefont {Schindler}\ \emph {et~al.}(2013)\citenamefont
  {Schindler}, \citenamefont {Nigg}, \citenamefont {Monz}, \citenamefont
  {Barreiro}, \citenamefont {Martinez}, \citenamefont {Wang}, \citenamefont
  {Quint}, \citenamefont {Brandl}, \citenamefont {Nebendahl}, \citenamefont
  {Roos} \emph {et~al.}}]{schindler2013quantum}%
  \BibitemOpen
  \bibfield  {author} {\bibinfo {author} {\bibfnamefont {Philipp}\ \bibnamefont
  {Schindler}}, \bibinfo {author} {\bibfnamefont {Daniel}\ \bibnamefont
  {Nigg}}, \bibinfo {author} {\bibfnamefont {Thomas}\ \bibnamefont {Monz}},
  \bibinfo {author} {\bibfnamefont {Julio~T}\ \bibnamefont {Barreiro}},
  \bibinfo {author} {\bibfnamefont {Esteban}\ \bibnamefont {Martinez}},
  \bibinfo {author} {\bibfnamefont {Shannon~X}\ \bibnamefont {Wang}}, \bibinfo
  {author} {\bibfnamefont {Stephan}\ \bibnamefont {Quint}}, \bibinfo {author}
  {\bibfnamefont {Matthias~F}\ \bibnamefont {Brandl}}, \bibinfo {author}
  {\bibfnamefont {Volckmar}\ \bibnamefont {Nebendahl}}, \bibinfo {author}
  {\bibfnamefont {Christian~F}\ \bibnamefont {Roos}},  \emph {et~al.},\
  }\bibfield  {title} {\enquote {\bibinfo {title} {A quantum information
  processor with trapped ions},}\ }\href {\doibase
  https://doi.org/10.1088/1367-2630/15/12/123012} {\bibfield  {journal}
  {\bibinfo  {journal} {New Journal of Physics}\ }\textbf {\bibinfo {volume}
  {15}},\ \bibinfo {pages} {123012} (\bibinfo {year} {2013})}\BibitemShut
  {NoStop}%
\end{thebibliography}%
\end{document}